\documentclass[pdftex]{pasj01} 
\usepackage[dvipdfmx]{}
\usepackage{pslatex} 
\usepackage{ulem}

\Received{2020/06/02}
\Accepted{}
 
\usepackage{color}
\usepackage{lscape}
\usepackage{enumerate}
\newcommand{\twelveco}{\mbox{$^{12}$CO}} 
\newcommand{\thirteenco}{\mbox{$^{13}$CO}} 
\newcommand{\ceighteeno}{\mbox{C$^{18}$O}} 
\newcommand{\twelvecoh}{\mbox{$^{12}$CO($J$=2--1)}} 
\newcommand{\twelvecohh}{\mbox{$^{12}$CO($J$=3--2)}} 
\newcommand{\thirteencoh}{\mbox{$^{13}$CO($J$=2--1)}} 
\newcommand{\twelvecol}{\mbox{$^{12}$CO($J$=1--0)}}

\newcommand {\msun}{\mbox{$M_\odot$}}
\newcommand {\lsun}{\mbox{$L_\odot$}}
\newcommand {\kms}{\mbox{km~s$^{-1}$}}
\newcommand {\kkms}{\mbox{K~km~s$^{-1}$}}

\newcommand {\nhtwo}{\mbox{$N_\mathrm{H_2}$}}

\newcommand {\hone}{\mbox{H{\sc i}}}
\newcommand {\htwo}{\mbox{H{\sc ii}}}

\newcommand {\ctwo}{\mbox{[C{\sc ii}]}}

\begin{document} 

\title{Cloud-cloud collisions and triggered star formation}

\author{Yasuo \textsc{Fukui}\altaffilmark{1}, Asao \textsc{Habe}\altaffilmark{2}, Tsuyoshi \textsc{Inoue}\altaffilmark{1}, Rei \textsc{Enokiya}\altaffilmark{1,3}, and Kengo \textsc{Tachihara}\altaffilmark{1}}
\altaffiltext{1}{Department of Physics, Graduate School of Science, Nagoya University, Furo-cho Chikusa-ku Nagoya, 464-8602, Japan}
\altaffiltext{2}{Faculty of Science, Department of Physics, Hokkaido University, Kita 10 Nishi 8 Kita-ku, Sapporo, Hokkaido 060-0810, Japan}
\altaffiltext{3}{Department of Physics, Faculty of Science and Technology, Keio University, 3-14-1 Hiyoshi, Kohoku-ku, Yokohama, Kanagawa 223-8522, Japan}
\email{fukui@a.phys.nagoya-u.ac.jp, enokiya@keio.jp}


\maketitle

\begin{abstract}
 Star formation is a fundamental process for galactic evolution. One issue over the last several decades has been determining whether star formation is induced by external triggers or is self-regulated in a closed system. 
The role of an external trigger, which can effectively collect mass in a small volume, has attracted particular attention in connection with the formation of massive stellar clusters, which in the extreme may lead to starbursts.
Recent observations have revealed massive cluster formation triggered by cloud-cloud collisions in nearby interacting galaxies, including the Magellanic system and the Antennae Galaxies as well as almost all well-known high-mass star-forming regions such as RCW~120, M20, M42, NGC~6334, etc., in the Milky Way.
Theoretical efforts are laying the foundation for the mass compression that causes massive cluster/star formation. Here, we review the recent progress on cloud-cloud collisions and triggered star-cluster formation and discuss the future prospects for this area of research.
\end{abstract}

\section{Introduction}\label{intro}
Star formation is one of the most fundamental processes in galactic evolution.
Most investigations of star formation have focused on self-regulated star formation in a turbulent molecular cloud (for reviews, see, e.g., \cite{mck07,zin07}), and theories based on turbulent molecular clouds explain the observed properties of star formation, including the stellar initial mass function (IMF; see e.g., \cite{kru09,mck03,bon98}).
However, It is not clear if the theories can explain the mechanism responsible for very active star formation, like the starbursts. 

The mechanism of high-mass star formation has been an issue of keen interest in the last few decades.
Infrared dark clouds---including hub-filament systems with moderate high-mass star formation \citep{per13} or other dense cloud cores---are considered promising candidates for high-mass star formation (e.g., \cite{men05}).
Theoretical studies of high-mass star formation have considered two possibilities.
One is a self-gravitating, massive star-cloud system, which has been shown to form a rich stellar cluster of 2000 $\msun$ similar to the Orion Nebula Cluster (ONC).
Sophisticated numerical techniques incorporating feedback have been employed (e.g., \cite{bon03,dal15}), and it has been demonstrated that the inhomogeneous cloud evolution with multiple stellar condensations can lead to the formation of a rich cluster in a few million years (Myr).
However, it remains to be explained how single, isolated O-stars with smaller system mass are formed, since such O-stars are numerous in the Galaxy \citep{asc18}.
The other possibility is a massive, compact cloud core that contains 100 $\msun$ within a 0.1 pc radius, which has a mass column density around 1 g~cm$^{-2}$.
Numerical simulations with these initial conditions have successfully demonstrated that such a cloud core can lead to the formation of two $\sim$30 $\msun$ O-stars \citep{kru09}.
 Subsequently, similar simulations for a more massive cloud of 1000 $\msun$, which adopted the initial condition with a mass column density of 1 g~cm$^{-2}$, showed the formation of a cluster similar to the ONC \citep{kru12}. 
An issue that still remains to be addressed is how such a high column density is produced, since the core/cloud-formation process---which may form numerous low-mass stars prior to high-mass star formation---is beyond the scope of the simulations.

Recent observational studies have provided evidence that cloud-cloud collisions (CCCs) trigger high-mass star formation in the Milky Way.
Table~1 lists more than 50 high-mass star-forming regions for which observational evidence of high-mass star formation triggered by a CCC has been reported.
They include major $\htwo$ regions like M42 and M17 in the solar neighborhood, the young massive cluster R136 in the Large Magellanic Cloud (LMC) and the massive open cluster NGC~604 in M33, as well as the massive (candidate) clusters in the Antennae Galaxies.
These objects suggest the important role of CCCs in forming massive clusters as well as isolated high-mass stars.

The aim of this article is to summarize recent observational and theoretical results on CCCs and discuss the role of the collision process in star formation, with an emphasis on high-mass star formation.
The article is organized as follows.
Section 2 describes theories of CCCs, including historical background, and Section 3 discusses the essential observational signatures and statistical properties of CCCs.
Section 4 presents the theories of the compression layer in a CCC, together with the relevant physical processes, with an emphasis on the magnetic field.
Section 5 describes individual CCCs, including massive clusters and galactic interactions.
Section 6 summarizes the article.

\section{Theories of CCCs}
\subsection{Historical Background}
Triggered star formation has been reviewed previously, covering several mechanisms including CCCs, and $\htwo$-driven compression (e.g., \cite{elm98}).
Around 1940, interstellar absorption lines at optical wavelengths were discovered, and this opened the door to investigations of discrete interstellar clouds.
From the velocity separations of the clouds, 15--20 $\kms$, \citet{oor54} estimated that such clouds are often undergo collisions.
This was the first indication of the important role of CCCs in determining the mass spectrum of interstellar clouds.
The first numerical simulations of colliding interstellar clouds were carried out by \citet{sto70a}.
The simulations were one-dimensional, since the computer power was very limited then.
\citet{sto70a} assumed two identical interstellar clouds of neutral hydrogen with densities of 5--10 H$_{atm}$~cm$^{-3}$, colliding at a speed of 10 $\kms$. These clouds were the standard model described in the textbook by \citet{spi68}. \citet{sto70a} and \citet{sto70b} showed that both colliding clouds are destroyed in the collision, since the collision speed is highly supersonic, and the duration of the collision is too short for the growth of self-gravitational instability.

In the 1970s, many molecular clouds were observed by millimeter-wave telescopes (e.g., \cite{dam01} and references therein, \cite{sol80}).
\citet{lor76} obtained observational evidence for a collision between molecular clouds in NGC~1333 and discussed the possible connection between the collision and high-mass star formation.
This work demonstrated that theoretical studies of CCCs needed to be extended to molecular clouds.
In the 1980s, supercomputers emerged, and \citet{smi80} numerically simulated one-dimensional colliding flows, including radiative gas cooling, dissociation, and the formation of molecules, and he showed that a dense molecular-gas layer is formed in a CCC.
\citet{smi80} also discussed the gravitational instability of the dense molecular gas layer and showed that a slower collision speed favors gravitational instability of that layer.
Two-dimensional numerical simulations of collisions between two dense molecular clumps were performed by \citet{gil84}.
He showed that dense, gravitationally unstable gas layers were formed by the collisions.
He discussed the conditions in which such dense layers are formed by CCCs, and he proposed a simple model to explain the formation of gravitationally unstable regions.
Three-dimensional simulations were carried out using smoothed particle hydrodynamics (SPH) (\cite{lat85}; \cite{lat88}; \cite{nag87}).
Due to the limits of supercomputer power in those days, the number of SPH particles was limited.
\citet{lat85} simulated head-on collisions of identical clouds and of non-identical clouds, and \citet{lat88} simulated off-center collisions of clouds.
Since the number of SPH particles was at most 2000, these simulations were limited to small clouds.
\citet{nag87} simulated collisions of identical clouds with a larger number of SPH particles than \citet{lat85} and \citet{lat88}, assuming the clouds to be isothermal and in gravitational equilibrium before the collision.
By employing 4000 and 8000 SPH particles, they also found that a gravitationally unstable layer is formed in a CCC.
In spite of these efforts, there were only a few attempts to test these theories by observations of the molecular clouds until the 2010s.

\citet{hab92} extended SPH simulations to head-on collisions between non-identical clouds, which is more realistic for collisions that have a wide range of sizes.
Their goal was to investigate whether oblique shock waves between non-identical clouds can create a converging flow, which would help to form massive dense cores in the shocked layer efficiently.
By using an axisymmetric SPH code, they achieved the high spatial resolution needed to resolve a gravitationally unstable massive core.
They showed successfully that non-identical clouds facilitate the formation of a gravitationally unstable massive core, as compared with the identical clouds. 

Subsequently, numerical simulations of collisions of non-identical clouds were performed by \citet{ana10}, \citet{tak14}, \citet{tak18}, and \citet{shi18}.
\citet{ana10} elucidated the role of the non-linear thin-shell instability in the formation of dense clumps in colliding clouds.
\citet{tak14} and \citet{tak18} demonstrated the important roles of internal turbulence in the pre-collision clouds and of the collision speed.
Prior to these studies, \citet{kim96} had simulated collisions of identical clouds, assuming a turbulent, clumpy cloud distribution, and they showed that many dense clumps are formed by a CCC, but they did not take self-gravity into account.
The simulations by \citet{tak14} and \citet{tak18}, which did incorporate self-gravity, predicted that the dense cores formed by a CCC have a power-law mass function.
\citet{shi18} studied star formation in colliding clouds, and they included UV feedback from the newly formed stars by employing a sink-particle model developed by \citet{fed10}.
They showed that the UV feedback triggers star formation by compressing the gas around the sink particles.
In a high-velocity CCC, the magnetic field plays a crucial role in the compressed layer created by the CCC.
\citet{ino13} and \citet{ino18} performed high-resolution, three-dimensional MHD simulations and concluded that the magnetic field---amplified by shock compression in the CCC---can explain massive-core/star formation in the collision interface, as we discuss in more detail in Section 4.
A number of other numerical simulations elaborating CCCs have been performed more recently (e.g., \cite{bis17,kob18,li18,whi18,wu18,sak20,dob20,lio20}).
These recent studies are used for comparison with recent observations of CCCs.

In this paper, we focus particularly on the role of ``head-on'' and ``fast'' molecular-cloud collisions in high-mass star formation.
This is because many high-mass star-forming molecular clouds have consistently been interpreted as sites of ``fast'' gas collisions, with relative velocities of $\sim$10 $\kms$.
In addition, the molecular-gas structures around high-mass stars can be well explained by the results of the ``face-on'' CCC simulations by \citet{hab92} and \citet{tak14}.
Note that slow CCCs also induce star formation to some extent, while oblique CCCs may have negative effects on star formation due to the induction of shear flows.
However, we do not discuss slow CCCs or oblique CCCs in detail in this review, because observations suggest that fast and head-on CCCs are the ones associated with high-mass star formation.
We only note that there is a case of an oblique collision where the formation of an early B star may have been triggered (e.g., NGC2068 and NGC2071, \cite{fuj20a}).

\subsection{``Complementary Distribution'' and ``Bridge'' as Observational Signatures of a CCC}
Here we discuss the distribution and kinematics of the gas in a CCC based on numerical simulations and summarize the observable signatures of a CCC.
Collisions between two clouds of exactly the same size are expected to be very rare.
Numerical simulations of CCCs by \citet{hab92}, \citet{ana10}, \citet{tak14}, and \citet{tak18} assume that a small spherical cloud collides head-on with a large spherical cloud; hereafter, we call this the ``Habe-Ohta model''.
In these simulations self-gravity is included, but the magnetic field is not taken into account.

\begin{figure*}
 \begin{center}
  \includegraphics[width=16cm]{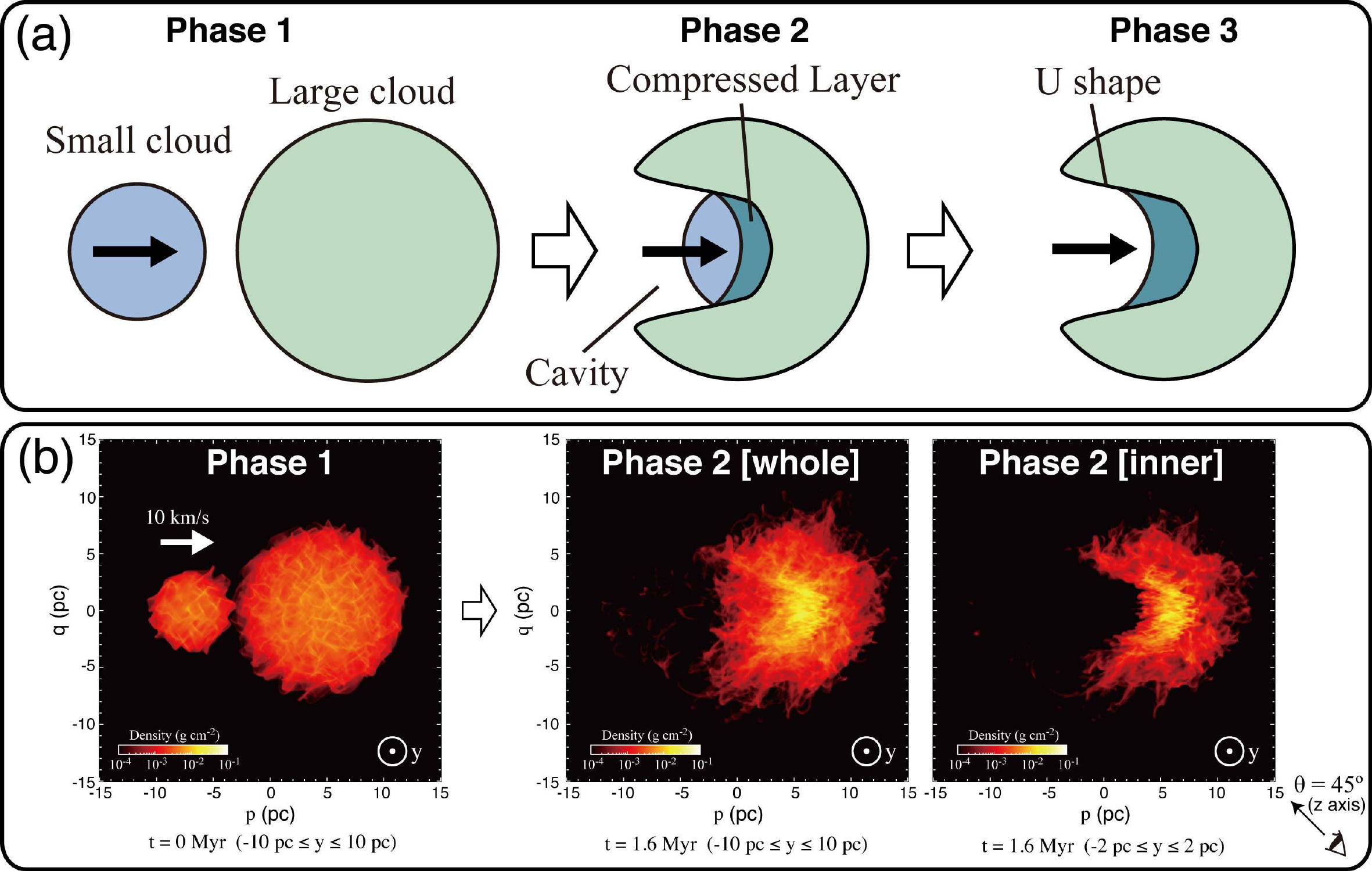} 
 \end{center}
\caption{(a) Schematic picture of three evolutionary epochs of the Habe-Ohta model of a cloud-cloud collision between two spherical clouds with different sizes. When the small cloud drives into the large cloud, a U-shaped cavity is created in the large cloud, and the small cloud streams into the compressed layer formed at the collision interface. (b) Surface density plots of the 10 $\kms$ collision model calculated by \citet{tak14}. The leftmost panel shows the top view of the two clouds prior to the collision (Phase 1), while the middle and right panels show snapshots at 1.6 Myr after the onset of the collision (Phase 2). The integration ranges in the $y$ axis are $-$10 pc to $+$10 pc for the middle panel and $-$2 pc to $+$2 pc for the right panel. The eye symbol and arrow define the viewing angle used in analyzing the synthetic $\twelvecol$ data presented in Figure~2. Figures are adapted from \citet{fuk18a} and reproduced with permission from AAS.}
\label{HO}
\end{figure*}

Figure~1a shows schematic drawings of a side view of the Habe-Ohta model at three epochs, based on the numerical simulations by \citet{tak14}: the initial phase (Phase 1), an intermediate phase (Phase 2), and the final phase (Phase 3).
First, in Phase 1 a small cloud and a large cloud are approaching each other.
Once a collision occurs, a compressed layer is formed at the interface layer between the two clouds, creating a U-shaped cavity in the large cloud in Phase 2.
The diameter of the cavity is nearly equal to that of the small cloud.
The gas in the two clouds streams into the compressed layer during the collision.
Finally in Phase 3, the small cloud has fully merged into the compressed layer, while the U-shaped cavity in the large cloud remains open in the direction of incidence of the small cloud.
The strongest compression takes place in the compressed layer inside the U shape, where dense cores and star(s) will be formed by gravitational instability if the gas column density of the compressed layer becomes large enough.

Figure~1b shows the simulation results in a top view of the projected density distribution in $p$-$y$-$q$ coordinates, where the collision happens in the $p$-$q$ plane, with the collision direction along the $q$ axis.
The simulation parameters are given in Table~2.
Note that the simulations take into account realistic turbulence and density inhomogeneities in the initial clouds.
The initial collision velocity of 10 $\kms$ between the two clouds is supersonic for a cloud sound speed less than 1 $\kms$.
In the course of the collision the small cloud is decelerated by the collisional interaction, and the collision velocity decreases to $\sim$7 $\kms$ at 1.6 Myr.
Phase 1 is prior to the onset of the collision, when the time is 0 Myr.
Phase 2 is at an elapsed time of 1.6 Myr, by which time the small cloud has penetrated into the large cloud and created a U-shaped cavity in the large cloud.
Two images are shown for Phase 2: Phase 2 [whole] shows all of both clouds, while Phase 2 [inner] shows those parts of the clouds only within $y$=$\pm$2 pc in the $p$-$q$ plane. Phase 2 [inner] shows the cavity more clearly than Phase 2 [whole].

In order to understand the details of the observed velocity distributions of the colliding clouds, synthetic observations of the $\twelvecol$ emission were made \citep{fuk18a,fuk18b,tor17a}, and they are presented as velocity-channel distributions.
Generally, colliding clouds are observed along a line of sight at an angle $\theta$ to the collision direction.
In Figures 2b--2i we show the cloud distributions at 1.6 Myr, corresponding to Phase 2, for an angle $\theta$=45 degrees as a typical case.
The $x$-$y$-$z$ coordinates used in Figure~2 are defined by rotating the $p$-$q$ plane of Figure~1b counterclockwise by 45 degrees (see also Figure~2a).
At the low velocities of $\sim$$-$5 $\kms$ to $-$2 $\kms$, only the small cloud is observed (Figures 2b, 2c, and 2d), while at the high velocities of $\sim$0 $\kms$ to $+$2 $\kms$, only the large cloud is observed (Figures 2g, 2h, and 2i).
However, we also find emission at the intermediate velocities of $\sim$$-$2 $\kms$ to 0 $\kms$ between the two clouds (Figures 2e and 2f).
The cavity in the large cloud is clearly seen in Figure 2g around $x$$\sim$2 pc.
The cavity is displaced by $\sim$3 pc from the peak of the small cloud located at $x$$\sim$5 pc, as shown in an overlay of the two clouds (Figure 2i).
Although it varies within the clouds, a typical CO line profile often shows a single skewed peak  (Figure 2j).
The position-velocity diagram shows a single broad cloud spanning $\sim$5 $\kms$ in a V-shape (Figure 2k), whereas it initially consisted of two discrete clouds.
The single broad cloud is due to the intermediate-velocity gas produced by the collisional interaction. This is the bridge feature discussed as a characteristic feature of a CCC (e.g., \cite{haw15}; Figure 10--14 of \cite{tor17a}).

\begin{figure*}
 \begin{center}
  \includegraphics[width=16cm]{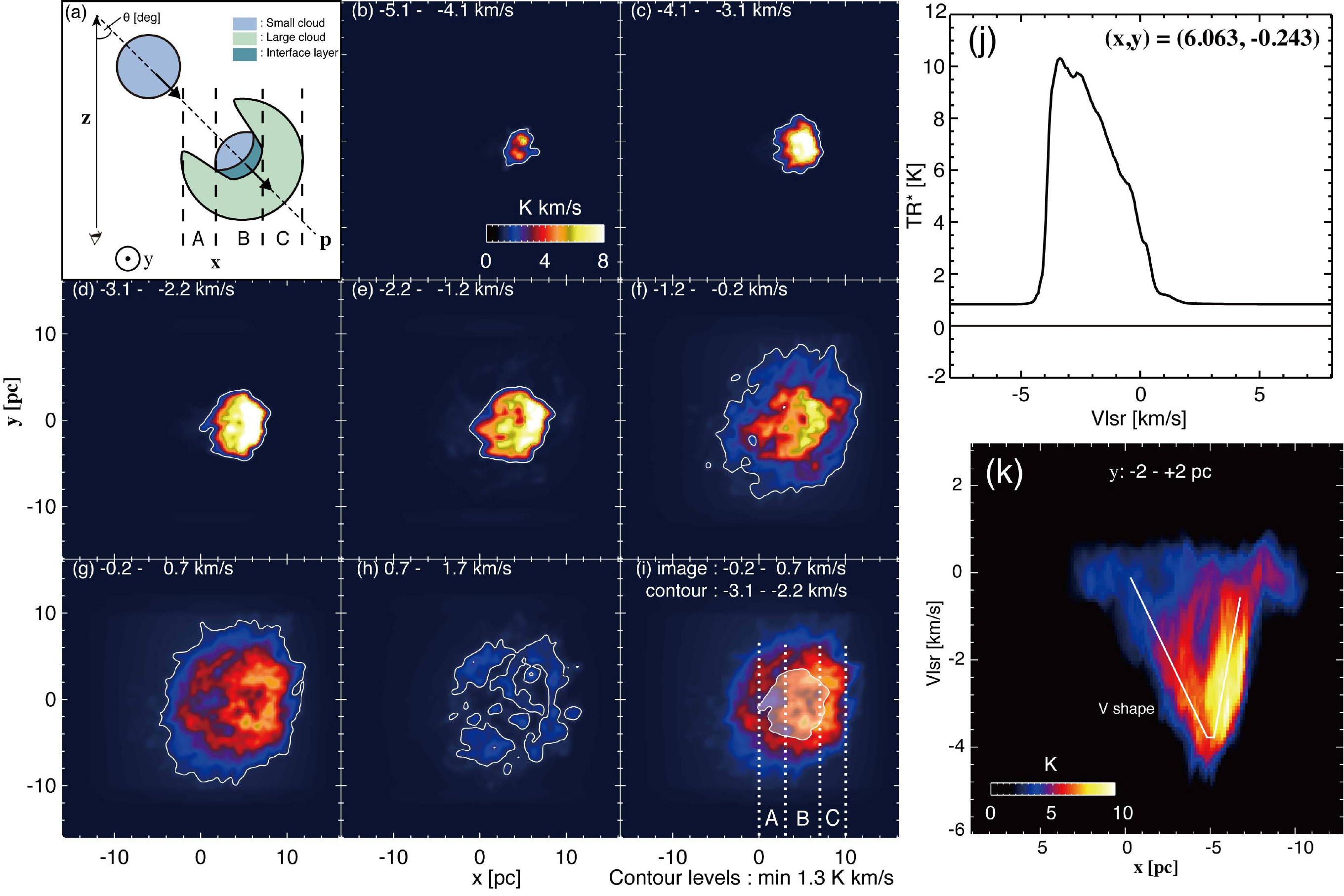} 
 \end{center}
\caption{Synthetic observations of $\twelvecol$ emission based on the numerical simulations by \citet{tak14} observed at an angle of the relative motion to the line of sight $\theta$ = 45$^{\circ}$ (see the rightmost panel of Figure~1b). Panel (a) shows the definition of the $x$-$y$-$z$ coordinates used to generate the synthetic data. Panels (b)--(h) show the velocity-channel distributions at intervals of 0.93 $\kms$ in velocity. Panel (i) shows the complementary distribution between the large cloud---the image in panel (g)---and the small cloud with the contour in panel (c) at 4 $\kkms$. Panel (j) shows a spectrum toward the interface layer and panel (k) shows the position-velocity diagram integrated over the $y$ range of $-$2 pc to $+$2 pc. Figures are adapted from \citet{fuk18a} and reproduced with permission from AAS.}
\label{synth}
\end{figure*}

We pay particular attention to three characteristic features of a CCC, which are summarized as follows:

\begin{enumerate}[i)]
 \item \uline{Complementary distribution with displacement}: The simulations show that the cavity formed in the large cloud has a density distribution that is complementary to that of the small cloud in three-dimensional space. We find such a complementary distribution, with a displacement between the cavity and the small cloud (Figure 2j). The displacement is due to the projection effect, and it disappears if $\theta$$\sim$0 degree. An algorithm has been developed to optimize the displacement to fit the complementary distribution in a CCC \citep{fuj20b}. 
 \item \uline{Bridge}: The simulations show intermediate-velocity gas formed by the collisional interaction, which makes the two initial clouds appear as a single-peaked continuous cloud. The bridge sometimes has a V-shape, with the small cloud at the tip of the ``V'' and two bridges linking it to the large cloud.
 \item \uline{U shape}: The U-shaped cavity in Phase 2 and Phase 3 is clearly presented by \citet{hab92}, as reproduced in Figure~1 (Phase 2 and Phase 3). The U shape results from the directed compression of the large cloud caused by the small cloud, and it is characterized by a highly directed density distribution that is densest at the bottom of the U shape.
\end{enumerate}

\begin{figure}
 \begin{center}
  \includegraphics[width=8cm]{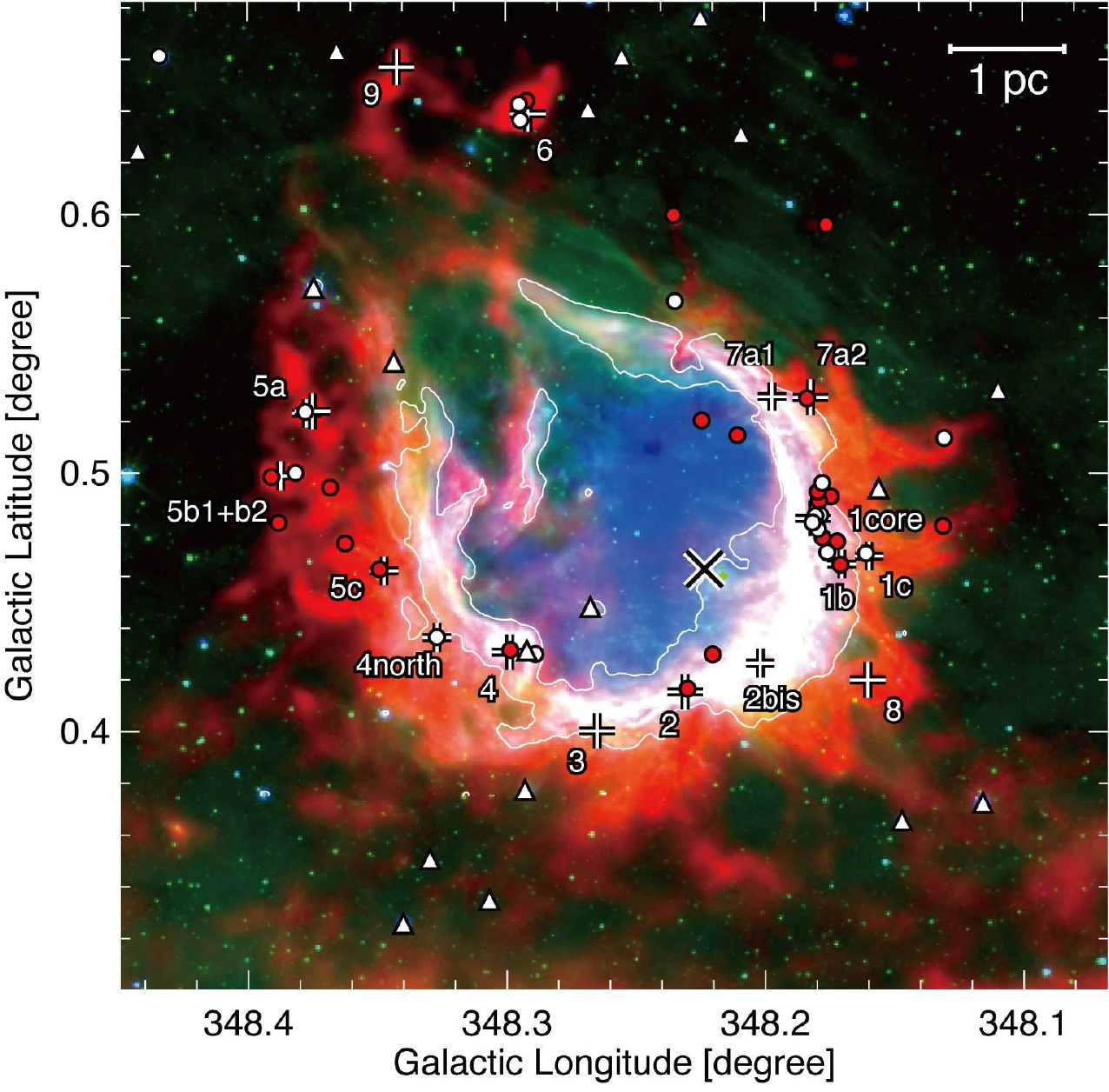} 
 \end{center}
\caption{Color composite image of RCW~120. Green, blue, and red show the Spitzer/IRAC 8 $\mu$m data \citep{ben03}, Spitzer/MIPS 24 $\mu$m data \citep{car09}, and Herschel/SPIRE 250 $\mu$m data \citep{zav10}. The large cross indicates the position of the exciting star, and the filled red circles, filled white circles, and filled white triangles indicate the positions of Class I, intermediate Class I-Class II or flat-spectrum, and Class II YSOs identified by \citet{deh09}. The small crosses and labels indicate the cold-dust condensations identified from 870 $\mu$m observations by \citet{deh09}. The white contours show the outline of the 8 $\mu$m ring, where the Spitzer 8 $\mu$m image is median-filtered with a 9$\arcsec \times 9\arcsec$ window. Adapted from \citet{tor15} with permission from AAS.}
\label{120}
\end{figure}


\section{Observed Properties of CCCs}

We summarize the main observational properties of CCC candidates by highlighting a few representative objects.
In addition, the statistical properties of CCCs are presented in Table~1 based on more than 50 CCC samples.
More detailed observational data are described in Section 5.

\subsection{Characteristic Signatures of a CCC}

The CCC candidates are often observed in millimetric CO emission, mainly toward $\htwo$ regions or reflection nebulae, where OB stars including OB-star candidates have already formed (Table~1).
These observations indicate that high-excitation molecular lines are not useful for tracing a CCC.
However, the CO emission, which samples a wide range of gas densities, is useful as an overall tracer of a CCC.
The high optical depth of the $\twelveco$ emission sometimes masks the collision signatures, so isotopic CO molecules are employed for various column-density ranges: $\twelveco$ for $\nhtwo$ $<$ 10$^{22.5}$ cm$^{-2}$, $\thirteenco$ for 10$^{22.5}$ cm$^{-2}$ $\le$ $\nhtwo$ $\le$ 10$^{23.0}$ cm$^{-2}$, and $\ceighteeno$ for $\nhtwo$ $>$ 10$^{23.0}$ cm$^{-2}$.
Other tracers, like $\ctwo$ may also be useful as tracers of CCCs \citep{haw18}, although their spatial coverage is limited.

The Habe-Ohta model of a CCC shows that several observational signatures are useful for identifying and characterizing a CCC: i) the displaced complementary distributions of the two clouds, ii) the bridge feature linking them, and iii) the U-shaped cavity.
In order to illustrate these signatures, we describe below three objects that have been formed by CCCs.

\subsubsection{RCW~120: A ``U-shaped Cavity'' and Evidence for Directional Gas Compression}
RCW~120 is an $\htwo$ region ionized by a single O-star (Figure~3).
It shows a clear shell-like shape, and the usual interpretation is that it is an $\htwo$ region bubble driven by the O-star ionization \citep{deh05}.
\citet{hos05} showed that gas is compressed by expanding $\htwo$ gas and becomes, gravitationally unstable, if the magnetic field is not taken into account.
However, it has been questioned whether such a bubble can be formed by the O-star, because compression by the $\htwo$ gas turns out to be difficult, due to the magnetic pressure acting against the compression, as shown by recent magnetohydrodynamic simulations \citep{inu15}.

\begin{figure*}
 \begin{center}
  \includegraphics[width=16cm]{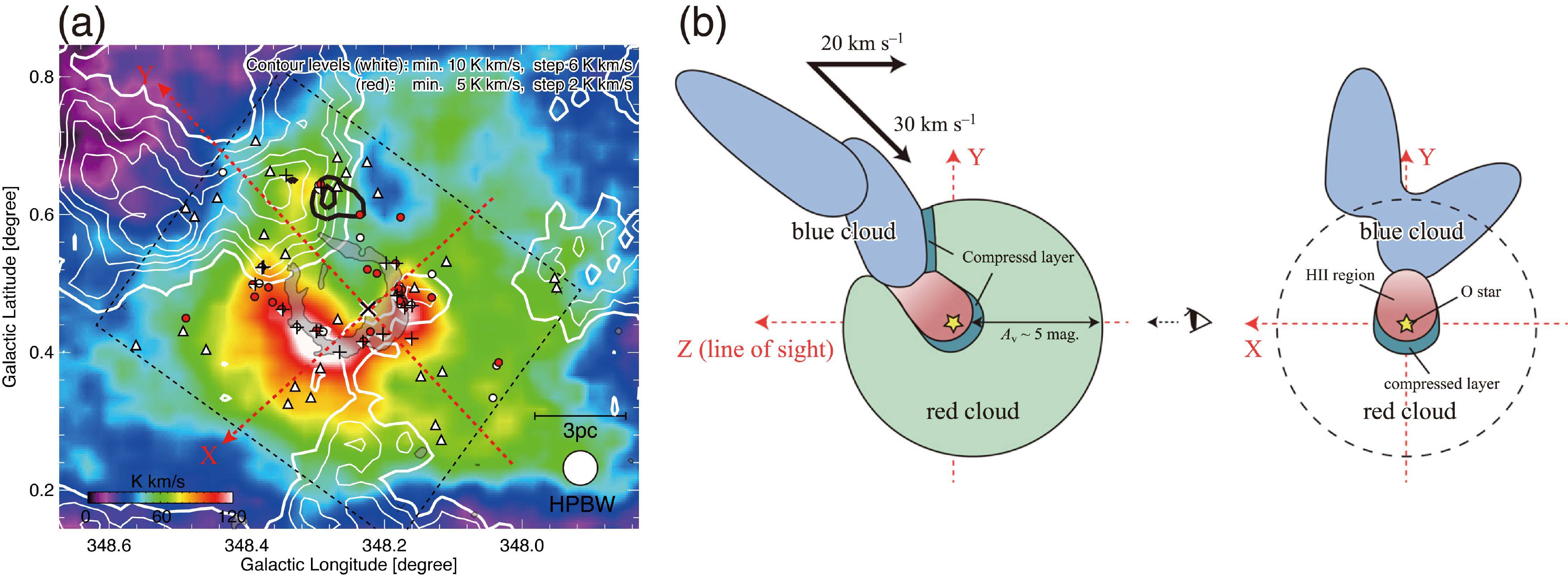} 
 \end{center}
\caption{(a) A comparison of the red cloud (image) and the blue cloud (white contours) associated with RCW~120, as observed with the NANTEN2 telescope in $\twelvecol$. The bridging feature at $-$23 $\kms$ to $-$20 $\kms$ is plotted with thick black contours. The YSOs, dust condensations, and 8$\mu$m ring are plotted in the same manner as in Figure~3, where the region that was used for the YSO identifications is shown by black dashed lines. (b) Schematic illustrations of the evolution of the CCC in RCW~120 as presented in the $Z$--$Y$ plane and the $X$--$Y$ plane, where the $Z$-axis is along the line of sight. The origin of the coordinate system is taken at the exciting O-star. Figures are adapted from \citet{tor15} and reproduced with permission from AAS.}
\label{120sche}
\end{figure*}

\citet{tor15} made a detailed analysis of the CO data and found two CO clouds with a velocity separation of 20 $\kms$, as shown in Figure~4. They therefore proposed the alternative scenario that a CCC formed the bubble and triggered the formation of the O-star in the collisionally compressed layer.
If the bubble were driven by the $\htwo$ region, it would be symmetric; however, it is actually asymmetric, showing a U shape open toward the northeast in the galactic coordinates (Figure~4).
Further, the location of the O-star is not at the center of the bubble, but instead is in the southwestern side close to the bottom of the U shape.
The asymmetric morphology is consistent with a collisionally created, U-shaped cavity, with the small cloud having collided from the northeast.
The collision timescale is estimated to be $\sim$0.15 Myr for $\theta$ = 45 degrees, from the ratio of the 3 pc bubble size to the 20 $\kms$ velocity separation.
They suggested that the small cloud collided from the opening side of the bubble, and the remnant of the small cloud is found outside RCW~120 \citep{tor15}.
According to this interpretation, RCW~120 shows the typical U-shape of Phase 3 in the Habe-Ohta model.
The model also explains the formation of the O-star, which is not explained in the $\htwo$-driven bubble scenario.
Similar U shapes are found in other CCC-candidate bubbles, including RCW~79 \citep{oha18a}, S44 \citep{koh18b}, S36 \citep{tor17b}, etc., as shown in Figure|~5.
It is remarkable that all of these objects show anisotropic density distributions as traced by CO, which is consistent with the U-shaped cavity.

\begin{figure*}
 \begin{center}
  \includegraphics[width=16cm]{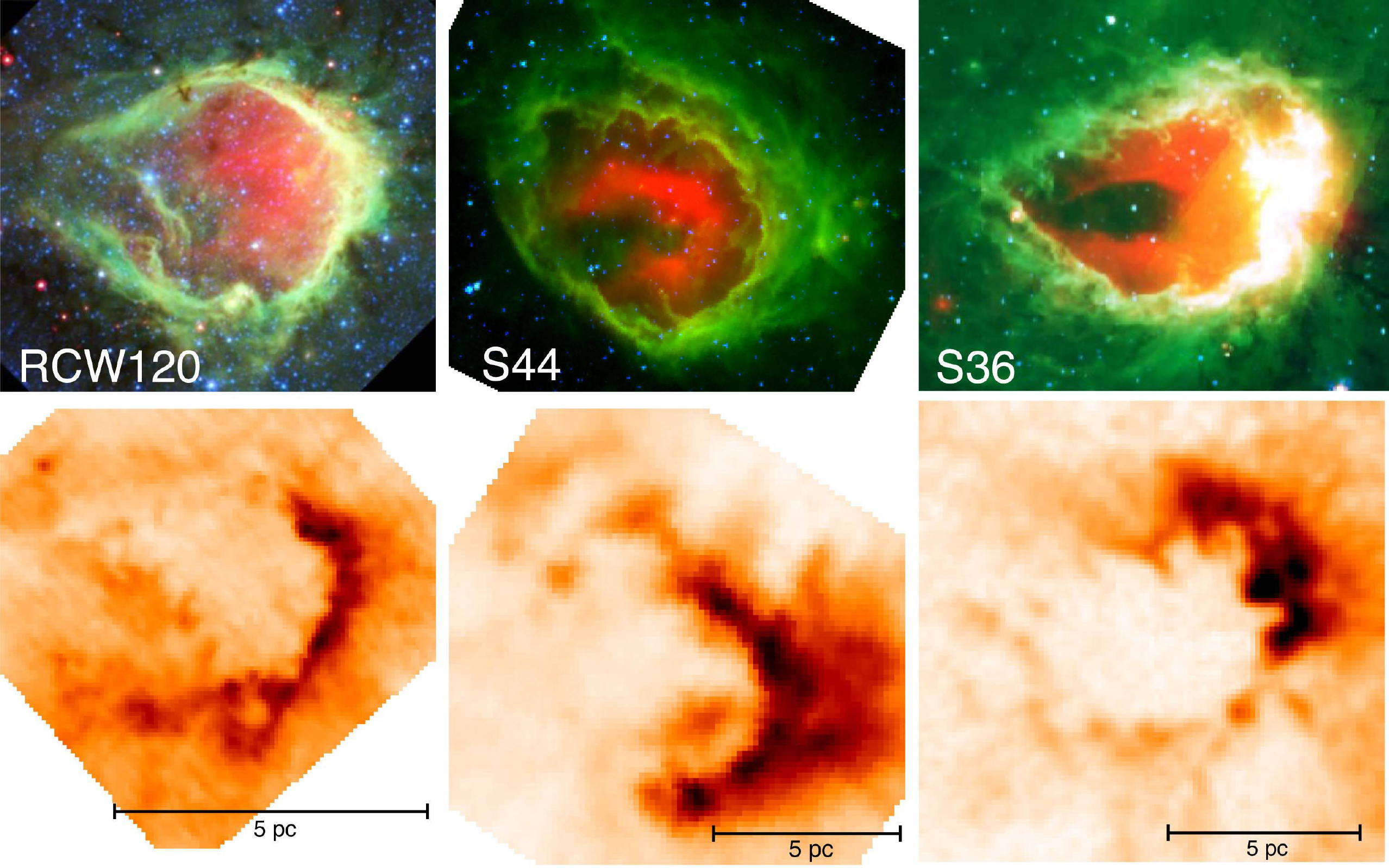} 
 \end{center}
\caption{Typical U-shaped bubbles seen in the Galactic disk at infrared wavelengths (upper panels; Three-color composite images with 24 $\mu$m in red, 8 $\mu$m in green, and 3.6 $\mu$m in blue) and at $\twelvecol$ emission (lower panels).}
\label{Ushapes}
\end{figure*}

\subsubsection{M43: ``Complementary Distribution and Displacement''}
By analyzing CO data for the Orion A cloud, \citet{fuk18a} presented a scenario in which two clouds with a projected velocity separation of 4 $\kms$ collided to trigger the formation of NU Ori, the B3 star exciting M43.
The small cloud in this case has a key-like shape, and the cavity in the large cloud has a keyhole shape that is complementary to the small cloud but displacement by 0.3 pc, as shown by the arrow in Figure~6.
Assuming an angle of roughly 30-60 degrees between the collision path and the line of sight, the collision timescale is estimated to be $\sim$0.1 Myr, with an uncertainty of a factor of $\sim$2.
The key-like cloud shape is unique, and the complementary distribution is remarkable.
A V-shape is also found in the position-velocity diagram, but the bridge was not found, due to the small velocity dispersion.

\begin{figure}
 \begin{center}
  \includegraphics[width=8cm]{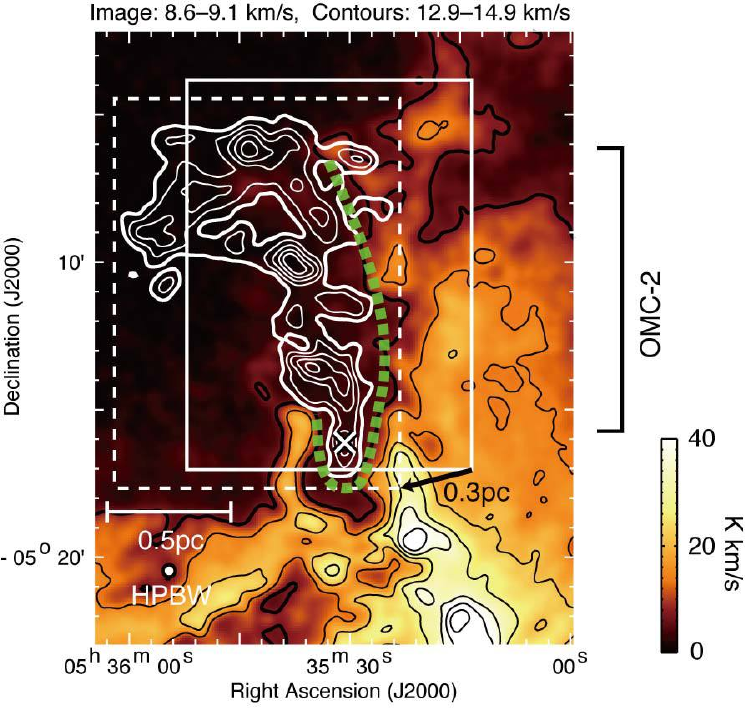} 
 \end{center}
\caption{Complementary distributions of the two clouds toward OMC-2. The image with black contours indicates the blue-shifted cloud and the white contours represent the red-shifted cloud. The black arrow indicates the displacement vector that provides the best complementary fit between the image and the white contours. The contact surface between the complementary distributions of the two clouds is indicated by the green dashed line. The velocity ranges for the blue-shifted and red-shifted clouds are 8.6--9.1 $\kms$ and 12.9--14.9 $\kms$, respectively. The lowest level and interval of the white contours are 13 $\kkms$ and 7 $\kkms$, while those of the black contours are 7 $\kkms$ and 7 $\kkms$. NU Ori [(R.A., decl.) = (5h35m31s, $-$5$^{\circ}$16$\arcmin$03$\arcsec$)] is plotted with a white cross. Figures are adapted from \citet{fuk18a} and reproduced with permission from AAS.}
\label{key}
\end{figure}

\begin{figure*}
 \begin{center}
  \includegraphics[width=16cm]{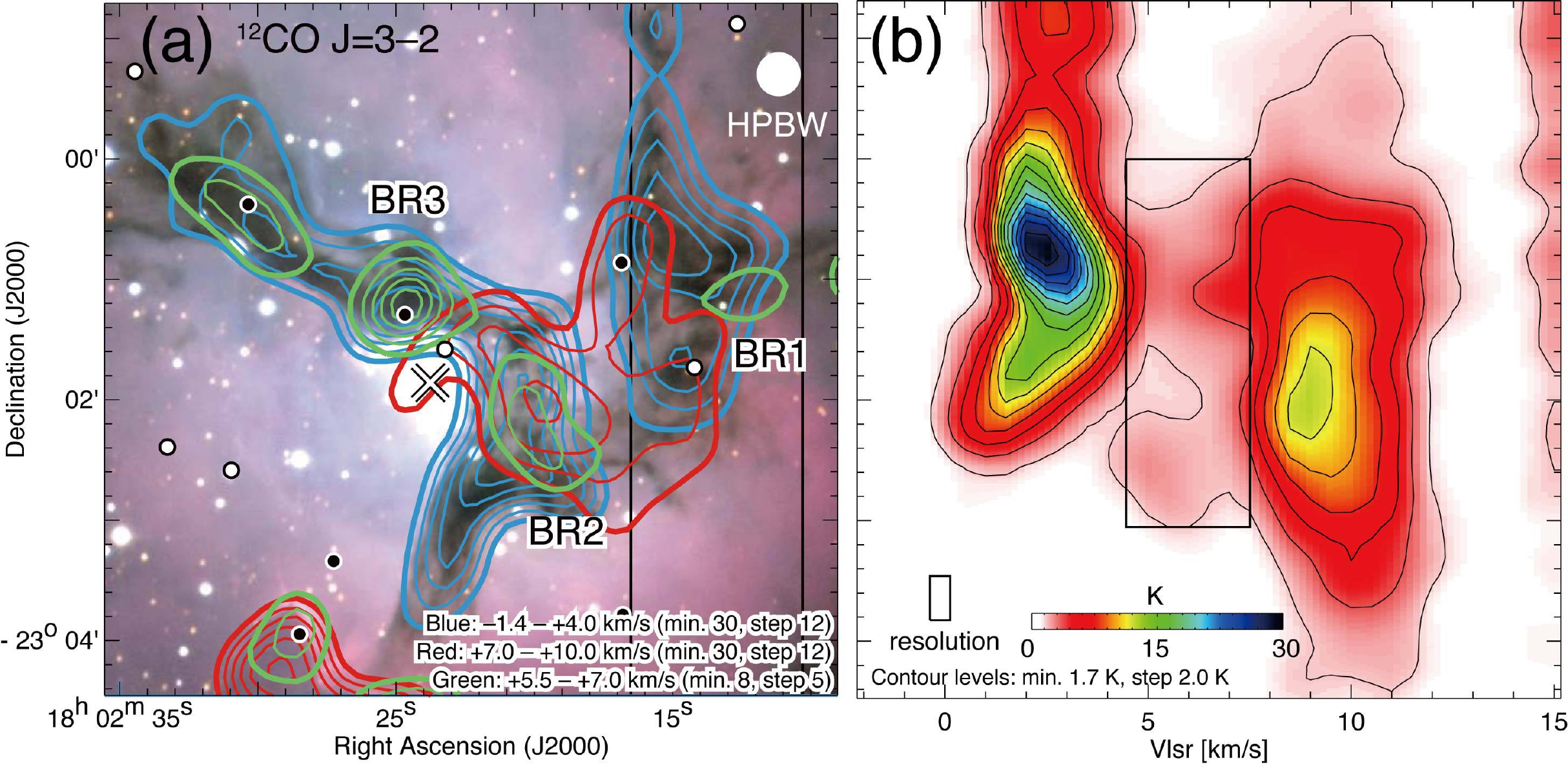} 
 \end{center}
\caption{(a) Contour maps of the two colliding clouds observed in $\twelvecohh$ are shown superimposed on an optical image of M20 (credit: NOAO). The exciting O7.5 star (HD~164492 A) is depicted by a cross, while class I/0 and class II young stars identified by the Spitzer color--color diagram \citep{rho06} are plotted with filled black circles and filled white circles, respectively. The 2 $\kms$ cloud and cloud C from \citet{tor17a} are plotted as blue contours and red contours, respectively. The bridge features BR1, BR2, and BR3 identified by \citet{tor17a}, are indicated by green contours. The velocity range and the contour levels are shown in the bottom right of the panel. The black lines indicate the integration ranges for the declination-velocity diagram shown in Figure~7b. (b) Declination-velocity diagram for the $\twelvecohh$ emission integrated over the ranges shown in Figure~7a with black lines. The bridge feature is indicated by the black rectangle. Figures are adapted from \citet{tor17a} and reproduced with permission from AAS.}
\label{M20}
\end{figure*}

\begin{figure*}
 \begin{center}
  \includegraphics[width=14cm]{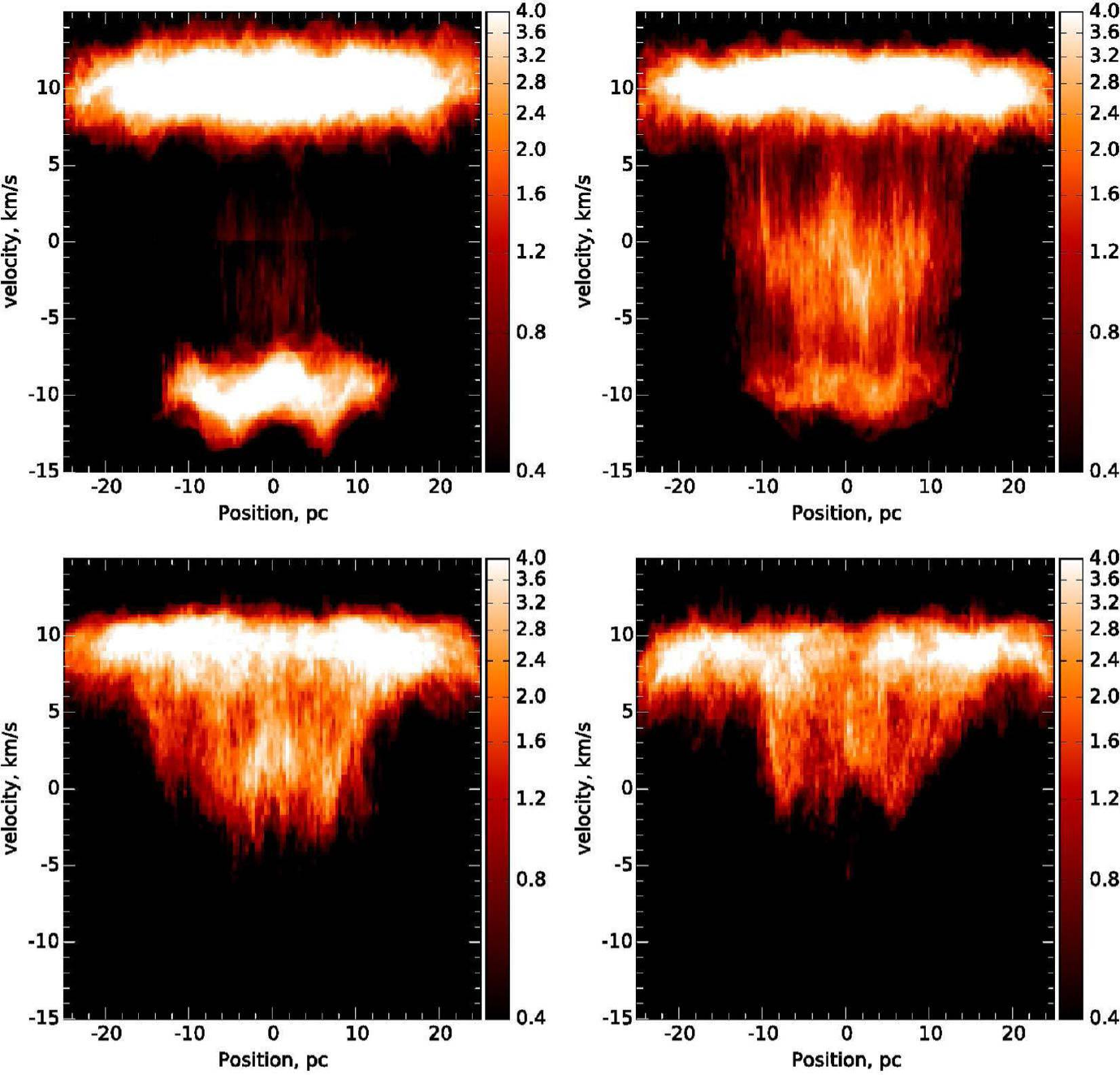} 
 \end{center}
\caption{$\twelvecol$ synthetic p-v diagrams of snapshots from the 20 $\kms$ collision model \citep{haw15}. The panels are for snapshots at 0.4 Myr (top left), 2 Myr (top right), 2.4 Myr (bottom left), and 4.4 Myr (bottom right). Adapted from \citet{haw15} with permission.}
\label{howth}
\end{figure*}

\subsubsection{M20: A ``Bridge'' Connecting the Colliding Clouds}
\citet{tor11} and \citet{tor17a} showed that the exciting star of M20 was formed by a CCC with a velocity separation of 8 $\kms$.
Distributions of the colliding red-shifted cloud, blue-shifted cloud, and the exciting star are shown in Figure~7a.
The separation is relatively large, and the two clouds are linked at 6 $\kms$ by diffuse emission, as shown in Figure~7b. This bridge feature is another observational signature of a CCC.
The collision path is nearly along the line of sight, making the projection effect small.
\citet{haw15} studied the physical properties of the bridge based on the numerical simulations by \citet{tak14}.
Figure~8 shows the time evolution of the bridge linking the two colliding clouds according to the numerical simulations.
It shows that the gas clouds exchange momentum with each other and that the interacting gas acquires velocities intermediate between those of the two clouds.
We note that bridges are not always observable, since the projection effect can reduce the observed velocity separation between the two clouds to a value smaller than the linewidths of the individual clouds.

\subsection{Statistics of the Cloud Properties in CCCs}
Here, we summarize the statistical properties of those CCC candidates (Table~1) that have been published.

Figure~9 shows histograms of the column density, collision velocity, size, and mass for CCC candidates in the Galaxy.
A typical collision velocity is 10 $\kms$, while in the inner disk the velocity tends to be larger, 15--20 $\kms$.
The masses of the colliding clouds are typically 10$^{3}$--10$^{5}$ $\msun$, and the sizes are 1--6 pc.
We infer that the velocity is determined by the cloud-to-cloud velocity dispersion of the molecular clouds in the galactic disk, which is mainly caused by supernovae (SN) feedback in addition to the acceleration by the gravitational field of the disk.
Observational evidence for the formation and acceleration of the molecular clouds by the collective effects of SN is obtained in Galactic supershells (e.g., \cite{fuk99}).
It is possible that the initial collision velocity may be $\sim$50 \% larger than this value due to the projection effect and to gas-dynamical interactions between the colliding clouds, as shown by the numerical simulations \citep{tak14}.

Given the size of the small cloud, the typical area involved in a collision where an O-star is formed is estimated to be $\sim$1 pc$^{2}$.
If we count only the O-star(s), the star-formation efficiency (SFE) in a CCC is estimated to be 10 \% for a 15$\msun$ O star and column density 1 $\times$ 10$^{22}$ cm$^{-2}$.
The SFE of a CCC is not extremely high, and this figure seems fairly consistent with the theoretical expectations as discussed in Section 4.
It is possible that a CCC may also trigger the formation of low-mass stars, increasing the SFE upward.

\begin{figure*}
 \begin{center}
  \includegraphics[width=14cm]{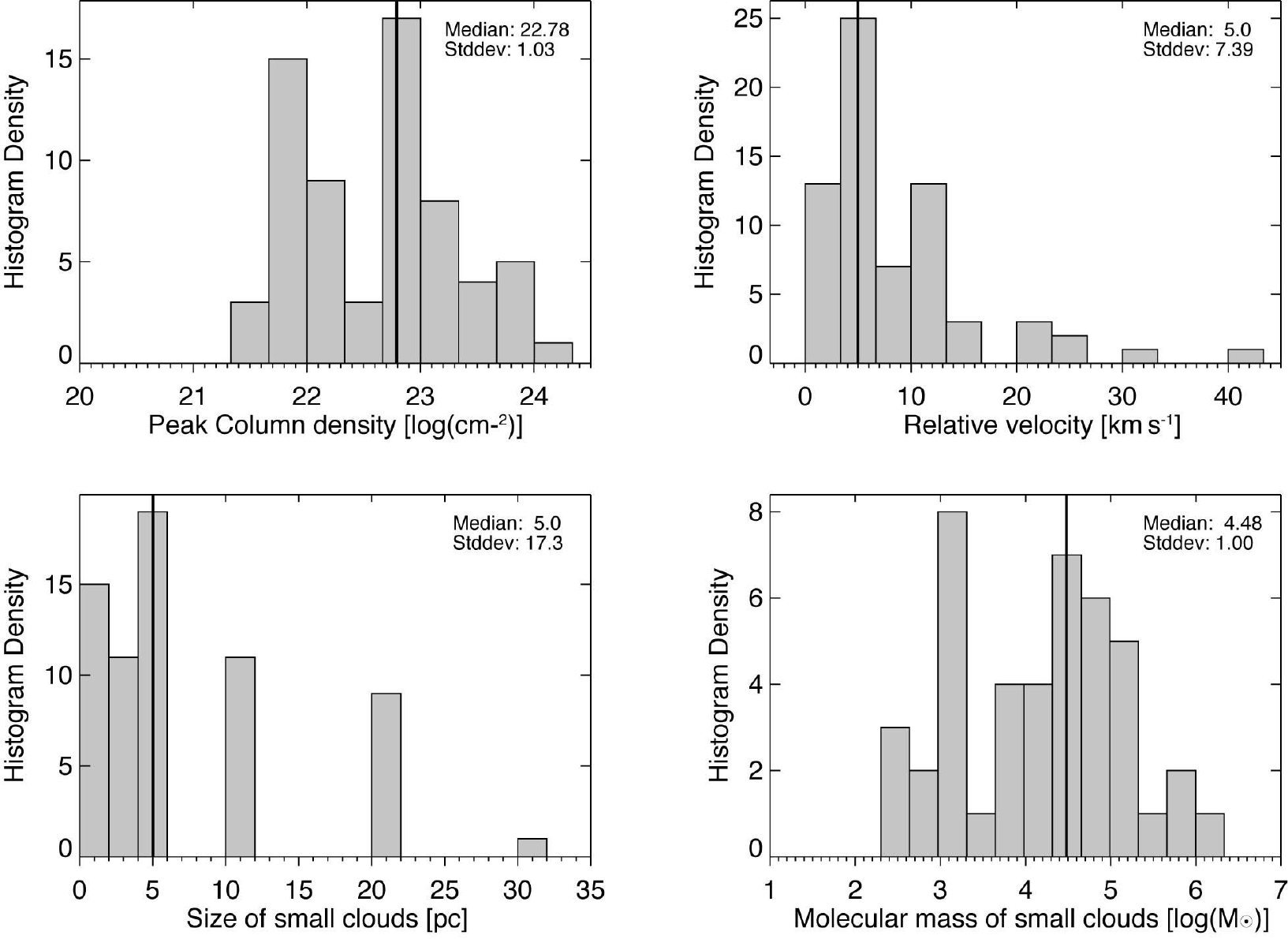} 
 \end{center}
\caption{Histograms of physical parameters for the Galactic CCC objects. Panels (a)--(d) correspond to the peak column density, relative velocity, size of the small cloud, and molecular mass of the small cloud, respectively. The median value and standard deviation are indicated at the top-right of each panel.}
\label{histo}
\end{figure*}

\begin{figure}
 \begin{center}
  \includegraphics[width=8cm]{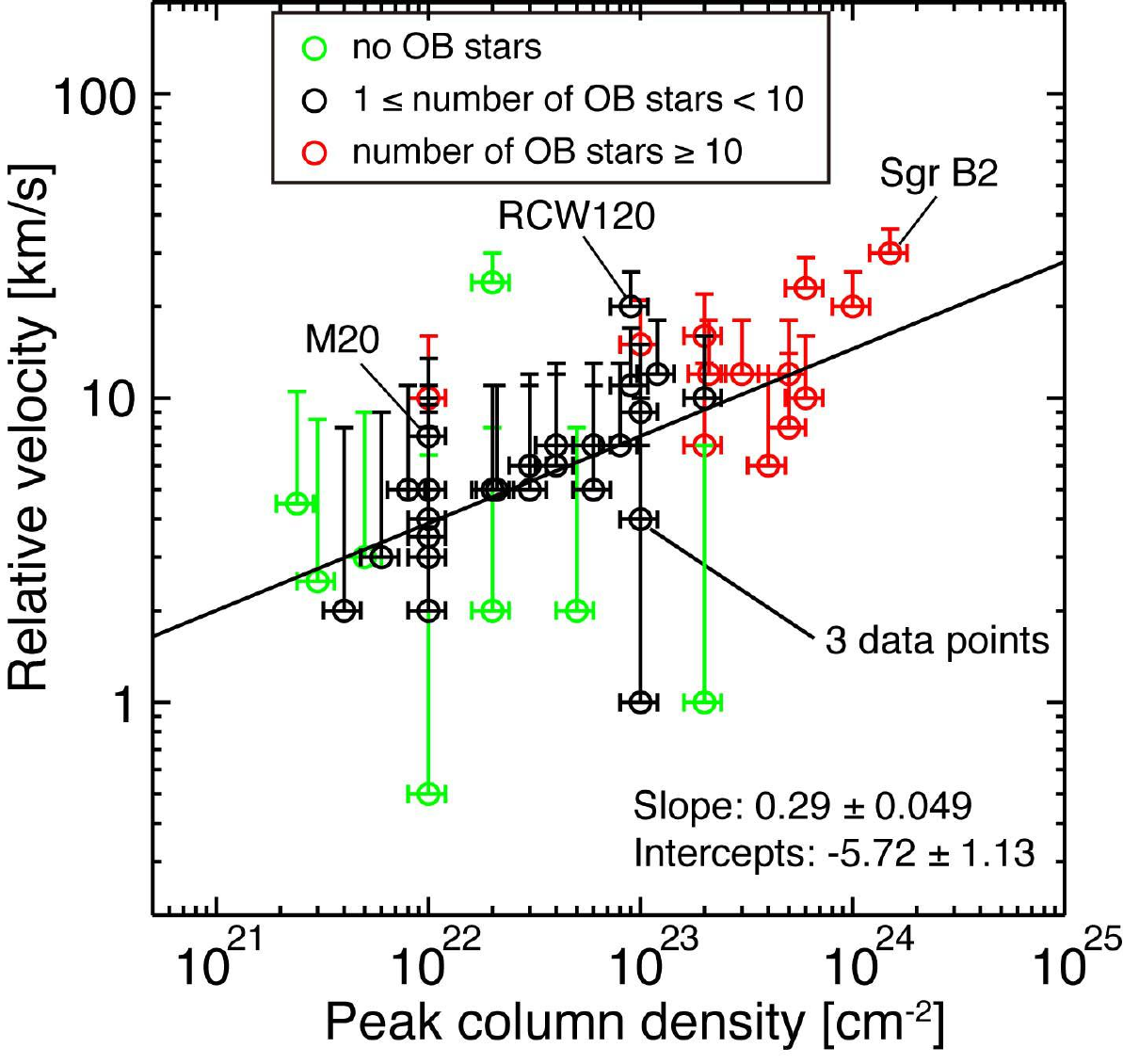} 
 \end{center}
\caption{Scatter plot of the peak column density and relative velocity of colliding clouds in the Galactic sources on a double-logarithmic scale. The black, red, and light-green symbols, respectively, indicate CCCs associated with clusters having less than 10 O- and early B-type stars, more than 10 O- and early B-type stars, and without any O- or early B-type stars. The black line is the best fit to the black and red symbols using a least-squares method. Adapted from \citet{eno19} and reproduced with permission.}
\label{NH2_V}
\end{figure}

\begin{figure}
 \begin{center}
  \includegraphics[width=8cm]{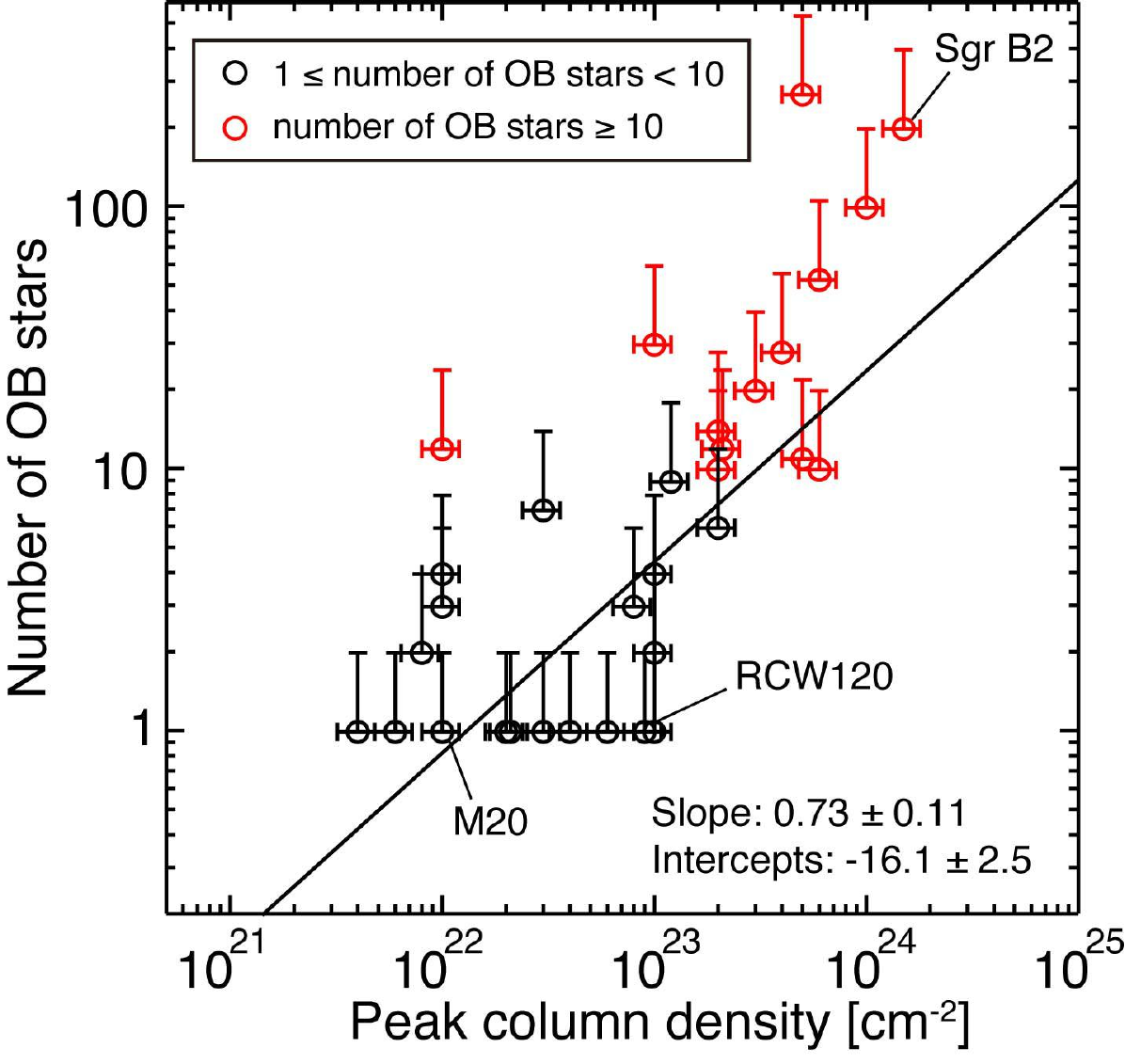} 
 \end{center}
\caption{Scatter plot of the peak column density and the number of O- and early B-type stars for colliding clouds in the Galactic sources on a double-logarithmic scale. The black line is the best fit to the black and red symbols using a least-squares method. Adapted from \citet{eno19} and reproduced with permission.}
\label{NH2_num}
\end{figure}

Figure~10 shows a scatter plot of the collision velocity and the column density, and Figure~11 is a scatter plot of the number of high-mass stars and the column density, as compiled by \citet{eno19}.
The trend that the velocity increases with the column density is in part ascribed to the massive clouds in the Central Molecular Zone (CMZ), which has both a high column density and large velocity.
The number of O-stars formed by a CCC is correlated with the column density.
Figure~11 shows that the formation of an O-star occurs for column densities larger than 1 $\times$ 10$^{22}$ cm$^{-2}$, and the formation of more than 10 O-stars requires column densities higher than 10$^{23}$ cm$^{-2}$. Below 1 $\times$10$^{22}$ cm$^{-2}$ no high-mass stars are formed.
The CMZ clouds do not follow these thresholds, which seems unusual, possibly due to the high turbulent pressure acting against the self-gravity.
The slope of the best-fit line seems slightly shallower than the data as a result of a large number of samples at the number of OB stars = 1. These points correspond nearby ($<$ a few kpc) $\htwo$ regions. Detections of high-column-density CCCs by higher angular-resolution observations will allow us to make more reliable result of the fitting, although the present value of the slope seems to be correct within the errors.
We note that Figure~11 also suggests that column density is a good parameter for characterizing the formed stars, and that other parameters may not be so important.
For instance, although the angle of obliquity of a CCC affects the formed stars, the trend in Figure~11 seems to be relatively independent of this parameter.

The CCC candidates also allow one to gain an insight into the overall collision frequency.
The colliding clouds are associated with $\htwo$ regions that are concentrated in the spiral arms in the Galactic disk.

Let us suppose that the clouds have 10$^{4}$ $\msun$ and there are 10$^{5}$ such clouds in total in the disk.
The collision frequency of a cloud is thus estimated to be one per 100 Myr if the distribution is uniform over the disk.
In the disk, the clouds are concentrated in the arms, where the collision frequency increases to one every 10 Myr, which is consistent with the numerical simulations of galactic gas dynamics \citep{tas09,tas11,fuj14,dob15}.
The CCCs therefore occur within the typical 20--30 lifetime of a giant molecular cloud (GMC; \cite{kaw09}), indicating that a GMC experiences a few collisions in its lifetime.
If we assume that a single collision triggers the formation of a 15 $\msun$ star (a late O-star), the total star-formation rate is estimated to be 0.15 $\msun$~yr$^{-1}$ (=15 $\msun$ $\times$ 10$^{5}$/10$^{7}$ yr), which corresponds to $\sim$10 \% of the total star-formation rate of $\sim$1--2 $\msun$/yr in the Galaxy \citep{mur10}.
This rate corresponds to the formation of  one O-star every 100 yr, which is somewhat smaller than the rate of SNe, which occur every 30--50 yr in the Galaxy \citep{van91}.
If nearly half of the SNe are assumed to be of the core-collapse type, it is possible that a substantial fraction of the O-stars may be formed by CCCs.
The collision frequency is therefore large enough to explain the high-mass star-formation rate.
If CCCs are the main initiating mechanism of high-mass star or cluster formation, the triggered star formation can explain empirical star formation laws such as the Kennicutt-Schmidt law \citep{sch59,ken98}.
\citet{tan00} showed that his star formation law, which assumes CCCs as the major mechanism of disk star formation is in agreement with the disk averaged data of \citet{ken98} for the case of flat rotation curves, and \citet{suw14} and \citet{aou20} suggested the validity of his model from observations of spiral galaxies. These works suggest the important role of CCCs for numerical simulations and star formation in external galaxies.

We next remark on the completeness of the CCC candidates (Table~1).
Among the colliding clouds, we observe relatively close clouds---within $\sim$5 kpc from the Sun---which suffer less contamination in the Galactic disk.

This bias is consistent with the fact that most of the CCC candidates (Table~1) are located on the near side of the disk, and it suggests that the actual rate of high-mass star formation by CCCs may be more than tripled this estimate.
It is also probable that there may still be a number of $\htwo$ regions to be searched for CCCs, because Table~1 is not complete for  known $\htwo$ regions.
The actual number of CCC candidates may thus be far more than $\sim$150, corresponding to at least 25 \% of that of the $\sim$600 Spitzer bubbles \citep{chu06}. It may also be possible that in some relatively old $\htwo$ regions the signatures of CCCs may be difficult to trace due to ionization.


\section{Theory of the Gas Dynamics of the Collisional Compression}
A collision between two clouds compresses the gas layer into a thin layer.
The details of such layers were investigated in the numerical MHD simulations by \citet{ino13}, \citet{ino18}, and \citet{fuk20}, who used the results to make synthetic observations. 
We here review the theoretical results on the collisional compression.

\subsection{Shock Compression of the Interface Layer}\label{secPIL}
Observations show that a CCC happens at a highly supersonic speed, typically $v_{\rm col}\simeq 10$ km s$^{-1}$.
Such a collision flow inevitably causes strong shock compression.
In molecular clouds, because radiative cooling is efficient, we often treat a cloud as an isothermal gas. This leads to a shock-compression ratio $r\simeq M_{\rm s}^2$ if we neglect the effect of the magnetic field, where $M_{\rm s}$ is the sonic Mach number.
Given that the typical sound speed is 0.2 km s$^{-1}$ ($M_{\rm s}\sim 50$ for $v_{\rm col}\simeq 10$ km s$^{-1}$), the density of the shocked layer is $n=n_0\,M_{\rm s}^2\sim 10^6$ cm$^{-3}$ for an initial cloud density $n_0=10^3$ cm$^{-3}$.
Since the thermal Jeans mass is a decreasing function of density, the shocked gas created by the cloud collision is sometimes mentioned as a favorable site for low-mass star formation.

However, the effect of a magnetic field can drastically change the physical state of the shocked cloud.
The density of the post-shock gas is not drastically enhanced, as discussed above, because magnetic pressure prevents over-contraction behind the shock.
Let us calculate the shock-compression ratio for the isothermal MHD equations.
For simplicity, we assume a non-oblique shock and neglect the component of the magnetic field that is parallel to the shock normal; this hardly affects the compression ratio, particularly for super-Alfv\'enic fast shock waves.
Under this assumption, the jump conditions for the isothermal MHD equations become
\begin{eqnarray}
\rho_1\,v_1&=&\rho_2\,v_2,\label{eq:b1}\\
\rho_1\,(v_1^2+c_{\rm s}^2)+\frac{B_1^2}{8\pi}&=&\rho_2,(v_2^2+c_{\rm s}^2)+\frac{B_2^2}{8\pi},\label{eq:b2}\\
B_1\,v_1&=&B_2\,v_2,\label{eq:b3}
\end{eqnarray}
where the subscripts 1 and 2, respectively, denote the pre-shock and post-shock variables.
By solving the above equations, we obtain the compression ratio as follows
\begin{equation}
\frac{\rho_2}{\rho_1}=-\left\{ \frac{1}{2}+\left( \frac{c_{\rm s}}{c_{\rm A,1}} \right)^2 \right\}+
\left[ 
\left\{ \frac{1}{2}+\left( \frac{c_{\rm s}}{c_{\rm A,1}} \right)^2 \right\}^2+
2\left( \frac{v_1}{c_{\rm A,1}} \right)^2
\right]^{1/2},\label{ratio}
\end{equation}
where $c_{\rm A,1}=B_1/\sqrt{4\pi\,\rho_1}$ is the pre-shock Alfv\'en velocity.
For a super-Alfv\'enic shock, $M_{\rm A}\equiv v_1/c_{\rm A,1}\gg1$ in a low-plasma-$\beta$ cloud with $c_{\rm A,1}/c_{\rm s}\gg 1$, and the compression ratio is thus reduced to 
\begin{equation}\label{eqMA}
\frac{\rho_2}{\rho_1}\simeq \sqrt{2}\,M_{\rm A}=17\,\left(  \frac{v_{\rm sh}}{10\mbox{ km s}^{-1}} \right)\,\left(  \frac{B_{1}}{10\,\mu\mbox{G}} \right)^{-1}\,\left(  \frac{n_{1}}{300\mbox{ cm}^{-3}} \right)^{1/2}.
\end{equation}
This compression ratio is applicable insofar as the pos-tshock magnetic pressure dominates the thermal pressure: $B_{\rm 2}^2/8\pi> \rho_2\,c_{\rm s}^2$.
By rewriting this inequality using the solutions of eqs. (\ref{eq:b1})--(\ref{eq:b3}), we obtain the range of the magnetic field strengths for which we can apply the MHD formula for the compression ratio---eq.~(\ref{eqMA})---to be
\begin{eqnarray}
B_{1}&>&\sqrt{16\,\pi\,\rho_{0}}\frac{c_{\rm s}^2}{v_{\rm sh}}\nonumber\\
&\simeq& 0.2\,\,\mu\mbox{G}\,\left( \frac{v_{\rm sh}}{10\mbox{ km s}^{-1}} \right)^{-1}
\left( \frac{n_{0}}{10^3\mbox{ cm}^{-1}} \right)^{1/2}\left( \frac{c_{\rm s}}{0.2\mbox{ km s}^{-1}} \right)^2.
\end{eqnarray}
This small value indicates that we almost always have to consider the effect of the magnetic field in a molecular cloud, and the shock compression is controlled by the Alfv\'en Mach number from eq.~(\ref{eqMA}).
Note that, in the above discussion, we consider only the effect of the magnetic-field component perpendicular to the shock-propagation direction.
Thus, the range of upstream magnetic-field strengths given above should be applied to the perpendicular component of the magnetic field in the general case.

\subsection{Structure Formation in the Interface Layer}
The shock compression creates a dense gas sheet in which gravity can activate star formation.
Such a process has been studied by many authors.
\citet{hun86}, \citet{whi94a}, and \citet{whi94b} studied the gravitational fragmentation of a shocked-compressed sheet using linear stability analysis and hydrodynamic simulations.
Although their studies are pioneering and insightful, they only considered the growth of the gravitational instability from a flat, non-turbulent sheet.
\citet{whi94b} estimated that the gravitational instability starts to grow after a time
\begin{eqnarray}\label{eq:W94}
t_{\rm start}&\simeq& \left( \frac{c_{\rm s}}{G\,\rho_1\,v_{\rm col}} \right)^{1/2}\nonumber\\
&\sim& 0.5\mbox{ Myr}\,\left( \frac{c_{\rm s}}{0.2\mbox{ km s}^{-1}} \right)^{1/2}\left( \frac{n_{1}}{300\mbox{ cm}^{-3}} \right)^{-1/2}\left( \frac{v_{\rm col}}{10\mbox{ km s}^{-1}} \right)^{-1/2},
\end{eqnarray}
following the initial impact of the clouds, where $v_{\rm col}$ is the relative velocity of the collision.
This timescale seems to be short enough to work, but in reality a kind of (magneto-) hydrodynamical instability grows immediately behind the shock, and it dominates dynamical structure formation inside the shock-compressed layer (see \cite{ino13}; \cite{vai13}; and Abe et al. 2020 for more detail).

Recent studies of molecular-cloud formation have revealed that molecular clouds are highly inhomogeneous and turbulent from the beginning (e.g., \cite{koy02}, \cite{hen08}, \cite{hei08}, \cite{vaz07}, \cite{ino08}, \cite{ino12}).
Thus, the collision of two uniform clouds---assumed for simplicity in many historical studies---is not a realistic setting.
By using three-dimensional, isothermal MHD simulations, \citet{ino13} showed that dense, filamentary structures are formed in the shock-compressed layer produced in a fast CCC, irrespective of the effects of self-gravity.
In Figure 12, we show the face on view snapshot of the column density of the shock compressed layer simulated by Abe et al. (2020), where we confirm the formation of very high column density filamentary blobs.
The line mass of the most massive filament created in this result is larger than 100 $M_{\rm sun}$ pc$^{-1}$.

\begin{figure}
 \begin{center}
  \includegraphics[width=8cm]{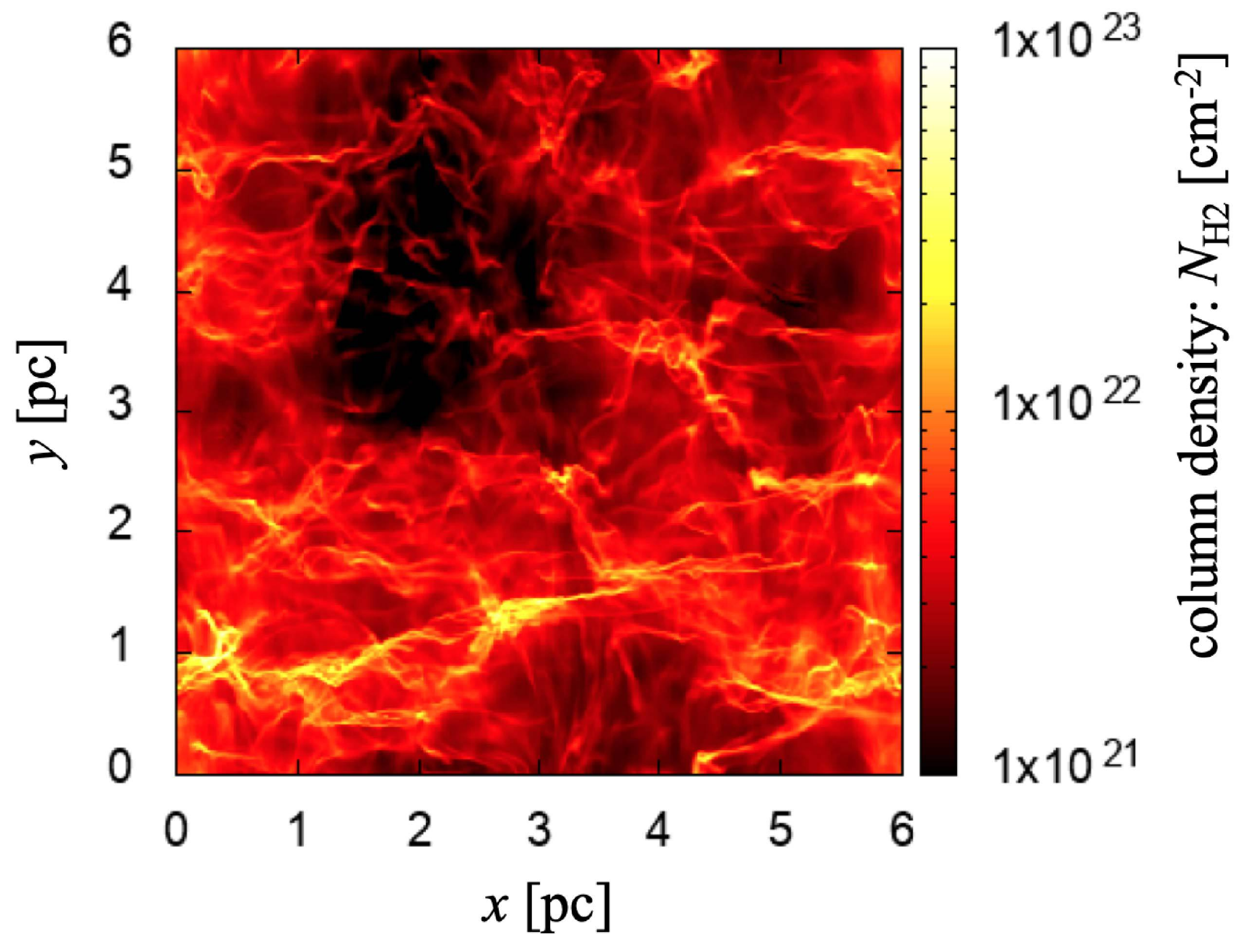} 
 \end{center}
\caption{Face-on snapshot of the column-density structure of the shock-compressed layer.
In this simulation by Abe et al.~(2020), two inhomogeneous clouds with $\langle n\rangle_{1}=300$ cm$^{-3}$ and $B_{y,1}=10\,\mu$G collide with a relative velocity of $10$ km s$^{-1}$ along the $z$-axis.
The most massive filament has a line-mass greater than 100 $M_{\rm sun}$ pc$^{-1}$.}
\label{fI1}
\end{figure}

\begin{figure}
 \begin{center}
  \includegraphics[width=7cm]{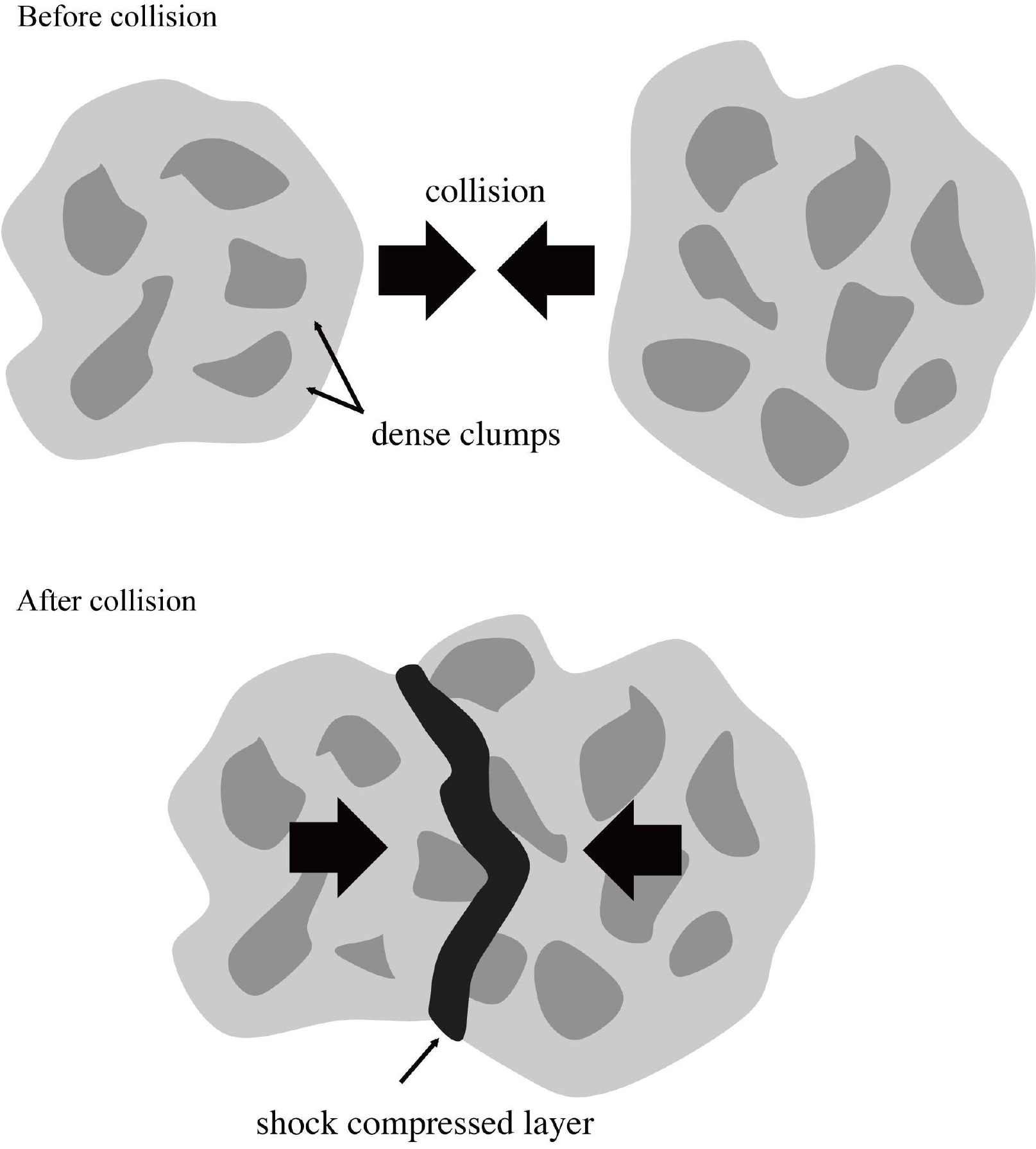} 
 \end{center}
\caption{Schematic illustration of clumpy clouds before ({\it top}) and after ({\it bottom}) collision.}
\label{fI2}
\end{figure}

\begin{figure*}
 \begin{center}
  \includegraphics[width=15cm]{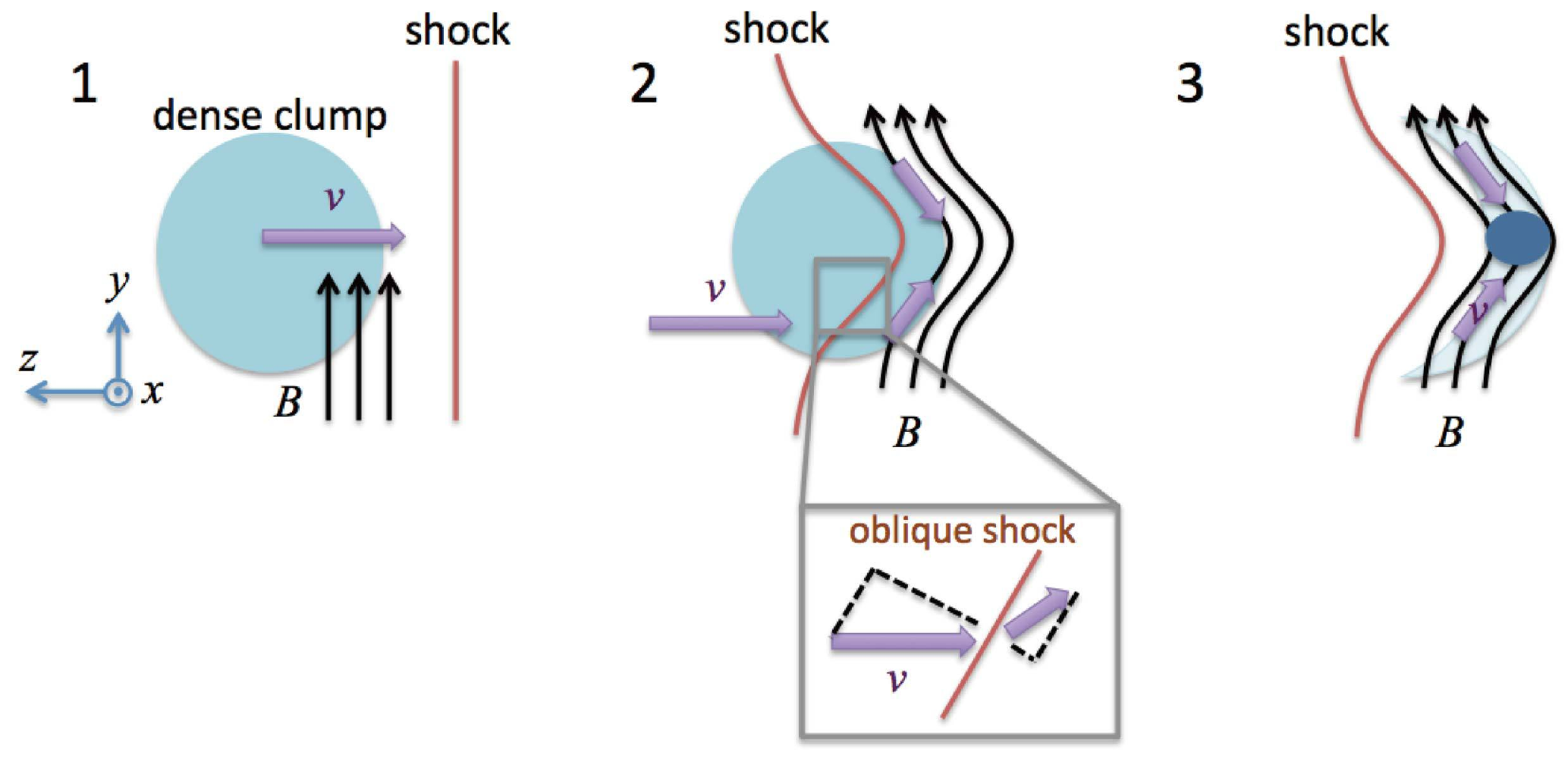} 
 \end{center}
\caption{Schematic illustration of the evolution of one dense clump swept up by a shock wave (from Figure 3 of \cite{ino18}). Adapted from \citet{ino18} with permission.}
\label{fI3}
\end{figure*}

How are these dense filaments formed?
The magnetohydrodynamical flow induced at the shock front plays a crucial role in the formation process:
As illustrated in Figure~13, if we consider the collision of highly inhomogeneous ($\Delta \rho/\rho\ge1$) flows, clumps embedded in the colliding clouds will be shock-compressed.
When a shock wave hits a dense clump, the shock front is deformed, because the shock speed is decelerated in the dense region.
At an oblique shock wave, the gas flow is kinked, as depicted in Figure 14.
This is because the velocity (or momentum flux) tangential to the shock is conserved, while the normal velocity stalls across the shock.
As a consequence of the kinked flow, focusing flows are generated behind the shock, which further compress the clump; i.e., the clump is compressed not only by the shock but also by the post-shock focusing flows.
Because of the enhanced magnetic field behind the shock, the focusing flows are induced only along the post-shock magnetic field.
This leads to the formation of filamentary structures perpendicular to the magnetic field (or perpendicular to the plane of the paper in Figure 14).
This mechanism of dense-filament formation was discovered by \citet{ino13} and \cite{vai13}; see also \cite{arz18} and \cite{kan20} for observational supports).
We emphasize that the filaments formed by this mechanism lie perpendicular to the magnetic field, at least when they are first created.
Abe et al. (2020) recently showed that the above oblique MHD-shock-induced filament formation dominates other filament formation mechanism when the shock velocity is larger than roughly 5 km s$^{-1}$.

\subsection{Massive Filaments and Cores}
If the line mass of a formed filament exceeds the critical value for gravitational instability, the filament starts to collapse and fragment.
For an unmagnetized filament, it is widely known that the critical line mass is given by \citep{sto63,ost64}
\begin{equation}\label{eq:S63}
\lambda_{\rm max}=2\,c_{\rm s}^{2}/G\,\simeq 19\mbox{ M}_{\rm sun}\,\mbox{pc}^{-1}\,(c_{\rm s}/0.2\,\mbox{km s}^{-1})^2.
\end{equation}
Recently, \citet{tom14} showed that the maximum line mass of a filament that is threaded by a magnetic field perpendicular to the filament is given by
\begin{equation}\label{eq:T14}
\lambda _{\rm max} \simeq 0.24 \Phi_{\rm cl}/G^{1/2}+1.66\,c_{s}^{2}/G,
\end{equation}
where $\Phi_{\rm cl}$ represents one half of the magnetic flux threading the filament per unit length (or one half of the magnetic field strength times the width of the filament).
It is clear that the first term in eq.~(\ref{eq:T14}) represents the contribution from the magnetic field, while the second term is a thermal contribution
\footnote{Note that eq.~(\ref{eq:T14}) is a fitting formula obtained from numerical solutions of static, magnetized filaments that is not optimized for the small-magnetic-field-strength regime.
This is the reason why the thermal term in eq.~(\ref{eq:T14}) is not equivalent to eq.~(\ref{eq:S63})}.
By comparing the two terms, we find that the critical line mass occurs when the magnetic-field strength is larger than the following value:
\begin{equation}\label{eq:Bcr}
B_{\rm cr}\simeq\frac{13.8\,c_{\rm s}^2}{G^{1/2}\,w}=70\,\mu\mbox{G}\,\left( \frac{w}{0.1\mbox{ pc}}\right)^{-1}\,\left( \frac{c_{\rm s}}{0.2\mbox{ km s}^{-1}}\right)^{2},
\end{equation}
where we have assumed the width of the filament to be $w=0.1$ pc, as suggested by observations \citep{arz11,arz19}.

The magnetic-field strength in the filament is a key parameter for evaluating the critical line mass.
In the case of a cloud collision, filaments are formed in the shock-compressed layer, and the magnetic field strengths in the filaments are approximately given by the shock-compressed value of the upstream magnetic field \citep{ino13,ino18}.
Note that the magnetic-field component parallel to the shock normal is not compressed by the shock, and thus the post-shock magnetic-field strength is determined by the amplification of the perpendicular components.
Because the perpendicular component of the magnetic field is compressed by the same factor as the volume density, the post-shock strength can be evaluated from eq.~(\ref{eqMA}) as
\begin{eqnarray}
B_{\rm filament}&\sim& B_{{\rm t},2}=\sqrt{2}\,M_{\rm A}\,B_{{\rm t}, 1}\simeq \sqrt{8\pi\,\rho_{1}}\,v_{\rm col}\nonumber\\
&=&170\,\mu\mbox{G}\,\left(\frac{n_1}{300\mbox{ cm}^{-3}}\right)^{1/2}\,\left(\frac{v_{\rm sh}}{10\mbox{ km s}^{-1}}\right),
\end{eqnarray}
where the shock velocity $v_{\rm sh}$ is approximately half of the relative collision velocity $v_{\rm col}$.

If we put the above filament magnetic-field strength estimate into eq.~(\ref{eq:T14}), we can estimate the critical line mass for the filaments:
\begin{eqnarray}\label{eqMfil}
\lambda_{\rm max}&\simeq& 45\mbox{ M}_{\rm sun}\,\mbox{pc}^{-1}\, (B_{\rm fil}/200 \mu\mbox{G})\,(w/0.1\, \mbox{pc})\nonumber\\
&&+15\mbox{ M}_{\rm sun}\,\mbox{pc}^{-1}\,(c_{\rm s}/0.2\mbox{ km s}^{-1})^2.
\end{eqnarray}
Again, we have assumed the width of the filament to be 0.1 pc, following observations that suggest that even massive filaments retain the constant width of 0.1 pc.
We can thus expect massive filaments containing more than 50 $M_{\rm sun}$ pc$^{-1}$ to be formed by a CCC.
Although the origin of the constant widths of filaments is not yet well-understood theoretically (e.g., \cite{fed16,nto16}), the above line-mass estimate is consistent with the filaments created by CCC simulations (\cite{ino18}; Abe et al. 2020).

As shown by \citet{ino18}, massive dense cores can be formed from a filament that undergoes global collapse (due to the growth of large-scale gravitational instability).
Since small-scale (filament-width scale) instabilities grow simultaneously with the large-scale instability, the massive cores formed in their simulations experience coalescence of low-mass cores.
This picture of high-mass star formation is consistent with recent observational suggestions, e.g., by \citet{per13}, \citet{per14}, and \citet{tok19}.
Because the collapsing filament is massive---due mostly to the strong magnetic field---the mass-accretion rate onto the central massive core can be large.
\citet{ino13} speculated that the mass-accretion rate may be as large as 
\begin{eqnarray}
\dot{M} &\sim& \frac{c_{\rm A}^3}{G}=\frac{B_{\rm fil}^3}{G^{1/2}\,(4\pi\rho_{\rm fil})^{3/2}}\nonumber\\
&\sim& 7\times10^{-4}\mbox{ M}_{\rm sun}\,\mbox{yr}^{-1}\,\left( \frac{B_{\rm fil}}{100\,\mu\mbox{G}} \right)^3\,\left( \frac{n_{\rm fil}}{10^4 \mbox{ cm}^{-3}} \right)^{-3/2}, 
\end{eqnarray}
where we have simply replaced the sound speed in the classical mass-accretion-rate formula by the Alfv\'en velocity, which is much faster than the sound speed.
The result of the simulation by \citet{ino18} showed that the mass-accretion rate is larger than $10^{-4}$ M$_{\rm sun}$ yr$^{-1}$ and that this results in the formation of a core as massive as $\sim$100 M$_{\rm sun}$.

A 20 $\msun$ star is formed in $\sim$2$\times$10$^{5}$ yr by a constant mass-accretion rate of 10$^{-4}$ $\msun$/yr.
Synthetic observations indicate that a filament forms multiple dense cores, with typical separations of 0.1 pc. The most massive one will become an O-star first, and it will ionize the surroundings within a radius of 1 pc in 1 Myr.
The outcome is a single O-star together with lower-mass stars, because the accretion flows are terminated by ionization prior to the formation of multiple high-mass stars.
If the initial gas density is higher, the formation of multiple high-mass stars in a pc-scale region may become possible, because each potential O-star can grow into an actual O-star more rapidly than the ionization by its neighbors can terminate its growth.
In such a case, the outcome will be an O-star cluster similar to a super star cluster (SSC) containing tens of-O stars.
Rapid star formation in a CCC is supported by observations of the star-formation history in the Galaxy, although such observations are not easy to carry out.
Such an effort by \citet{kud12} using the HST and VLT found that the duration of star formation is on the order of 10$^{5}$ yr for two SSCs, Westerlund 1 and NGC 3603, which lends support to the hypothesis of rapid star formation.

\subsection{The Core Mass Function Driven by CCCs}
\citet{fuk20} used the results of \citet{ino13} to make synthetic observations of the collision-compressed layer.
Table~3 lists the physical parameters adopted in the simulations.
Figure~15 shows the density, velocity fields, and pressure of the transition layer for a pixel size of 0.015 pc; in the collision, the velocity drops suddenly from the initial velocity of 10 $\kms$ to 1.5 $\kms$, and the density increases by more than a factor of 10.
The simulation shows that massive, dense cores are formed in the filament.
The most massive core has a mass of 60 $\msun$ and a size of 0.1 pc, and it tends to be gravitationally unstable, leading to high-mass star formation.
The unstable dense cores are found at high column densities---above 10$^{23}$ cm$^{-2}$---in the filaments (see Figure~11 of \cite{fuk20}).

The mass and size are similar to the dense cores observed in typical high-mass star-forming regions, which are typically expressed as ``100 $\msun$ within 0.1 pc'' or ``column mass of $\sim$1 g cm$^{-2}$''  (e.g., the sub-mm protocluster IRAS 05358+3543: \cite{beu07,mck03}).

\begin{figure*}
 \begin{center}
  \includegraphics[width=14cm]{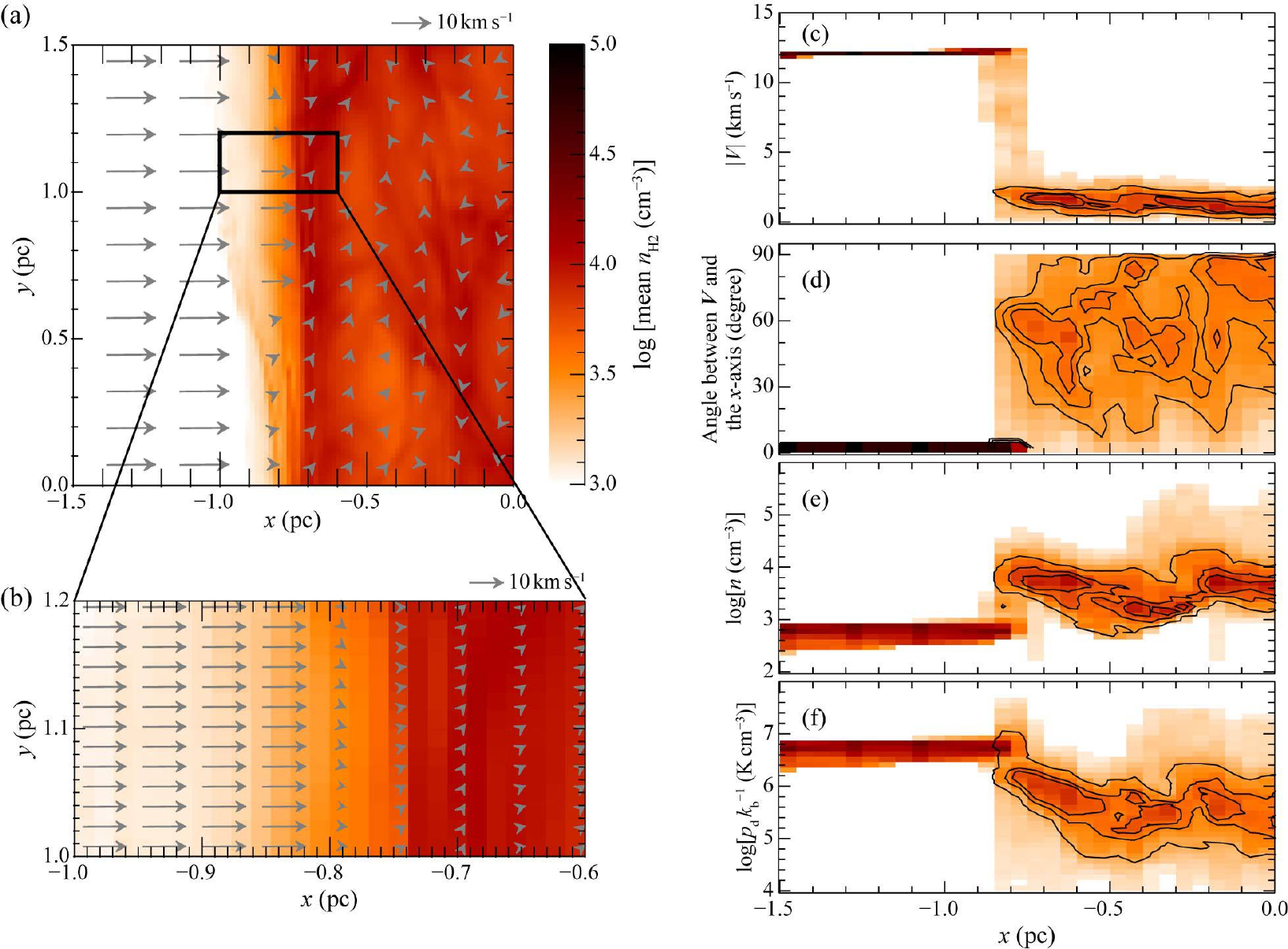} 
 \end{center}
\caption{(a) Projected velocity vectors (arrows) overlaid on the distribution of mean density (image) for $|z|$ $<$ 0.08 pc at $t$ = 0.7 Myr. (b) Close-up view of (a). (c) Distribution of the size of the velocity vector ($|V|$). (d) Angle between the velocity vector and the x-axis. (e) The density n and (f) dynamic pressure. The contours in panels (c)--(f) contain 30, 60, and 90 \% of the data points but exclude gas with the initial conditions ($|V|$ $\gtrsim$ 12 $\kms$ and angle $\sim 0^{\circ}$). Adapted from \citet{fuk20} with permission.}
\label{synth_MHD}
\end{figure*}

\begin{figure*}
 \begin{center}
  \includegraphics[width=11cm]{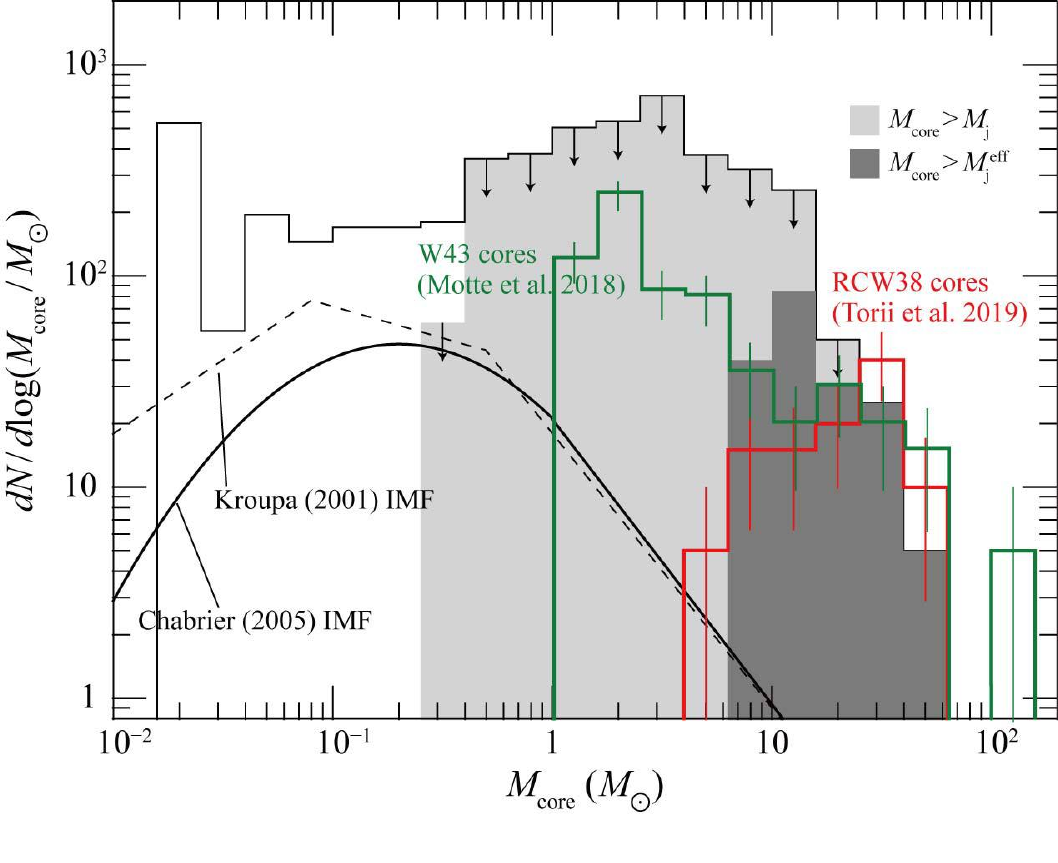} 
 \end{center}
\caption{Core mass function from a numerical simulation at $t$ = 0.7 Myr superimposed on those for RCW~38 cores \citep{tor20} and W43 cores \citep{mot18a} are superimposed.The error bars correspond to $\sqrt{N}$ statistical uncertainties. The dashed line shows the single-star IMF of \citet{kro01}, and the solid curve shows the system IMF by \citet{cha05}. Adapted from \citet{fuk20} with permission.}
\label{synth_CMF}
\end{figure*}

The core mass function shown in Figure~16 consists of two parts: The high-mass part cover the range 6--60 $\msun$, and most of those cores are gravitationally unstable.

Follow-up simulations \citep{ino18} have confirmed that the massive cores collapse into denser cores under their own self-gravity in $\sim$1 Myr.
On the other hand, the low-mass part of the function---less than 6 $\msun$---consists of cores for which the virial mass is larger than the gas mass, and the low-mass part provides upper limits for the star-forming cores.
One CCC therefore produces a top-heavy core mass function.
Simulations of protostellar collapse that assume the initial condition 100 $\msun$ within 0.1 pc show that the SFE is about 70 \% \citep{kru09}.
The stellar-mass function also may be given by the core mass function, if we assume that the stellar mass is 70 \% of the core mass.

The total mass of the massive cores amounts to 400 $\msun$ at an epoch of 0.7 Myr.
In the compressed layer, the total mass of gas denser than 10$^{4}$ cm$^{-3}$ is 18000 $\msun$, and the mass that is denser than 10$^{5}$ cm$^{-3}$ is 2900 $\msun$.
For these two cases, the SFE in a CCC is estimated to be 1.6 \% and 10 \%, respectively, by taking the ratio of the stellar mass (70 \% of the core mass) and the gas mass of the compressed layer.
The SFE may increase further by accumulating more mass, if the simulation is continued for 1 Myr or longer, but once an O-star is formed, its ionization will limit the mass growth of the dense cores.
We infer that the SFE is not extremely high in spite of the strong collisional compression.

A comparison of the theoretical core mass function with the observations of RCW~38 and W43 Main is shown in Figure~16.
The W43 dust cores observed with Atacama Large Millimeter/submillimeter Array (ALMA; \cite{mot18a}) are compared with the synthetic observations, and the agreement between them is good.
Gravitational stability is not tested by this comparison, because there is no velocity information.

The cores in RCW~38 were observed with ALMA in $\ceighteeno$ emission, with a mass-detection limit of 6 $\msun$ \citep{tor20}.
The mass range observed matches the high-mass part of the core mass function.
Most of the $\ceighteeno$ cores are found to be gravitationally unstable in RCW~38.
The core properties---i.e., the ranges of density, size, and mass---seem to be in accord with the simulations, but the detection limit needs to be lowered in the future for a better comparison.
In summary, the dense cores observed in the state-of-the-art ALMA observations are reasonably consistent with the simulated core mass function.

The stellar IMF has been extensively studied in the literature (e.g., \cite{kro01,cha03} and references therein).
Figure~16 compares the stellar initial mass functions and the theoretical core mass function formed in a CCC.
If we assume that the SFE of the cores is 0.7, the core mass roughly corresponds to the stellar mass.
Figure~16 indicates that CCCs are responsible mainly for the stellar-mass range 4 $\msun$--40 $\msun$, i.e., the high-mass tail of the mass distribution.
This demonstrates that the impact of a CCC lies in the formation of stars more massive than 4 $\msun$, for which the observed sample of stars is limited to well-studied clusters, including rich clusters like the ONC and NGC~3603 in the Galaxy, and R136 in the LMC (e.g., \cite{kro01}).
All three clusters are regions triggered by CCCs according to observational studies of the molecular/atomic clouds \citep{fuk18a, fuk14, fuk17}.
In the Galaxy, the total mass of these high-mass stars is $\sim$200--400 $\msun$ (20 $\msun$ times 10--20), a very small fraction of all the stars, which have an average total mass of $\sim$10$^{8}$ $\msun$ in an area of 1 kpc$^{2}$.
A CCC produces a top-heavy stellar mass function locally within an area of 1 pc$^{2}$, but it does not affect the IMF of the field stars.

\subsection{Low-Mass Star Formation by CCCs}
The triggering of low-mass star formation by CCCs is another matter of interest. However, it has not been well-discussed, since low-mass stars are believed to form spontaneously by self-gravity without any external trigger.
Although the universality of the IMF is under intense debate, if a CCC results in a top-heavy core mass function, as discussed in this section, low-mass star formation may not be enhanced as dramatically by a CCC as high-mass star formation is.

NGC~1333 is one of the rare low-mass star-forming regions where a CCC has been suggested \citep{lor76}.
In this cloud, a mild collision with a moderate column density and a small collision velocity ($\sim$ 2 $\kms$) is supposed to be taking place, and no massive stars earlier than B5-type have been formed.
Based on $\hone$ and OH observations, the idea of collisions of expanding shells with the ambient interstellar medium near NGC~1333 and IC~348 has been proposed \citep{san74}.
The young star-cluster in NGC~1333, on the other hand, shows the signature of a burst of formation.
The high fraction of class 0/I objects with respect to class II/III YSOs (39/98) identified by the Spitzer Space Telescope \citep{gut08} implies that triggering of low-mass star formation occurred $\lesssim$ 1 Myr ago.
\citet{hac17} pointed out that NGC~1333 consists of many filamentary clouds (also called ``fibers'' in their terminology), and dense cores are located at the junctions of multiple fibers, suggesting that collisions of the fibers are responsible for the formation of dense cores and thus for the active star formation in this cloud.
Since filaments are commonly seen as the building blocks of interstellar clouds, collisions among the filaments are expected as a natural consequence of a CCC.
Similar mechanisms of dense core formation by colliding filaments have been reported in  LDN~1641N and Serpens South \citep{nak12,nak14}.

From the statistics of star formation triggered by CCCs, it is suggested that low-mass star formation may be enhanced rather than producing high-mass stars by CCCs with low initial column density and/or low collision velocity (see Section 3.2), but more observational samples are needed to establish this definitively.

\subsection{Star-Formation Efficiency in a CCC}
One may expect the SFE in a CCC to be enhanced by the strong compression. 
We explore this issue next based on both observations and theories. 
Let us suppose that the stellar mass of interest covers the wide range from 0.1 $\msun$ to 50 $\msun$.
Due to observational limitations, such a wide mass range can be observed with high sensitivity only at the small distance of 400--500 pc, beyond which observations of low-mass stars are limited by extinction and contamination in the Galactic disk.
The best region for this purpose is the Orion region.

The star-formation history in the ONC has been an issue of keen interest (e.g., \cite{kro01}; \cite{stu16}; \cite{fuk18a}).
Extensive spectroscopic and photometric studies have provided one of the best samples, which contains stellar parameters for 2000 members in the ONC.
The ages of the stars, estimated by comparison with theoretical evolutionary tracks, ranges from a few Myr to 0.1 Myr, while the accuracy of the method is limited by the parameters used in the stellar-evolution calculations \citep{hil97}.
A CCC took place in M42 0.1 Myr ago, and most of the low-mass star formation continued over a few Myr prior to the collision according to \citet{fuk18a}.
The CCC triggered the formation of 10 O-stars with a total mass of $\sim$200 $\msun$.
Given the molecular mass of the colliding clouds---1000 $\msun$ within 1 pc of the Trapezium stars---the star formation efficiency is 20\%, which is consistent with the theoretical estimate in Section 4.
The stellar mass of the ONC as a whole is 2000 $\msun$, and the molecular mass of the main CO ridge is 1.5 $\times 10^{4}$ $\msun$, indicating a SFE of $\sim$10 \%.
This shows that the CCC has not had much impact on the SFE.

\section{Observations of CCCs: Individual Regions}
\subsection{Formation of SSCs by CCCs}
\subsubsection{The CCCs in the Antennae Galaxies}
The Antennae Galaxies NGC~4038/NGC~4039 located 22 Mpc from the Sun are the best-known pair of interacting galaxies \citep{whi95,wil00}.
The galaxies consist of two interacting grand-design spirals, as shown by their overlapping disks. The two galactic nuclei are still separated, however, suggesting that the galaxies are still in an early stage of merging.
Early molecular observations detected massive molecular clouds in the overlap region, and it was suggested that the visible young clusters may have been formed by the interaction at a velocity of $\sim$100 $\kms$, \citep{wil00}.
The filamentary molecular distribution was resolved at 10 pc resolution by recent CO observations with ALMA \citep{whi14,fin19}. 

\begin{figure*}
 \begin{center}
  \includegraphics[width=14cm]{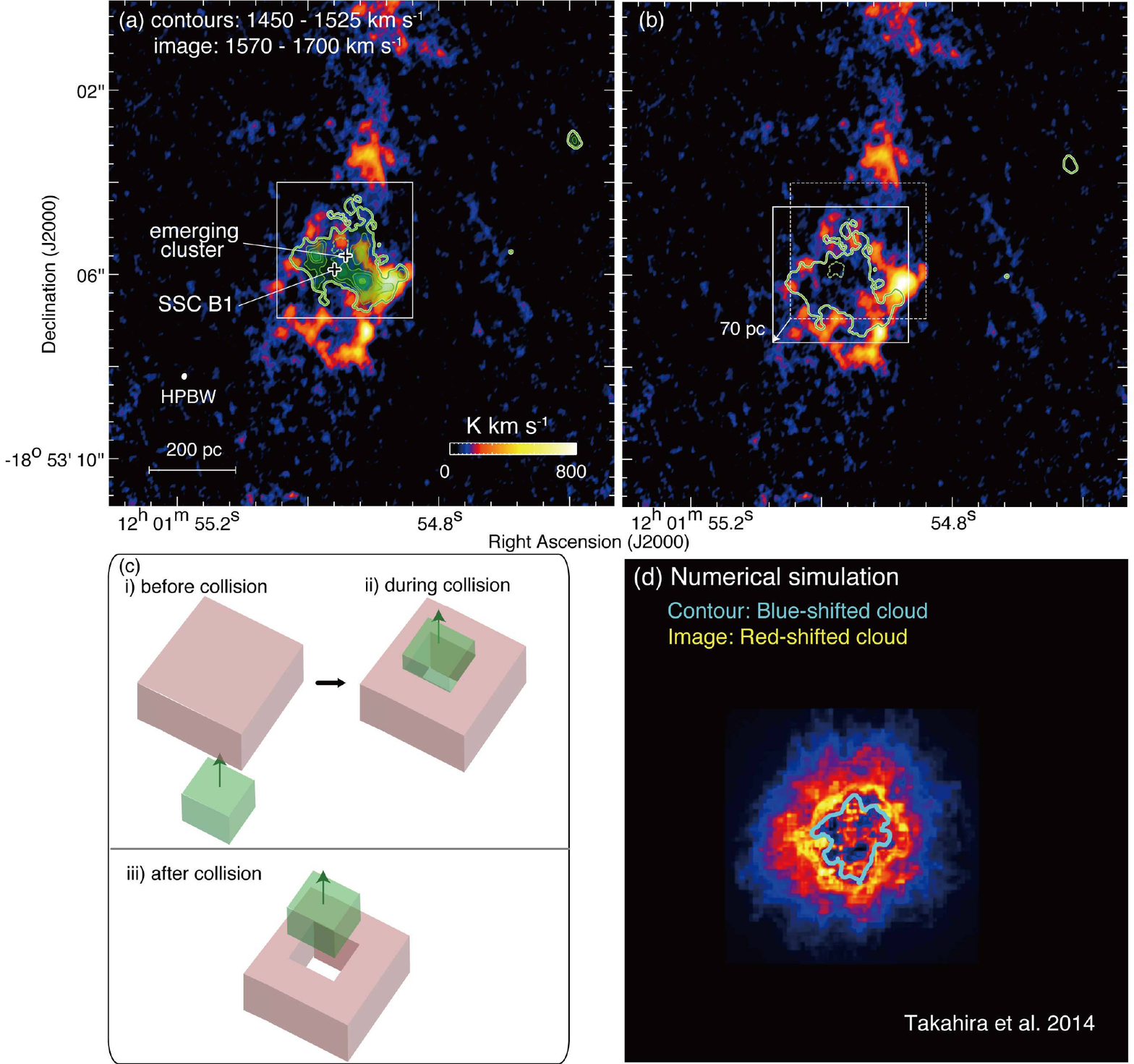} 
 \end{center}
 \caption{(a) CO intensity map for the blue-shifted cloud (contours) and the red-shifted cloud (image) in SGMC 4/5 in the Antennae \citep{tsu20b}. The black crosses indicate the positions of SSC B1 and the emerging cluster classified by radio continuum, CO and Optical/NIR data \citep{whi14}. (b) The same as (a), but with the contours of the blue-shifted cloud is displaced. The contour level is 300 $\kkms$. (c) Schematic view of the collision between the red-shifted and blue-shifted clouds. Sub-panel (I) corresponds to the clouds before the collision, sub-panel (II) corresponds to time during the collision, and sub-panel (III) corresponds to a period after the collision for SGMC 4/5. Panels (a) and (b) correspond to sub-panels (iii) and (ii), respectively. (d) Result from a synthetic observation based on the numerical simulation by \citet{tak14}. The contour and the image indicate the blue-shifted cloud and the red-shifted cloud, respectively. Adapted from \citet{tsu20b} and reproduced with permission.}
\label{ante}
\end{figure*}

\begin{figure*}
 \begin{center}
  \includegraphics[width=14cm]{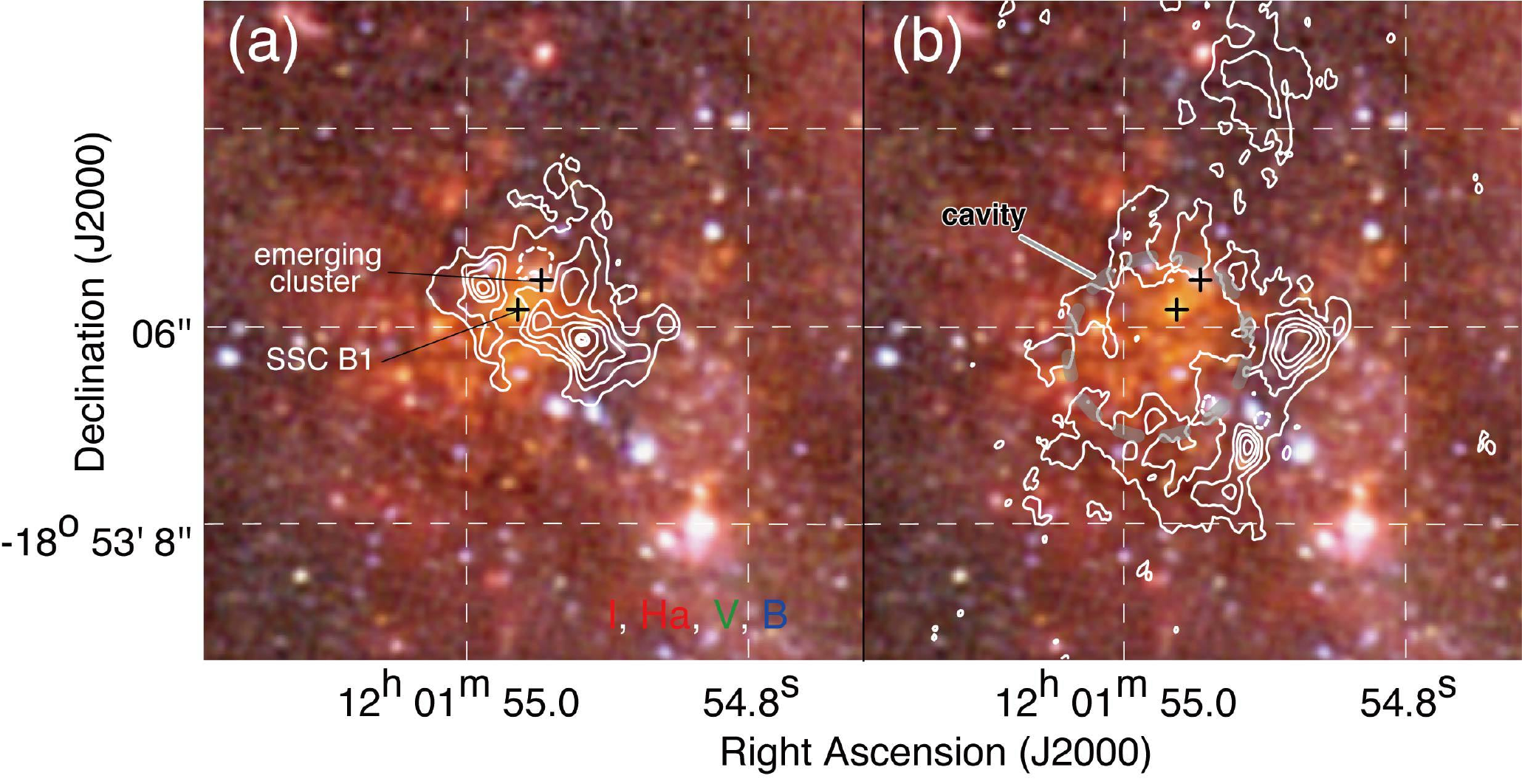} 
 \end{center}
\caption{(a) and (b) False-color image toward SGMC 4/5 obtained with HST overlaid with the integrated intensity distribution of $\twelvecohh$ for the blue-shifted cloud and the red-shifted cloud. The B-band image is shown in blue, the V-band image is in green, and a combination of the I-band and H$\alpha$ images is in red. The integrated velocity ranges for the left and right panels are 1450 to 1525 $\kms$, and 1570 to 1700 $\kms$, respectively. The contour levels are 300, 450, 600,750, 900,1060, 1070, and 1090 $\kkms$ for (a); and 120, 220, 320, 420, 520, 620, 720, 820, 920, and 1020 $\kkms$ for (b). The black crosses show the positions of SSC B1 and the emerging cluster classified by radio continuum, CO, and Optical/NIR data\citep{whi14}. Adapted from \citet{tsu20b} and reproduced with permission.}
\label{ante_hbl}
\end{figure*}

\citet{tsu20a} analyzed the ALMA CO data in the overlap region in the same manner at 50 pc resolution, and they found that the complementary distributions like those in all GMC complexes.
The ages of the clusters in the overlap region range from 1 Myr to 6 Myr.
In the older clusters, significant cloud dispersal by ionization makes it difficult to identify detailed collision signatures.
The tidal interaction between the Antennae Galaxies has been simulated numerically by \citet{ren15}, and they conclude that the onset of the interaction dates back probably some 100 Myr. 

\citet{tsu20b} analyzed the CO emission in the region of a SSC B1 in more detail and found observational signatures of CCCs in the overlap region.
Figure~17a shows the CO distribution at 10 pc resolution toward SSC B1 in two velocity ranges separated by 100 $\kms$, which are apparent in the double-peaked CO profile.
In Figure~17b the two distributions are clearly complementary to each other, and they coincide very well if a displacement of 70 pc is applied, as shown by an arrow. 

This analysis indicates that two clouds with sizes of 200--300 pc collided with each other and that the small cloud, which is moving toward us, created a cavity in the large cloud.
Figure~17c shows a schematic drawing of the collision, where the two clouds collided at their centers by chance.
Figure~17d shows a simulation of the two clouds in a head-on collision, which fits well with the observations.
The Antennae Galaxies are the first case in which massive clusters have been demonstrated to be formed by a CCC at 100 $\kms$, which was driven by the galactic collision.
Figure~18, an overlay of the CO clouds with an HST near-infrared image, clearly shows the correspondence between the intensity depression of the molecular clouds and the near-infrared members of the cluster.
The path length divided by the velocity gives a rough estimate of the collision timescale to be 1 Myr, to within a factor of $\sim$2, depending on the angle of the collision velocity vector to the line of sight.
This timescale is similar to the cluster age \citep{whi10}.
SSC B1 is the most luminous (10$^{7}$ $\lsun$) cluster in the overlap region, and its total mass is estimated to be 6.8 $\times 10^{6}$ $\msun$ \citep{whi14}.
The combined mass of the two clouds is 7 $\times 10^{7}$ $\msun$ and the SFE is about 10 \%.

\subsubsection{SSC Formation in the Magellanic System}
The Magellanic System---consisting of the LMC and the Small Magellanic Cloud (SMC)---is another pair of interacting galaxies in which the galactic interaction has accelerated the $\hone$ gas to produce colliding $\hone$ flows.
\citet{fuj90} proposed that a close encounter between the two galaxies 200 Myr ago removed the gas mainly from the SMC and that this gas is now falling down onto the LMC disk.
\citet{fuk17} proposed that R136, the most massive SSC in the Local Group, containing 10$^{5}$ $\msun$, was formed by a trigger due to the collision of the $\hone$ gas at 60 $\kms$.
This interaction resulted in the collisional triggering of the formation of the SSC R136.
Figure~19 shows the $\hone$ distribution of the two velocity components, which exhibits complementary distributions with displacement (Figures~19a, 19c, 19d), and the position--velocity diagram, which displays bridge features connecting the two (Figure~19b).
It is further becoming clear that at least 70 \% of the high-mass stars have been formed by the same event over the whole LMC disk.
In particular, colliding flows have been identified from the $\hone$ data in the N44, N11, and N79 regions \citep{tsu19}.

\begin{figure*}
 \begin{center}
  \includegraphics[width=14cm]{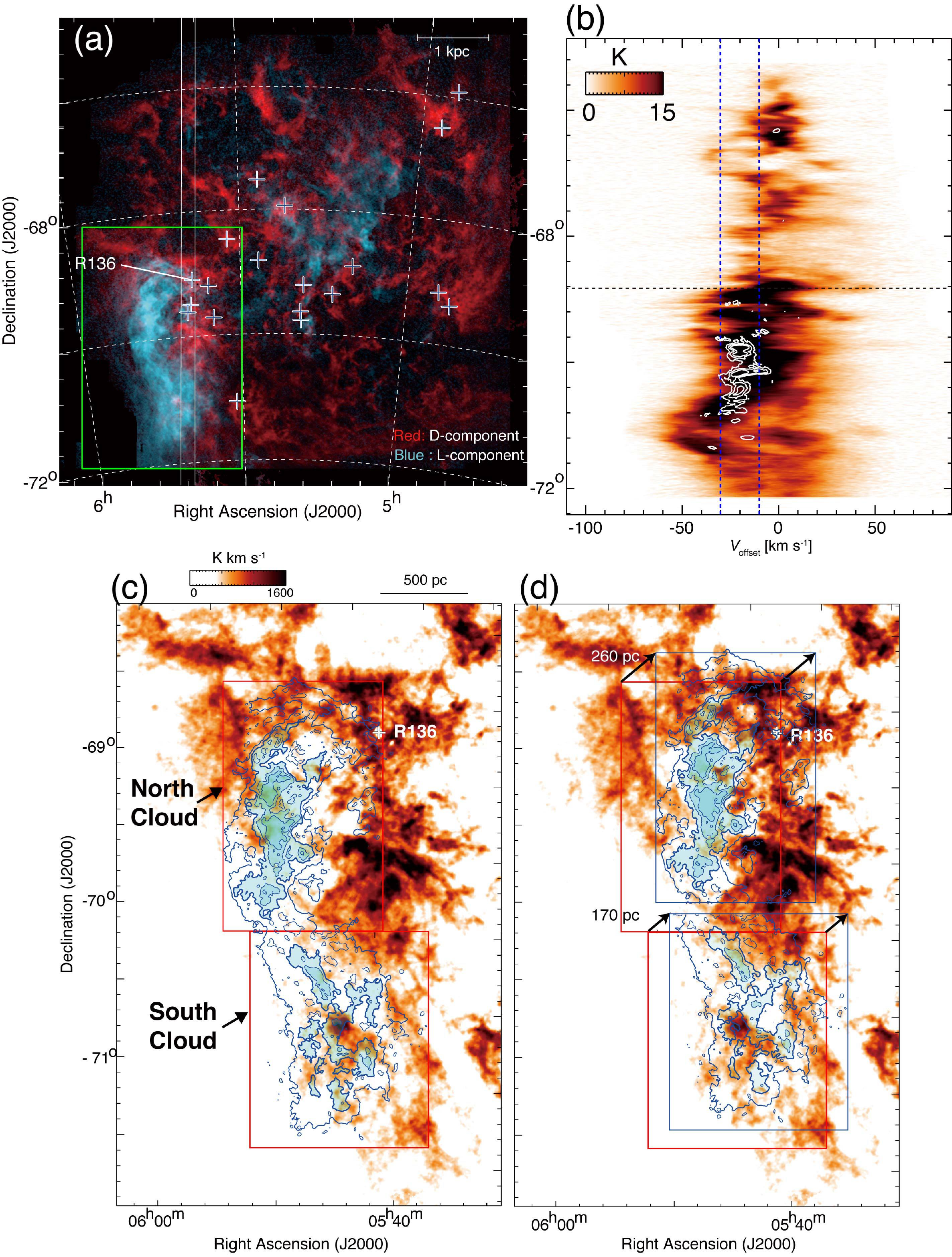} 
 \end{center}
\caption{(a) Two-color composite image of two velocity components of atomic hydrogen toward the LMC. Red corresponds to gas that extends  over the whole disk of the LMC (the D component) and blue to spatially confined gas at lower radial velocities (the L-component), The blue crosses indicate the positions of bright $\htwo$ regions. The white lines, and light-green rectangle indicate the integration ranges for Figure~19b and for the regions shown in figure~19c, respectively. (b) Declination-velocity diagrams of $\hone$ superposed on the CO contours. The position of R136 is indicated by the dashed horizontal line. The contour levels are 0.015, 0.03, and 0.05 K. The blue dashed vertical lines indicate the intermediate velocity range between the L- and D-components. (c) $\hone$ intensity map of the D-component superposed on the L-component's contours of the CO Arc and the Molecular Ridge. The contour levels are 500, 800, 1000, 1200, 1400, 1600, 1800, and 1900 $\kkms$. (d) The same image as (c) but with the contours of the L-component displaced. The projected displacements of the northern and the southern clouds are 260 pc and 170 pc, respectively, at a position angle of 320$^{\circ}$. The red boxes indicate the initial position of the L-component, and the blue boxes in (d) indicate the displaced positions. The blue shaded regions in (c) and (d) indicate where the $\hone$ intensity is greater than 800 $\kkms$. Figures are adapted from \citet{fuk17} and reproduced with permission.}
\label{LMC}
\end{figure*}

One of the theoretical features produced by a CCC is a filamentary gas distribution in the compressed layer \citep{ino13}.
We find such filamentary features in various objects that are undergoing a CCC, including M42, RCW~38, W43, etc.
One spectacular case is found also in N159, which is part of the CO ridge connecting to R136.
N159 is an active site of high-mass star formation that is likely driven by the colliding flows.
It consists of two regions separated by 50 pc, N159 E and N159 W.
Figures~20a and 20b show recent ALMA images of filamentary CO, with tens of well-developed CO filaments \citep{fuk19,tok19}.
The high directivity of these two regions in the north-south direction is remarkable, with a pivot toward the ionizing star and $\htwo$ region in the Papillon Nebula (N159E; see the white contour of Figure~20a) and toward a cluster of young high-mass stars (N159W; see the white contours of Figure~20b).
\citet{tok19} suggested that these massive filaments---with masses an order of magnitude larger than those of the massive filaments observed in the Galactic high-mass star-forming region NGC~6334 \citep{and16}---cannot be dynamically supported by accretion-driven MHD waves, which is a model proposed previously \citep{and16}.
However, these filaments are successfully explained in terms of a model developed in an MHD simulation of a CCC (Figure~20c; see \cite{ino18} for details).

\begin{figure*}
 \begin{center}
  \includegraphics[width=16cm]{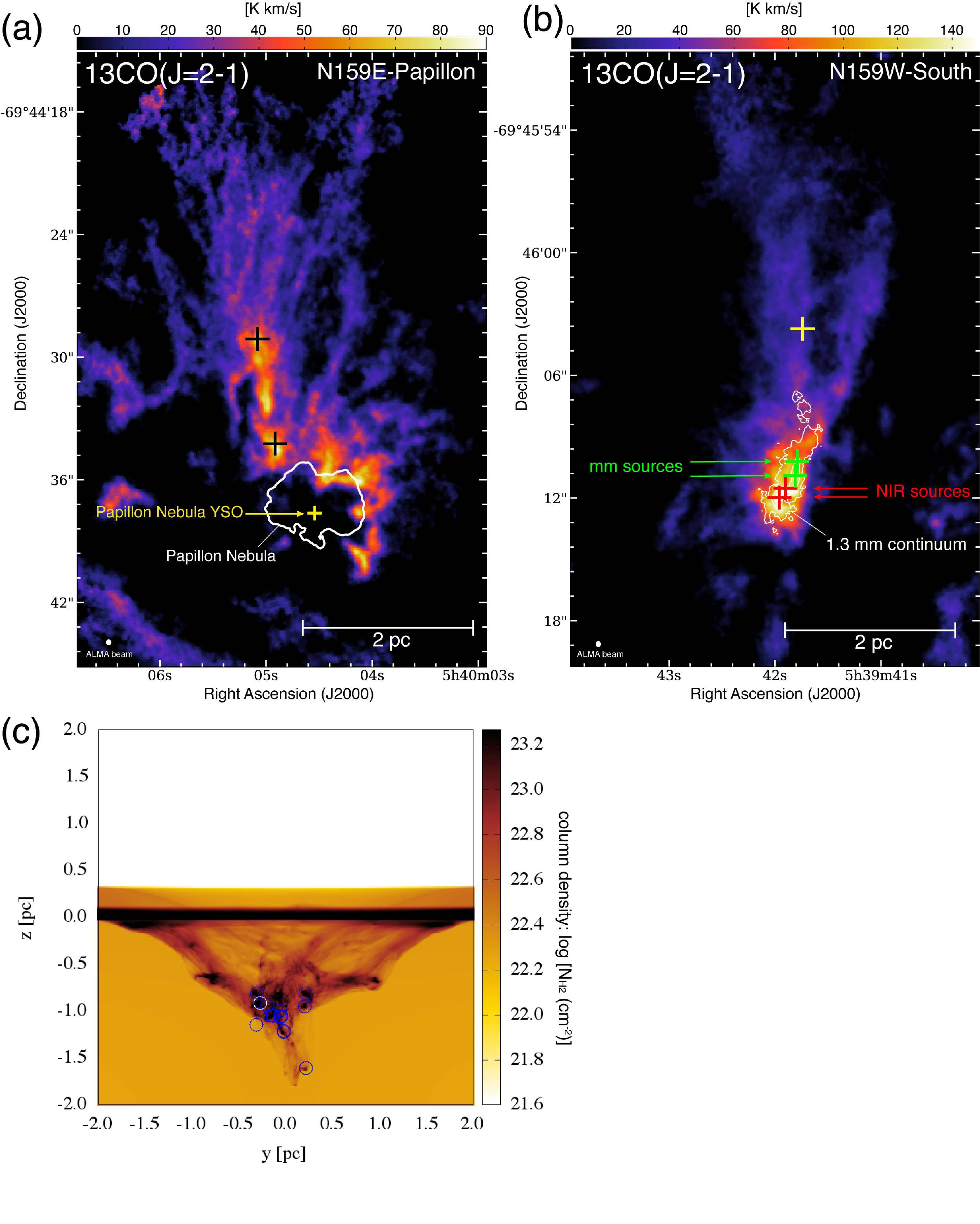} 
 \end{center}
\caption{(a) $\thirteencoh$ image toward the N159E-Papillon region obtained by ALMA Cycle 4 observations. The angular resolution is shown at the lower left corner. The yellow cross represents the position of the Papillon Nebula YSO, given by the 98 GHz continuum peak \citep{sai17}. The black crosses denote the positions of MMS-1 and MMS-2 (see \cite{fuk19} for details). The shape of the Papillon Nebula is traced by H$\alpha$ emission, as shown by the white contour in the panel. (b) $\thirteencoh$ distributions of the N159W-South region obtained by the ALMA Cycle 4 observations. The red, green, and yellow crosses are positions of the infrared sources, the dust-continuum peaks of MMS-3 and 4, and the median position between the red-shifted and blue-shifted outflow lobes (see \cite{tok19} for details), respectively. The white contours indicate the 1.3 mm continuum emission. The angular resolution, 0$\arcsec$.29 $\times$ 0$\arcsec$.25, is given by the white ellipse at the lower left corner. (c) Column density distribution of molecular hydrogen created by the collision of a small molecular cloud ($\sim$2 pc) and a large cloud ($>$4 pc) at $0.7$ Myr after impact at a relative velocity of 10 $\kms$. Circles mark the positions of sink particles (the white one indicates the position of the most massive sink $>$ 50 M$_{\rm sol}$). The simulation was performed by \citet{ino18}. Figures are adapted from \citet{tok19} and \citet{ino18} and reproduced with permission from AAS.}
\label{pck}
\end{figure*}

M33 also interacted with M31 in the past, and the interaction produced an $\hone$ bridge between the two galaxies \citep{loc12}.
\citet{tac18} showed evidence for two $\hone$ velocity components interacting toward NGC~604 in M33, a massive open cluster containing 10$^{6}$ $\msun$, comparable to R136.
It is possible that this interaction triggered the formation of NGC~604, as in R136.
Triggering by tidally driven $\hone$ flows at velocities higher than a few $\times$ 10 $\kms$ may be a common process of high-mass star formation in interacting galaxies.
Compared to the Antennae Galaxies, at 10--100 cm$^{-3}$ the gas is less dense and the velocity of 30--60 $\kms$ is somewhat smaller.
The dwarf galaxy NGC~4449 also shows signs of similar tidal interactions \citep{hun98}, and there are other dwarfs with galactic interactions, which are also candidates for tidal high-mass star formation \citep{lel14}.

\subsubsection{SSC Formation in the Milky Way Disk}
There are not many SSCs in the Milky Way; it harbors about ten SSCs, five of which are as young as 2 Myr \citep{por10}.
In SSCs older than a few Myr, ionization becomes significant, and the collision signatures of molecular gas may be destroyed rapidly, making the identification of a collision difficult.
These SSCs include Westerlund 2 \citep{fur09,oha10}, NGC~3603 \citep{fuk14}, RCW~38 \citep{fuk16}, DBS[2003]179 \citep{kuw20}, and Tr14 in Carina \citep{fuj20b}.
The SSCs in the Milky Way have masses of 10$^{4}$ $\msun$, one to two orders of magnitude smaller than the SSCs formed by galaxy interactions.
It seems impossible to identify parent clouds in more-evolved SSCs like Westerlund 1 with an age of 5 Myr.
It is notable that all the SSCs with remaining molecular gas within $\sim$5 pc of the cluster show collision signatures, lending support for the CCCs as an essential process in SSC formation. 
\\
\\
\uline{Westerlund 2} is a super SSC of 10$^{4}$ $\msun$ that is associated molecular clouds extending over the broad velocity range of 30 $\kms$.
The cluster harbors 20 O-stars and 10$^{4}$ lower-mass members \citep{whi04}.
Based on large-scale CO observations, \citet{fur09} proposed that a collision between two molecular clouds 2 Myr ago triggered the formation of the cluster.
The distributions of the two clouds, which have similar masses and densities, are complementary to each other, and they show bridge features connecting them in velocity, as shown in Figure~21.
The two clouds are shown to be associated with the cluster by the temperature increase in the molecular gas as well as by the morphological correspondence with the infrared emission from the $\htwo$ region \citep{oha10}.
The inner 10 pc of the cluster is already strongly ionized, which is consistent with the $\sim$5 $\kms$ velocity of the ionization front.
The two clouds each contain 10$^{5}$ $\msun$ and are typical GMCs, and the CCC is a typical case of SSC formation in the Milky Way.
An alternative model, which explained the cloud motion in terms of acceleration by the cluster \citep{dam07}, is not supported by the higher-resolution CO data of \citet{fur09}.

NGC~3603 is another SSC, which harbors more than 30 O-stars \citep{mof04}.
NGC~3603 has been extensively studied for the stellar-mass function, which has been found to be significantly top-heavy \citep{eis98,sto06,hir08}.
\citet{fuk14} discovered two molecular clouds with 20 $\kms$ velocity separation and showed that they are dynamically interacting, as shown by the bridge features connecting the two clouds.
The cloud properties are fairly similar to those in Westerlund~2, and the authors suggest that a CCC triggered the formation of NGC~3603 $\sim$1 Myr ago along with Westerlund~2.
\begin{figure*}
 \begin{center}
  \includegraphics[width=16cm]{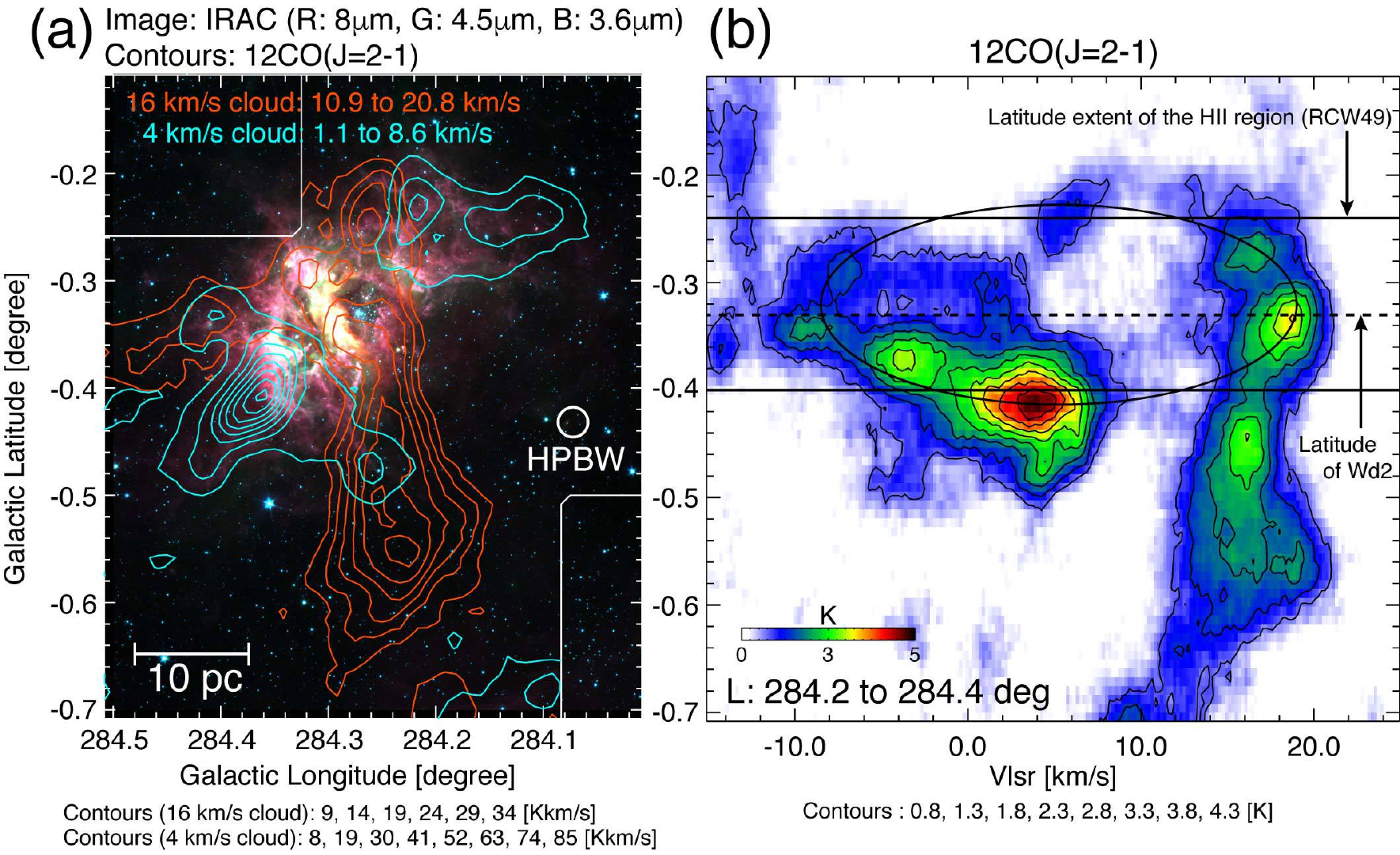} 
 \end{center}
\caption{(a) Contours of the red-shifted (16 $\kms$) and blue-shifted (4 $\kms$) clouds in Westerlund 2 superposed on a three-color composite image obtained with $Spitzer$ (red = 8.0 $\mu$m, green = 4.5 $\mu$m, blue = 3.6 $\mu$m). (b) Velocity-latitude diagram for $\twelvecoh$ emission integrated over the longitude range 284$\fdg$2--284$\fdg$4. The solid black oval shows the velocity signature consistent with expansion around the central cluster and/or bright central regions of RCW~49. The dotted line and solid lines mark the latitudes of Westerlund 2 and the bright ridge of infrared emission, respectively.}
\label{wd2}
\end{figure*}
\\
\\
\uline{The youngest SSC RCW~38}  RCW~38 is the youngest SSC in the Milky Way, with an age of 0.1 Myr. It contains 10$^{4}$ $\msun$, including 20 O-star candidates.
The SSC is still heavily embedded in molecular gas.
\citet{fuk16} detected and mapped two CO clouds toward RCW~38 with a 10 $\kms$ velocity separation, which are linked by a bridge feature toward the cluster.
The two clouds are named the ``ring'' and ``finger'' clouds from their morphology, as shown in Figure~22, and the ring cloud has a central cavity, inside of which the cluster is distributed.
The two clouds show an enhanced line-intensity ratio $\twelvecohh$/$\twelvecol$, indicating that they are associated with RCW~38.
The distributions of the finger and bridge show good spatial correlation with the 20 O-star candidates, and \citet{fuk16} proposed that the ring and finger clouds collided to form the SSC.
The column densities of the two clouds differ by an order of magnitude: the ring cloud has a column density of 10$^{23}$ cm$^{-2}$ while that of the finger cloud is 10$^{22}$ cm$^{-2}$.
The area of the collisional interaction is $\sim$0.5 pc in diameter and shows good correspondence with the O-stars.
The eastern part of the finger cloud is already ionized by the most luminous O5.5 star IRS2, which was probably formed by the collision of the already-ionized part of the clouds.
The collision timescale is estimated to be 0.1 Myr from the cloud size divided by the velocity separation.
This case suggests that the distribution of high-mass star formation is determined by the collision area, thanks to the very young age and on-going collision. 
\begin{figure}
 \begin{center}
  \includegraphics[width=7cm]{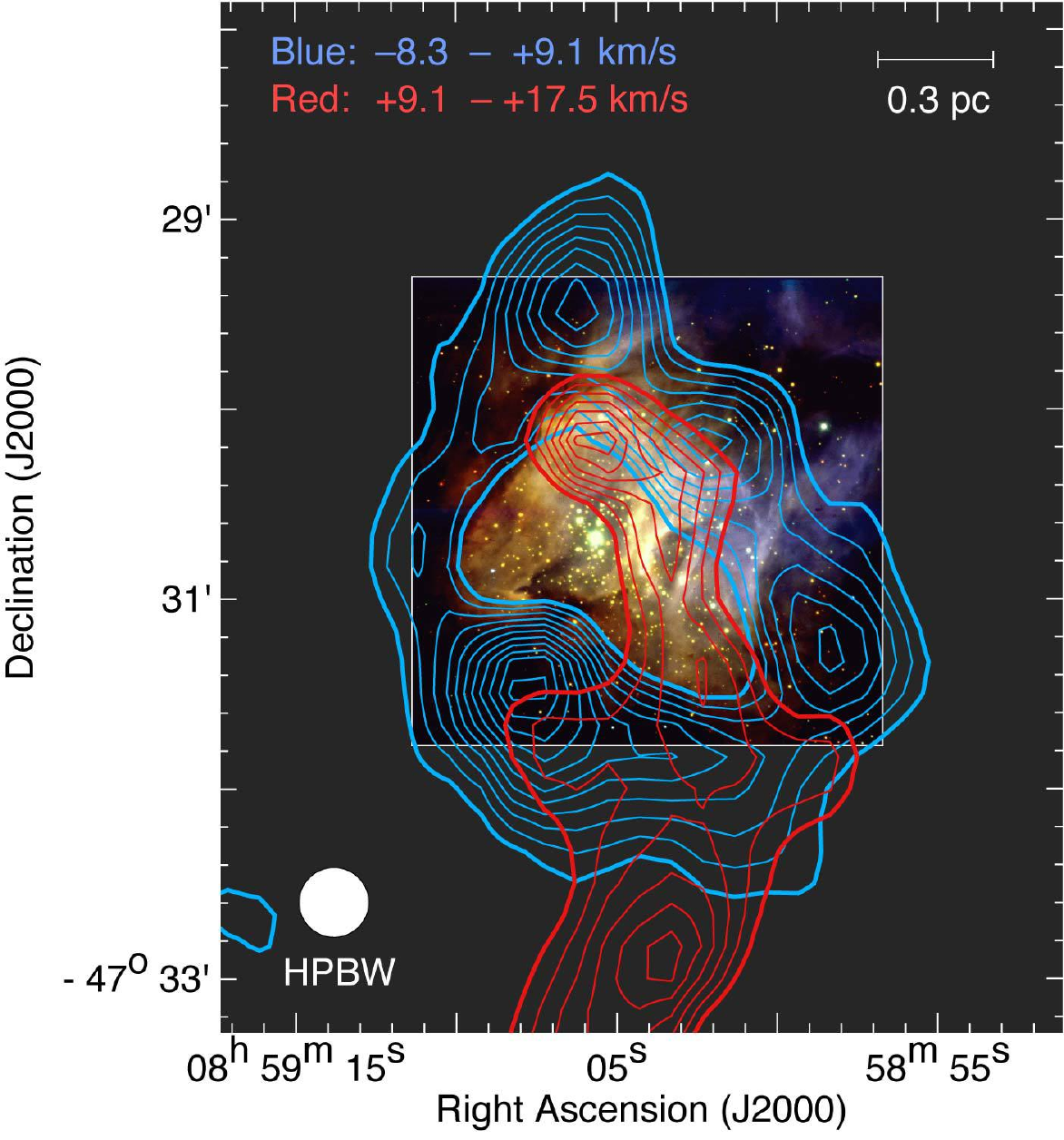} 
 \end{center}
\caption{$\twelvecohh$ distributions of the ring cloud (blue contours) and the finger cloud (red contours) superimposed on an optical image obtained with the VLT. The contours are plotted at every 15 $\kkms$ from 130 $\kkms$ for the ring cloud and at every 5 $\kkms$ from 20 $\kkms$ for the finger cloud. Adapted from \citet{fuk16} with permission from AAS.}
\label{38}
\end{figure}
\\
\\
\uline{Tr~14 in $\eta$ Carinae}    The $\eta$-Carinae region is one of the most active regions of star formation in the Milky Way.
It harbors several clusters, of which Trumpler 14 (Tr~14), with an age of 1 Myr, is the youngest.
\citet{fuj20b} carried out an extensive analysis of CO $J$=1--0 and 2--1 data obtained with Mopra and NANTEN2 and found that the molecular gas in the region consists of at least four components with different velocities over a range of 35 $\kms$.
In particular, Tr~14 is associated with the molecular gas showing the brightest CO emission, and a complementary distribution between the two clouds has been identified.
Similarly, Tr~15 and Tr~16 are associated with molecular clouds.
The gas has already been ionized, in a timescale of a few Myr, while colliding-cloud candidates with complementary distributions have been identified at the 5--10 pc scale.
\\
These SSCs---as well as some others, like DBS[2003]179 \citep{kuw20}---lend support to the hypothesis that a CCC is an essential process in triggering the formation of an SSC.
\\
\\
\uline{SSCs in The Central Molecular Zone} The velocity dispersion is very large in the CMZ in the Galactic Center, and star formation is strongly suppressed (e.g., \cite{lon13a}).
The cause of the large velocity dispersion is ascribed either to the gas motions driven by the bar potential \citep{bin91} or to the magnetic instabilities in the strong magnetic field \citep{suz15,kak18,fuk06}.
There are three SSCs in addition to Sgr~B2 that include many young high-mass stars.
Sgr~B2 is a typical case of a collision, as first suggested by \citet{has94} and as shown in Figure~23, and a CCC with the very high column density of 10$^{24}$ cm$^{-2}$ has recently been confirmed by Enokiya et al. (2020b submitted; for an alternative intepretation, see \cite{lon13b}).
It is probable that the collision triggered the active high-mass star formation in Sgr~B2.

Shock heating is not effective for collision velocity on the order of 10 $\kms$, since molecular cooling is very rapid, with cooling times on the order of 10$^{4}$ yr \citep{ino13}.
The molecular gas undergoing a CCC is therefore usually considered to be isothermal.
However, the CMZ is exceptional, due to the large velocity span of $\sim$100 $\kms$, and the temperature rise and shock chemistry---such as SiO overabundance---are used as shock tracers \citep{tor10, tsu15a, eno19}.
In galactic tidal interactions $\hone$ gas can collide at a high velocity, on the order of 100 $\kms$, which is able to heat the $\hone$ gas to high temperatures.
Such heating, which is caused by the galactic tidal interactions, is observed as thermal X-rays in the LMC (Knies et al. 2020 submitted), and it may serve as a shock tracer in interacting galaxies.

The 50 $\kms$ molecular cloud close to Sgr A was analyzed by \citet{tsu15b} in SiO and H$^{13}$CO$^{+}$ emissions, and they found the signatures of a CCC by using 45-meter data.
\citet{ueh19} used the ALMA data to follow up the CCC scenario.
The core mass function derived in the region of the CCC shows a clear top-heavy trend.
In addition, \citet{eno19} showed that the molecular clouds toward the footpoints of loops 1 and 2 and the $l$=1\fdg3 complex show collision signatures, but they are not associated with star formation, probably due to the low column density as compared with the extremely large turbulent-velocity field.
The Arches and Quintuplet clusters are associated with molecular gas, but it is not yet clear if they were formed by CCCs.
The large velocity dispersions and high cloud density in the CMZ make it difficult to disentangle the collision signatures, even though collisions are common phenomena, as shown above.
It may be possible that the Arches and Quintuplet were formed by CCCs a few Myr ago at the positions of Sgr~C and Sgr~B2 \citep{sto14}, respectively, but the two clusters have since moved more than 100 pc at their large velocities of 100--200 $\kms$, and the crowded environment makes the identification of their parent clouds uncertain \citep{sto14}.
\begin{figure}
 \begin{center}
  \includegraphics[width=8cm]{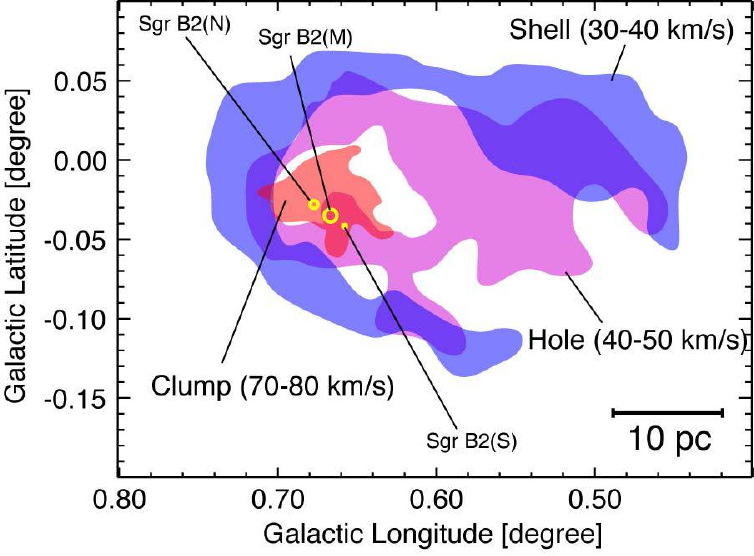} 
 \end{center}
\caption{A schematic picture of the kinematic features associated with the colliding clouds near the Sgr~B2 region constructed by the latest $^{13}$CO and C$^{18}$O ($J$=2--1) data obtained by \citet{gin16}. The red-shifted cloud (the clump) collided to the blue-shifted cloud (the shell) and generated three condensations of high-mass stars, namely Sgr~B2(N), Sgr~B2(M), and Sgr~B2(S), and formed the intermediate velocity feature (the hole).}
\label{b2}
\end{figure}

\subsection{Cluster Mass vs. Pressure in SSCs}
Figure~24 shows the masses of clusters as a function of the gas pressure in the clouds in the collision region, where the clouds in the Antennae Galaxies, the LMC, and Westerlund 2 are shown \citep{tsu20a}.
A theoretical study of cluster formation predicts a correlation between these parameters \citep{elm97}, and this seems to be consistent with Figure~24.
The theoretical pictures of cluster formation are, however, not necessarily the same as the current picture, which is modeled by supersonic motions over a short timescale under non-gravitational gas compression (Section 4). It is therefore preferable to construct a more-sophisticated model of cluster formation that fully incorporates the timescale in such a correlation analysis.
The plot is to be extended further at the low-mass and high-mass ends to cover more orders of magnitude in cluster mass. 
\begin{figure}
 \begin{center}
  \includegraphics[width=7cm]{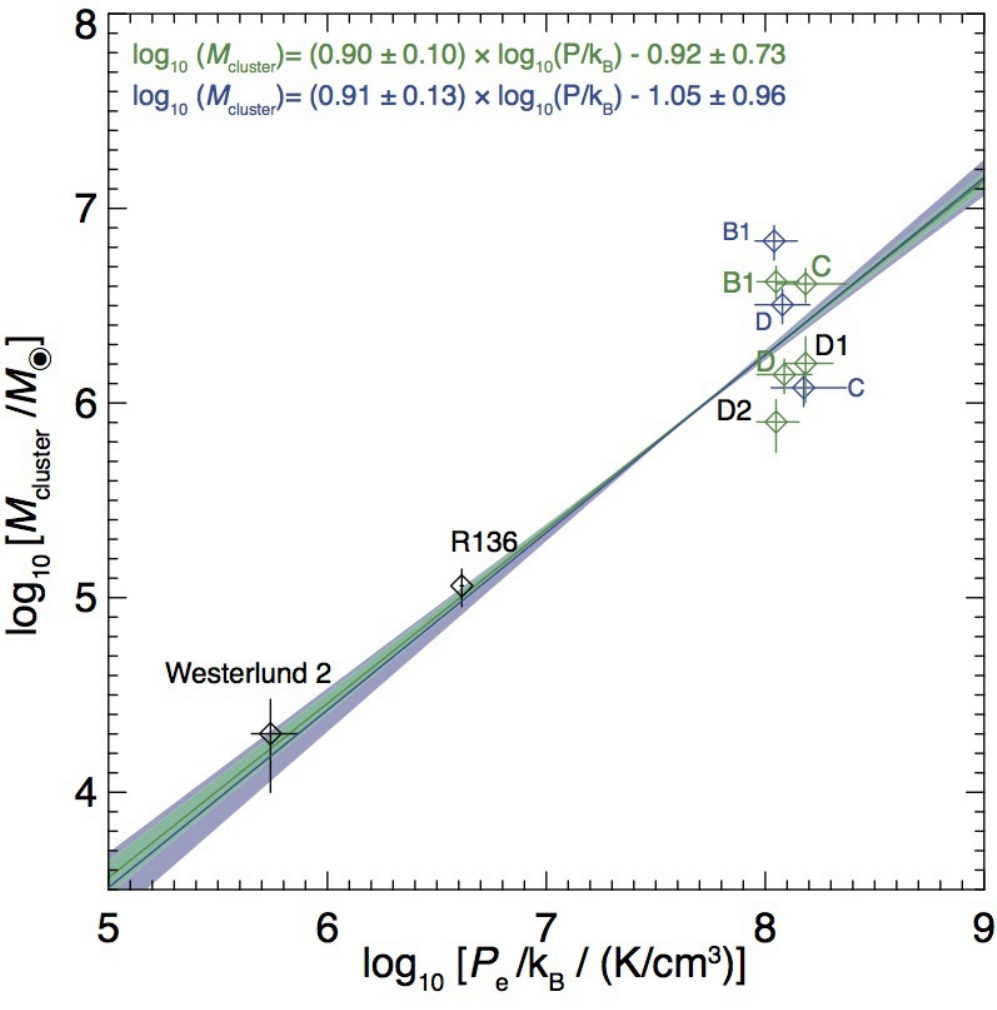} 
 \end{center}
\caption{Correlation plot between the external pressure ($P_\mathrm{e}$/$k_\mathrm{B}$) and total stellar mass on SSC ($M_\mathrm{cluster}$). The masses of SSCs in the Antennae cataloged by \citet{gil07} are shown as green diamonds. The blue diamonds represent the masses of SSCs B, C, and D cataloged by \citet{whi10}. Representative values of the $P_\mathrm{e}$/$k_\mathrm{B}$ and $M_\mathrm{cluster}$ are plotted as diamonds with error bars. The colored area and solid lines show various correlations and linear regressions from $\chi$$^{2}$ fittings, respectively. Light-green and blue indicate the results obtained using the values of cluster mass cataloged by \citet{gil07} and \citet{whi10}, respectively. Adapted from \citet{tsu20a} with permission.}
\label{Pex}
\end{figure}

\subsection{Formation of O-stars in the Milky Way}
$\htwo$ regions ionized by OB stars are primary areas in which CCCs may be occurring in the Milky Way.
Most $\htwo$ regions near the Sun are found to be associated with colliding clouds.
They include many Messier objects and Spitzer bubbles as well as other $\htwo$ regions.
One expects that in old $\htwo$ regions the parent clouds will already have been dispersed by stellar feedback, making it difficult to identify a CCC based on the molecular data.
We summarize some of the outstanding cases of CCCs below.

\subsubsection{M43: An $\htwo$ Region with a Single Early B Star}
The Orion Nebula M42 is the best-known $\htwo$ region in the Galaxy.
\citet{fuk18a} proposed that the molecular cloud, which looks like a single cloud, can actually be deconvolved into two velocity components with a 4 $\kms$ velocity separation, which show significantly different spatial distributions.
Figure~25 shows that the two components have distributions that are complementary to each other, as is typical of a CCC, and the authors argued that the blue-shifted cloud, including $\theta$1 Ori, the Trapezium stars, $\theta$2 Ori, and the Orion KL object---a high-mass protostar candidate---were formed by a CCC on a timescale of 0.1 Myr.
A detailed scenario for cluster formation is presented, in which the low-mass members of the cluster were formed prior to the CCC over a few Myrs, and the recent CCC triggered the formation of the nearly ten O-stars within 0.1 Myr.
\begin{figure*}
 \begin{center}
  \includegraphics[width=12cm]{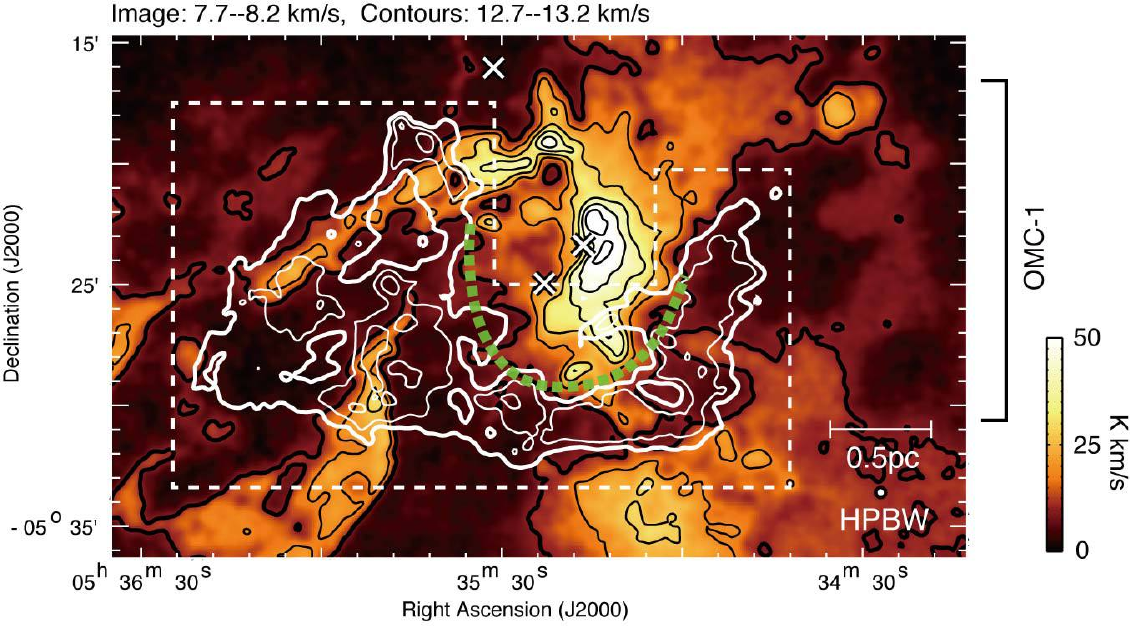} 
 \end{center}
\caption{Complementary distributions of the two velocity distributions toward OMC-2. The image with black contours and the white contours indicate the blue-shifted and red-shifted clouds, respectively. The contact faces of the two clouds, with complementary distributions, are indicated by the green dashed line. The velocity ranges for the blue-shifted and red-shifted clouds are 7.7--8.2 and 12.7--13.2 $\kms$. The lowest level is 5.5 $\kkms$, and the level internal to the white contours is 4.0 $\kkms$, while the corresponding levels for the black contours are 8 $\kkms$ and 7 \kkms. NU Ori [(R.A., decl.) = (5h35m31s, $-$5$^{\circ}$16$\arcmin$03$\arcsec$)], Ori C [(R.A., decl.) = (5h35m16s, $-$5$^{\circ}$23$\arcmin$23$\arcsec$)], and $\theta$2 Ori A [(R.A., decl.) = (5h35m23s, $-$5$^{\circ}$24$\arcmin$58$\arcsec$)] are plotted with white crosses. Adapted from \citet{fuk18a} and reproduced with permission.}
\label{M42}
\end{figure*}

\subsubsection{NGC6334: An $\htwo$ Region with a Massive Cluster}
NGC~6334 is an active region of high-mass star formation, with an $\htwo$ region distributed over 10 pc.
Six dust/gas condensations---named I(N) and I-V---have young ages, ranging from 10$^{4}$ to 10$^{6}$ yr, and they are associated with at least $\sim$10 O-star candidates, which are deeply embedded in the six condensations.
Toward this region, the molecular cloud at $-$4 $\kms$ cloud is lined up along the Galactic plane.
\citet{fuk18b} discovered another weak blue-shifted molecular cloud at $-$17 $\kms$, with a velocity separation of 10 $\kms$.
The secondary cloud, extending over the main cloud, shows bridge features connecting with the main cloud toward the condensations.
The bridge features suggest that a CCC is taking place.
The column density of the main cloud is 10$^{23}$ cm$^{-2}$, while that of the secondary cloud is 10$^{22}$ cm$^{-2}$.
The two clouds show similar peak positions, which the authors interpreted as due to the O-stars formed in the very early phase of this CCC because of the high density of the main cloud \citep{fuk18b}.

\subsubsection{W43: The Galactic Mini-Starburst}
W43 is a Galactic mini-starburst that is particularly active in high-mass star formation, including at least 50 O-stars (for a review see \cite{mot18b}).
The molecular cloud complex is a remarkably massive one in the Milky Way, with a total mass of 7.1 $\times$ 10$^{6}$ $\msun$. It extends over 150 pc and lies at a distance of 5.4 kpc \citep{bal10}.
\citet{sof19} and \citet{koh20} studied the molecular distribution in detail by using the CO $J$=1--0 data taken in the FUGIN project, which covers a large region with high resolution.
W43 has three components, each of which has two-to-four velocity components.
Some two or three of the clouds were found to be connected by bridge features, as shown in Figure~26, and a CCC is a plausible scenario for high-mass star formation, as shown by the CO distributions.
\citet{ngu11} and \citet{mot14} presented a picture that W43, located at the end of the Galactic central bar, was formed by an extensive accumulation of molecular gas a stagnation point driven by the bar potential.
W43 is probably the nearest object with such high activity, and it offers a possible link to distant starbursts.
\begin{figure*}
 \begin{center}
  \includegraphics[width=14cm]{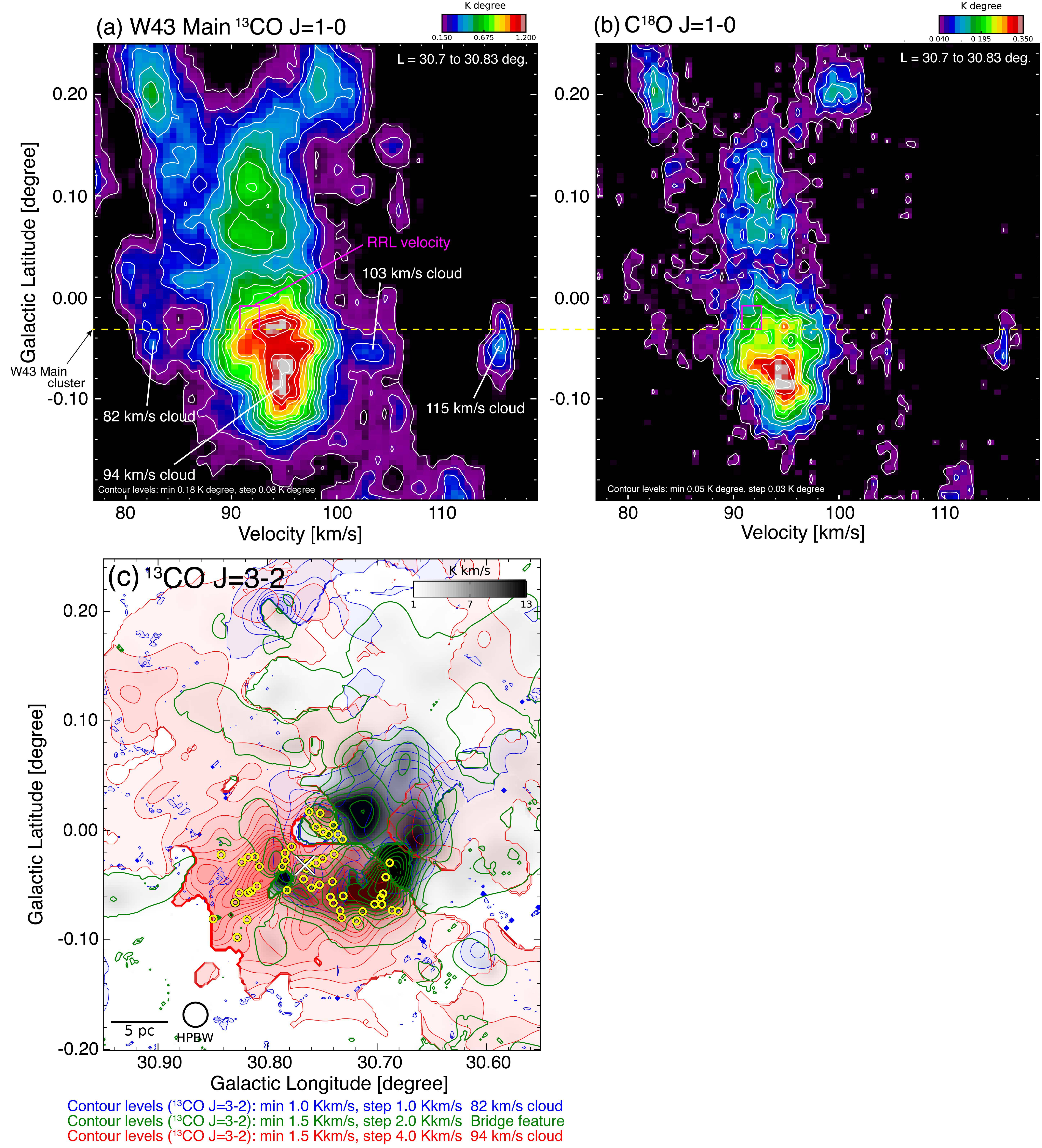} 
 \end{center}
\caption{Galactic latitude-velocity diagram of (a) $\thirteenco$ and (b) $\ceighteeno$ $J$ = 1--0 integrated over the latitude range from 30\fdg7 to 30\fdg83. The contour levels and intervals are 0.18 K degree and 0.08 K degree for (a) and 0.05 K degree and 0.03 K degree for (b). The black boxes show the radio recombination-line velocity (91.7 $\kms$) at ($l$, $b$) = (30\fdg780, $-$0\fdg020) from \citet{lui17}, with the resolution $\sim$1.86 $\kms$ $\times$ 82$\arcsec$. The yellow dashed line indicates the position of the W43 Main cluster \citep{blu99}. (c) The spatial distributions of bridges (green contours) superposed on the blue and red-shifted clouds (blue and red contours) in W43 Main. The yellow circles indicate the 51 protocluster candidates (W43 MM1-MM51) cataloged by \citet{mot03}. The white cross shows the position of the W43 Main cluster \citep{blu99}. Adapted from \citet{koh20} and reproduced with permission.}
\label{W43}
\end{figure*}

\section{Discussion and Concluding Remarks}
We have reviewed the observational and theoretical studies of star formation triggered by CCCs.

The article is based on the observational data for more than 50 CCC candidates published in recent decades, with the Habe-Ohta model of a head-on CCC was adopted as the typical picture.
The main points of this article are summarized as follows:
\\

\begin{enumerate}
  \item \uline{Frequency of CCCs in a galaxy}: It is becoming clear that a CCC is an important process in triggering star formation.
Numerical simulations of galactic-scale gas dynamics show that the mean free time between CCCs is $\sim$10 Myr.
This is consistent with the collision frequency of molecular clouds containing 10$^{4}$ $\msun$---one collision per 100 years over the Galaxy---which is similar to thefrequency of the core-collapse SNe.
If we assume that one 15 $\msun$ star is formed in such a collision, the O-star formation rate is estimated to be 0.15 $\msun$/yr, which is on the order of several \% of the total star-formation rate in the Galaxy.
Accordingly, we suggest that CCCs are the dominant mechanism of high-mass star formation.\\

  \item \uline{Cloud morphology and observable signatures}: Theories of two-cloud collisions \citep{hab92} show three observational signatures of CCC: i) Complementary distributions with displacement, ii) a bridge in velocity space between the two clouds, and iii) a U shape in the final phase.
The gas dynamics is highly directional due to the collision velocity, which makes a sharp contrast to the isotropic gas dynamics of a turbulent cloud without a collision.
These collision signatures are used to identify CCCs involving gas clouds over a large range of cloud sizes, from 1 pc to 1 kpc.
Identification is not straightforward, however, if the cloud morphology is irregular or the projection effect is significant. Stellar feedback, including ionization and protostellar outflows, also make it difficult to establish a collision firmly in some cases, but this weakness is being overcome with the increasing number of the samples.\\

  \item \uline{Observed star/cluster formation triggered by a CCC}: Observations show that more than 50 regions of star formation have been identified as sites of high-mass star formation triggered by CCCs \citep{eno19}.
They include SSCs, $\htwo$ regions ionized by O-star clusters and isolated O-stars.
SSCs with masses of 10$^{4}$ $\msun$ to 10$^{7}$ $\msun$ are located in interacting galaxies---the Antennae Galaxies, the Magellanic System, and M33---and there are several young SSCs in the Milky Way.
Many young $\htwo$ regions show the signatures of CCCs---for example, in the Orion A and Orion B regions, the Sagittarius Arm, and other sites of active star formation.
The large number of observed cases of collision lends support for CCCs as one of the dominant mechanisms of high-mass star formation.\\
  \item \uline{Observed physical parameters of SSCs in CCC in the external galaxies}: Collisions take place between both atomic and molecular clouds.
In collisions of dwarf galaxies (the LMC and M33), $\hone$ collisions are driven by galactic tidal interactions.
The close encounter between the galaxies strips $\hone$ gas, resulting in $\hone$ flows that subsequently collide at 30--60 $\kms$, triggering the formation of SSCs and high-mass stars.
Mergers between grand-design spiral galaxies also drives collisions between molecular clouds, which triggers the formation of massive SSCs, as in the Antennae Galaxies.
A preliminary analysis \citep{tsu20a} suggests a possibility that the masses of the SSCs are correlated with the pressure in the collision, as predicted theoretically \citep{elm97}.\\

  \item \uline{Observed physical parameters and threshold column density in CCC of the Milky Way}: In the disk of the Milky Way, CCCs have been identified between molecular clouds.
The typical cloud size, mass, and collision velocity are 1--5 pc, 10$^{3}$--10$^{5}$ $\msun$, and 5--15 $\kms$.
The formation of O-stars has a threshold value in molecular column density.
Formation of a single O-star takes place at column densities larger than 10$^{22}$ cm$^{-2}$, and the formation of more than ten O-stars requires column densities more than 10$^{23}$ cm$^{-2}$.
In the central region of the Milky Way, the threshold values seem to shift upward, and column densities even higher than $\sim$10$^{24}$ cm$^{-2}$ are required for high-mass star formation in order to compress gas with a large velocity dispersion.\\

   \item \uline{Gas compression and top-heavy core mass function in a CCC}: A collision is an efficient mechanism for the formation of dense, massive molecular cloud cores according to \citet{fuk20}.
In the compressed layer, dense gas is organized into filaments, and dense cloud cores are formed in the filaments.
The growth in mass of a filament is due to flow guided along the magnetic field.
The core mass function of the dense cores is top-heavy, with a significant fraction of the massive cores having 6--60 $\msun$ and sizes of 0.01--0.1 pc.
The mass function is markedly different from the IMF, which has a peak at a stellar mass of 0.1--1 $\msun$.
The core mass function is consistent with recent ALMA observations of RCW~38 and W43.
The collision area and column density determine the shape and the number of stars formed by a collisional trigger.
This allows a variety of cluster masses and shapes, as observed, ranging from a single O-star to more than 100 O-stars.
The SFE is estimated to be 1.6 \%--10 \% for gas density of 10$^{4}$ cm$^{-3}$--10$^{5}$ cm$^{-3}$, not particularly high in spite of the strong compression.
So, the impact of CCCs lies in the formation of high-mass stars.\\

   \item \uline{High-mass star formation in a cloud-cloud collision}: High-mass star formation in a cloud-cloud collision is a rapid process that is governed by the typical collision timescale of 10$^{5}$ yr.
The mass-accretion rate in the layer is enhanced to 10$^{-4}$ $\msun$/yr to 10$^{-3}$ $\msun$/yr \citep{ino13}.
A 30 $\msun$ star is formed in 10$^{5}$ yr by accretion at a constant rate of 3 $\times$ 10$^{-4}$ $\msun$/yr.
Such a short timescale is consistent with the short, 10$^{5}$ yr duration of star formation measured in the two SSCs NGC~3603 and Westerlund 1.\\

   \item \uline{Starbursts as an extreme case of high-mass star formation}: Recent observational studies show that CCCs are responsible for starbursts, including those in the Antenna Galaxies, R136 in the LMC, NGC~604 in M33, and Sgr~B2 in the Galactic Center.
It has also been shown that multiple CCCs are taking place and triggering the Galactic mini-starburst in W43.
It is important to accumulate high-resolution data on the molecular gas and young stars in active extragalactic star-forming regions in order to clarify the role of CCCs in even-more-active galaxies.
\end{enumerate}


\begin{ack}
We thank an anonymous referee for useful comments that helped improve the manuscript.
We are grateful to the following scholars for their valuable discussions on this work and contributions in the molecular observations of candidates for cloud-cloud collisions; Kazufumi Torii,  Akio Ohama, Toshikazu Onishi, Shinji Fujita, Mikito Kohno, Atsushi Nishimura, Katsuhiro Hayashi, Hidetoshi Sano, Kazuki Tokuda, Kazuyuki Muraoka, and Yusuke Hattori.
The authors would like to thank Enago (www.enago.jp) for the English language review.
This work is based [in part] on observations made with the Spitzer Space Telescope, which is operated by the Jet Propulsion Laboratory, California Institute of Technology under a contract with NASA.
This work was financially supported Grants-in-Aid for Scientific Research (KAKENHI) of the Japanese society for the Promotion for Science (JSPS; grant number society for 15K17607, 15H05694, and 20H01945). \end{ack}

\clearpage
\begin{figure}[!htbp]
  \begin{center}
    \includegraphics[width=16.5cm]{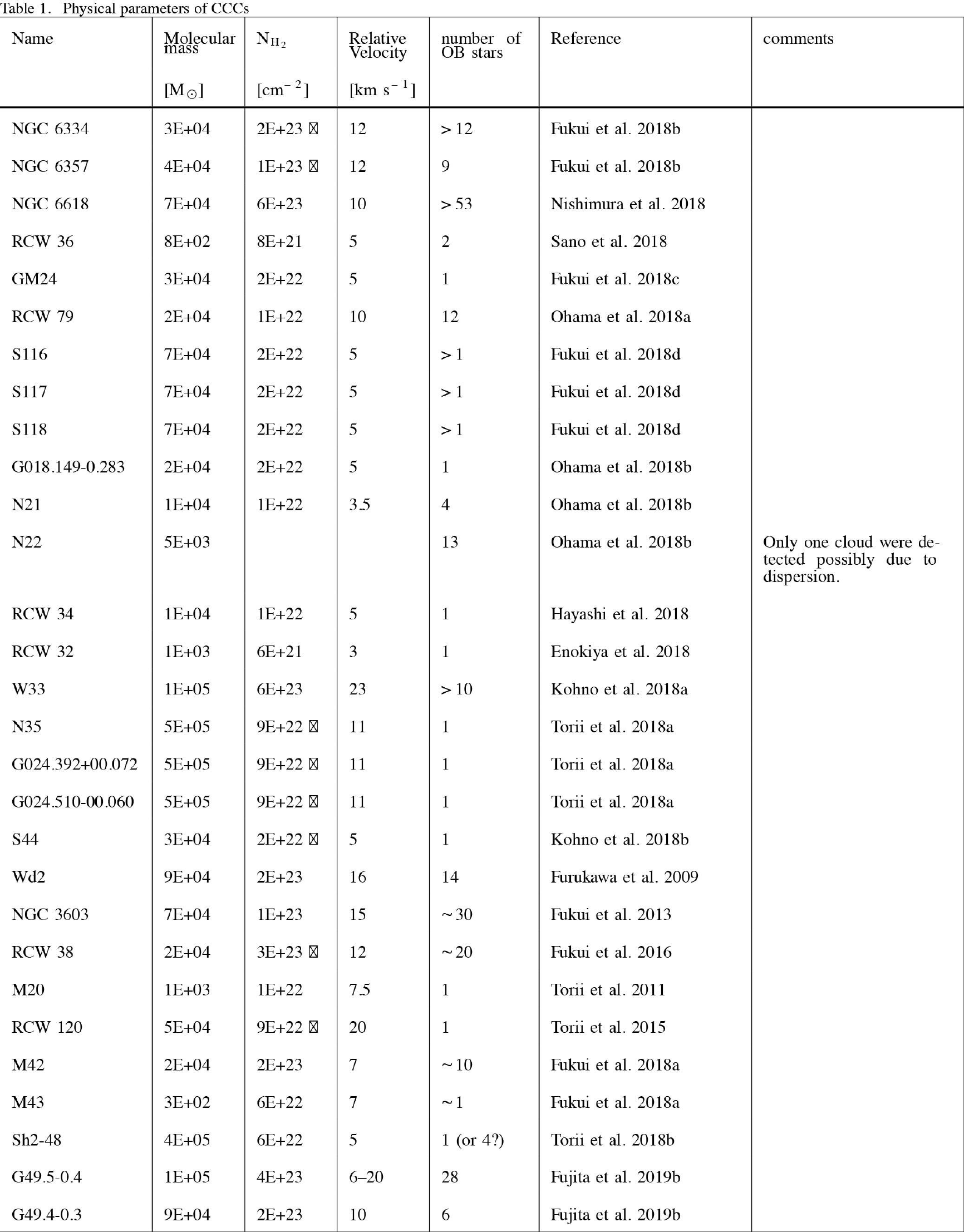}
  \end{center}
\end{figure}

\clearpage

\begin{figure}[!htbp]
  \begin{center}
    \includegraphics[width=16.5cm]{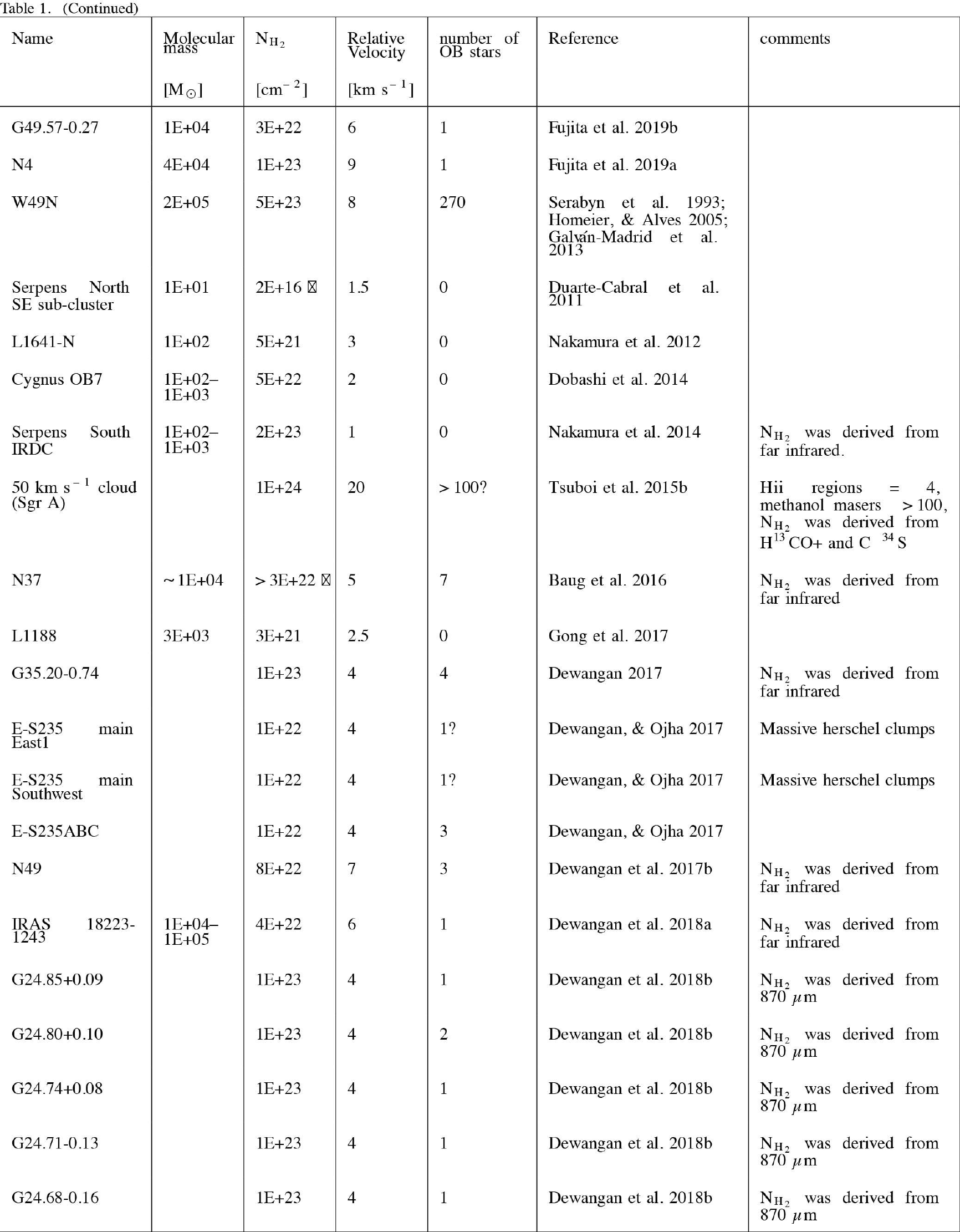}
  \end{center}
\end{figure}

\clearpage

\begin{figure}[!htbp]
  \begin{center}
    \includegraphics[width=16.5cm]{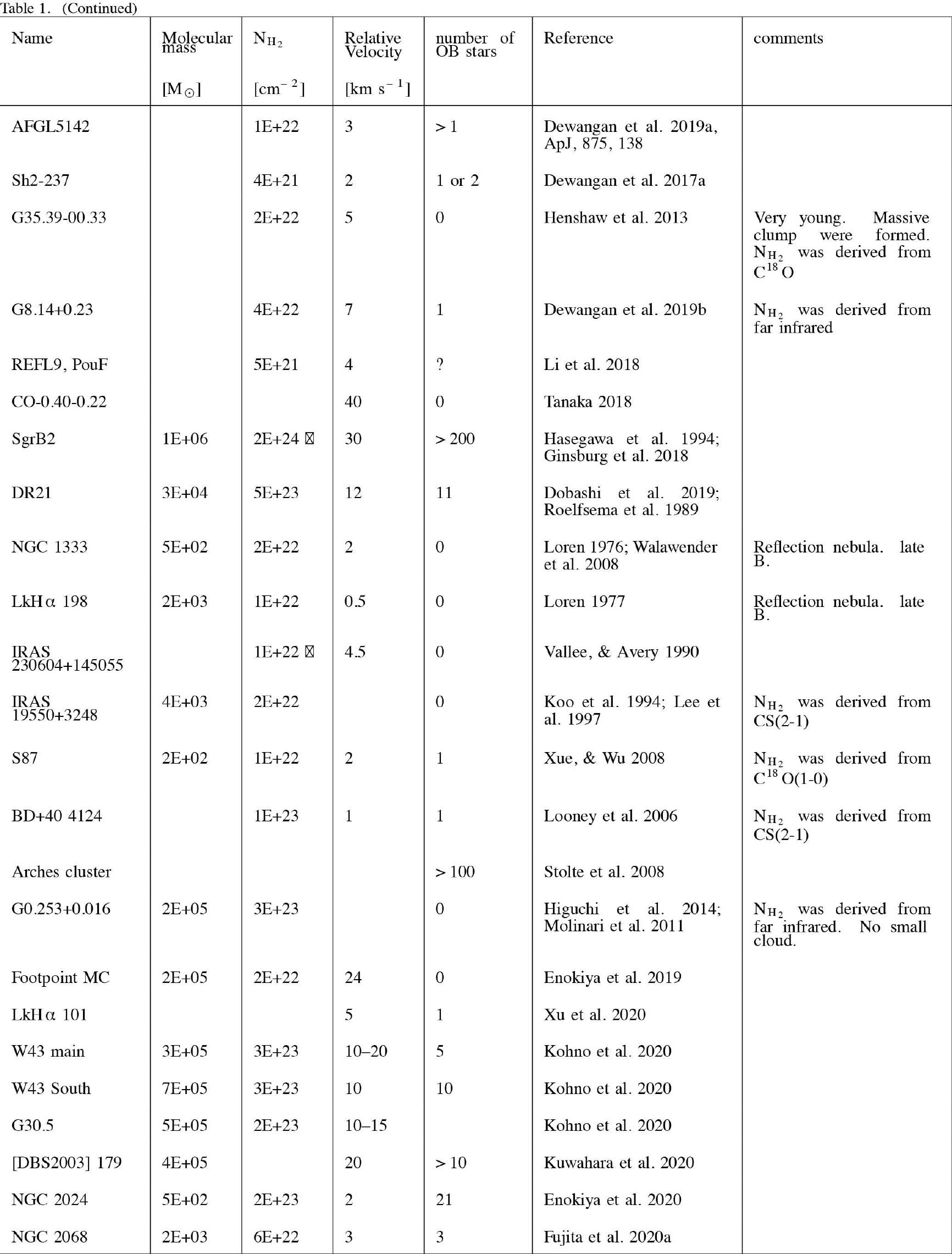}
  \end{center}
\end{figure}

\clearpage

\begin{figure}[!htbp]
  \begin{center}
    \includegraphics[width=16.5cm]{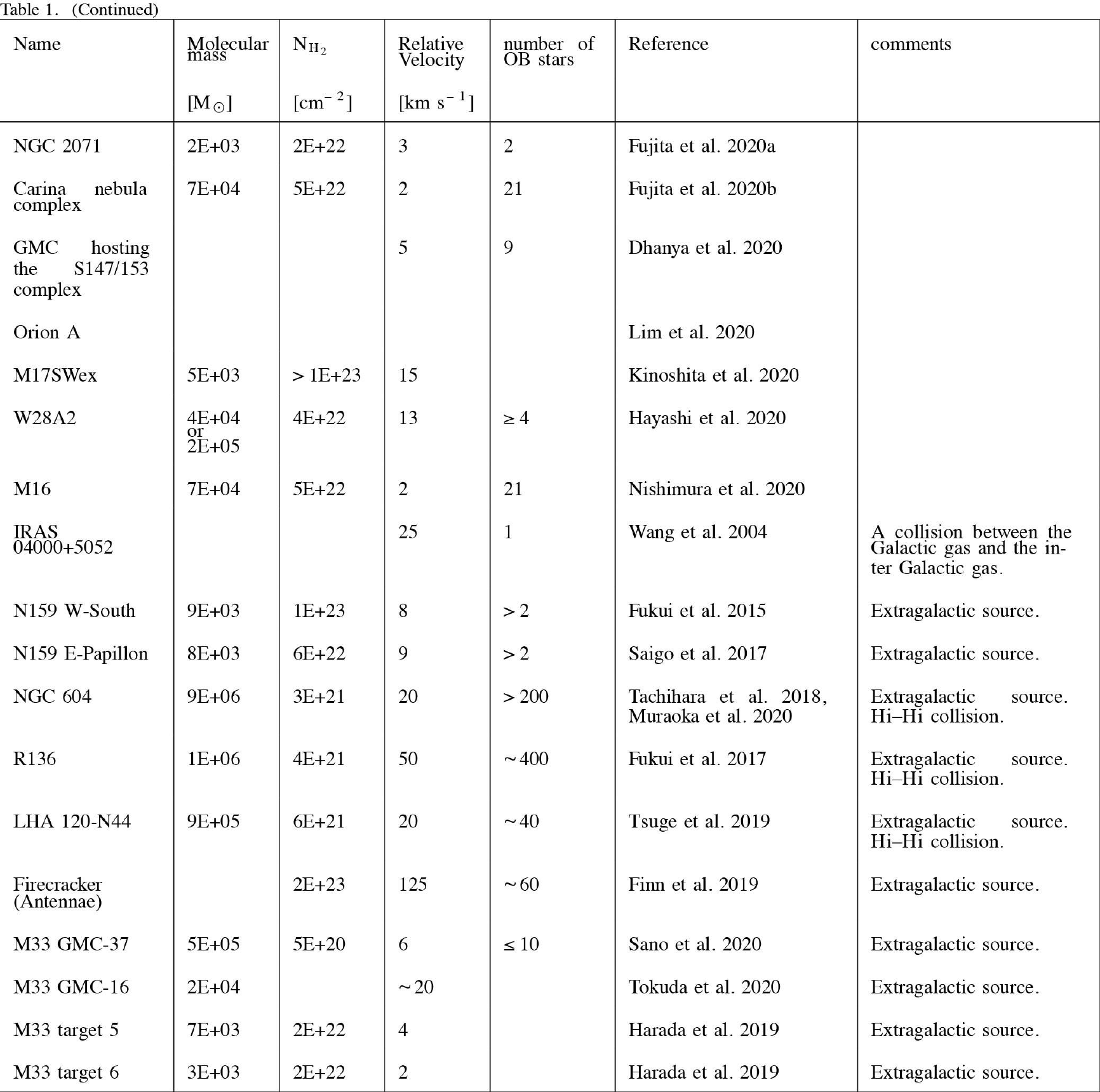}
  \end{center}
\end{figure}

\clearpage

\setcounter{table}{1}
\begin{table*}[h]	
\begin{center}
\tbl{Parameters of the Numerical Simulations \citep{tak14}.}{%
\begin{tabular}{@{}cccccc@{}} \noalign{\vskip3pt}
\hline\hline
\multicolumn{1}{c}{Box size [pc]} & $30 \times 30 \times 30$  &  &  &  \\ [2pt]
\noalign{\vskip3pt}
Resolution [pc] & 0.06 &  &  \\
Collsion velocity $V_0$ [km s$^{-1}$] & {10 (7)$^{\dag}$} & & \\
\hline
Parameter & The small cloud & The large cloud & note    \\
\hline
Temperature [K] & 120 & 240 &  \\
Free-fall time [Myr] & 5.31 & 7.29 &  \\
Radius [pc] & 3.5 & 7.2 &  \\
Mass [$M_{\odot}$] & 417 & 1635 &  \\
Velocity dispersion [km s$^{-1}$] & 1.25 & 1.71 &  \\
Density [cm $^{-3}$] & 47.4 & 25.3 & Assumed a Bonner-Ebert sphere  \\
\hline\noalign{\vskip3pt} 
\end{tabular}} 
\label{init} 
\begin{tabnote}
\footnotemark[$\dag$] The initial relative velocity between the two clouds is 10 $\kms$, but the collisional interaction decelerates the relative velocity to about 7 $\kms$ by 1.6 Myrs after the onset of the collision. The present synthetic observations correspond to a relative velocity $V_0=7$ $\kms$ at 1.6 Myrs. Adapted from \citet{fuk18c} with permission.
\end{tabnote} 
\end{center} 
\end{table*}

\begin{table*}
\tbl{Model Parameters \citep{ino13}.}{%
\begin{tabular}{lr@{\,}l}  
\hline\noalign{\vskip3pt} 
\multicolumn{1}{c}{Parameter} & \multicolumn{2}{c}{Value} \\  [2pt] 
\hline\noalign{\vskip3pt} 
$\langle n\rangle_{0}$ & 300 & cm$^{-3}$ \\
$\Delta n/\langle n\rangle_{0}$ & 0.33 & \\
$B_{0}$ & 20 & $\mu$G \\
$V_\mathrm{coll}$ & 10 & km\,s$^{-1}$ \\
Resolution & (8.0/512) & pc \\
\hline\noalign{\vskip3pt} 
\footnotemark[$\dag$] Adapted from \citet{fuk18a} with permission from AAS.
\end{tabular}}\label{modelparam}
\end{table*}

\clearpage




\begin{thebibliography}{}
\bibitem[Anathpindika(2010)]{ana10} Anathpindika, S.~V.\ 2010, \mnras, 405, 1431
\bibitem[Andr{\'e} et al.(2016)]{and16} Andr{\'e}, P., Rev{\'e}ret, V., K{\"o}nyves, V., et al.\ 2016, \aap, 592, A54
\bibitem[Aouad et al.(2020)]{aou20} Aouad, C.~J., James, P.~A., \& Chilingarian, I.~V.\ 2020, \mnras, 496, 5211
\bibitem[Arzoumanian et al.(2011)]{arz11} Arzoumanian, D., et al. 2011, \aap, 529, L6
\bibitem[Arzoumanian et al.(2018)]{arz18} Arzoumanian, D., et al. 2018, \pasj, 70, 96
\bibitem[Arzoumanian et al.(2019)]{arz19} Arzoumanian, D., et al. 2019, \aap, 621, 42
\bibitem[Ascenso(2018)]{asc18} Ascenso, J.\ 2018, The Birth of Star Clusters, 1
\bibitem[Bally et al.(2010)]{bal10} Bally, J., Anderson, L.~D., Battersby, C., et al.\ 2010, \aap, 518, L90
\bibitem[Baug et al.(2016)]{bau16} Baug, T., Dewangan, L.~K., Ojha, D.~K., et al.\ 2016, \apj, 833, 85
\bibitem[Benjamin et al.(2003)]{ben03} Benjamin, R.~A., Churchwell, E., Babler, B.~L., et al.\ 2003, \pasp, 115, 953
\bibitem[Beuther et al.(2007)]{beu07} Beuther, H., Churchwell, E.~B., McKee, C.~F., et al.\ 2007, Protostars and Planets V, 165
\bibitem[Binney et al.(1991)]{bin91} Binney, J., Gerhard, O.~E., Stark, A.~A., et al.\ 1991, \mnras, 252, 210
\bibitem[Bisbas et al.(2017)]{bis17} Bisbas, T.~G., Tanaka, K.~E.~I., Tan, J.~C., et al.\ 2017, \apj, 850, 23
\bibitem[Blum et al.(1999)]{blu99} Blum, R.~D., Damineli, A., \& Conti, P.~S.\ 1999, \aj, 117, 1392
\bibitem[Bonnell et al.(1998)]{bon98} Bonnell, I.~A., Bate, M.~R., \& Zinnecker, H.\ 1998, \mnras, 298, 93
\bibitem[Bonnell et al.(2003)]{bon03} Bonnell, I.~A., Bate, M.~R., \& Vine, S.~G.\ 2003, \mnras, 343, 413
\bibitem[Carey et al.(2009)]{car09} Carey, S.~J., Noriega-Crespo, A., Mizuno, D.~R., et al.\ 2009, \pasp, 121, 76
\bibitem[Chabrier(2003)]{cha03} Chabrier, G.\ 2003, \pasp, 115, 763
\bibitem[Chabrier(2005)]{cha05} Chabrier, G.\ 2005, The Initial Mass Function 50 Years Later, 41
\bibitem[Churchwell et al.(2006)]{chu06} Churchwell, E., Povich, M.~S., Allen, D., et al.\ 2006, \apj, 649, 759
\bibitem[Dale et al.(2015)]{dal15} Dale, J.~E., Ercolano, B., \& Bonnell, I.~A.\ 2015, \mnras, 451, 987
\bibitem[Dame et al.(2001)]{dam01} Dame, T.~M., Hartmann, D., \& Thaddeus, P.\ 2001, \apj, 547, 792
\bibitem[Dame(2007)]{dam07} Dame, T.~M.\ 2007, \apjl, 665, L163
\bibitem[Deharveng et al.(2005)]{deh05} Deharveng, L., Zavagno, A., \& Caplan, J.\ 2005, \aap, 433, 565
\bibitem[Deharveng et al.(2009)]{deh09} Deharveng, L., Zavagno, A., Schuller, F., et al.\ 2009, \aap, 496, 177
\bibitem[Dewangan et al.(2017a)]{dew17a} Dewangan, L.~K., Ojha, D.~K., Zinchenko, I., et al.\ 2017a, \apj, 834, 22
\bibitem[Dewangan et al.(2017b)]{dew17b} Dewangan, L.~K., Ojha, D.~K., \& Zinchenko, I.\ 2017b, \apj, 851, 140
\bibitem[Dewangan et al.(2018a)]{dew18a} Dewangan, L.~K., Ojha, D.~K., Zinchenko, I., et al.\ 2018a, \apj, 861, 19
\bibitem[Dewangan et al.(2018b)]{dew18b} Dewangan, L.~K., Dhanya, J.~S., Ojha, D.~K., et al.\ 2018b, \apj, 866, 20
\bibitem[Dewangan et al.(2019a)]{dew19a} Dewangan, L.~K., Ojha, D.~K., Baug, T., et al.\ 2019a, \apj, 875, 138
\bibitem[Dewangan et al.(2019b)]{dew19b} Dewangan, L.~K., Sano, H., Enokiya, R., et al.\ 2019b, \apj, 878, 26
\bibitem[Dewangan(2017)]{dew17} Dewangan, L.~K.\ 2017, \apj, 837, 44
\bibitem[Dewangan, \& Ojha(2017)]{dewo17} Dewangan, L.~K., \& Ojha, D.~K.\ 2017, \apj, 849, 65
\bibitem[Dhanya et al.(2020)]{dha20} Dhanya, J.~S., Dewangan, L.~K., Ojha, D.~K., et al.\ 2020, \pasj, doi:10.1093/pasj/psz137
\bibitem[Dobashi et al.(2014)]{dob14} Dobashi, K., Matsumoto, T., Shimoikura, T., et al.\ 2014, \apj, 797, 58
\bibitem[Dobashi et al.(2019)]{dob19} Dobashi, K., Shimoikura, T., Katakura, S., et al.\ 2019, \pasj, 58
\bibitem[Dobbs et al.(2015)]{dob15} Dobbs, C.~L., Pringle, J.~E., \& Duarte-Cabral, A.\ 2015, \mnras, 446, 3608
\bibitem[Dobbs et al.(2020)]{dob20} Dobbs, C.~L., Liow, K.~Y., \& Rieder, S.\ 2020, \mnras, 496, L1
\bibitem[Duarte-Cabral et al.(2011)]{dua11} Duarte-Cabral, A., Dobbs, C.~L., Peretto, N., et al.\ 2011, \aap, 528, A50
\bibitem[Eisenhauer et al.(1998)]{eis98} Eisenhauer, F., Quirrenbach, A., Zinnecker, H., et al.\ 1998, \apj, 498, 278
\bibitem[Elmegreen \& Efremov(1997)]{elm97} Elmegreen, B.~G., \& Efremov, Y.~N.\ 1997, \apj, 480, 235
\bibitem[Elmegreen(1998)]{elm98} Elmegreen, B.~G.\ 1998, Origins, 150
\bibitem[Enokiya et al.(2018)]{eno18}  Enokiya, R., et al.\ 2018, \pasj, 70, 49
\bibitem[Enokiya et al.(2019)]{eno19} Enokiya, R., Torii, K., \& Fukui, Y.\ 2019, \pasj, doi:10.1093/pasj/psz119
\bibitem[Enokiya et al.(2020)]{eno20} Enokiya, R., Ohama, A., Yamada, R., et al.\ 2020, \pasj, doi:10.1093/pasj/psaa049
\bibitem[Federrath et al.(2010)]{fed10} Federrath, C., Roman-Duval, J., Klessen, R.~S., et al.\ 2010, \aap, 512, A81
\bibitem[Federrath(2016)]{fed16} Federrath, C. 2016, \mnras, 457, 375
\bibitem[Finn et al.(2019)]{fin19} Finn, M.~K., Johnson, K.~E., Brogan, C.~L., et al.\ 2019, \apj, 874, 120
\bibitem[Fujimoto et al.(2014)]{fuj14} Fujimoto, Y., Tasker, E.~J., \& Habe, A.\ 2014, \mnras, 445, L65
\bibitem[Fujimoto \& Noguchi(1990)]{fuj90} Fujimoto, M., \& Noguchi, M.\ 1990, \pasj, 42, 505
\bibitem[Fujita et al.(2019a)]{fuj19a} Fujita, S., Torii, K., Tachihara, K., et al.\ 2019a, \apj, 872, 49
\bibitem[Fujita et al.(2019b)]{fuj19b} Fujita, S., Torii, K., Kuno, N., et al.\ 2019b, \pasj, 46
\bibitem[Fujita et al.(2020a)]{fuj20a} Fujita, S., Tsutsumi, D., Ohama, A., et al.\ 2020a, \pasj, doi:10.1093/pasj/psaa005
\bibitem[Fujita et al.(2020b)]{fuj20b} Fujita, S., Sano, H., Enokiya, R., et al.\ 2020b, \pasj, doi:10.1093/pasj/psaa078
\bibitem[Fukui et al.(1999)]{fuk99} Fukui, Y., Onishi, T., Abe, R., et al.\ 1999, \pasj, 51, 751
\bibitem[Fukui et al.(2006)]{fuk06} Fukui, Y., Yamamoto, H., Fujishita, M., et al.\ 2006, Science, 314, 106
\bibitem[Fukui et al.(2013)]{fuk13}  Fukui, Y., et al.\ 2013, \apj, 780, 36
\bibitem[Fukui et al.(2014)]{fuk14} Fukui, Y., Ohama, A., Hanaoka, N., et al.\ 2014, \apj, 780, 36
\bibitem[Fukui et al.(2015)]{fuk15} Fukui, Y., Harada, R., Tokuda, K., et al.\ 2015, \apjl, 807, L4
\bibitem[Fukui et al.(2016)]{fuk16} Fukui, Y., Torii, K., Ohama, A., et al.\ 2016, \apj, 820, 26
\bibitem[Fukui et al.(2017)]{fuk17} Fukui, Y., Tsuge, K., Sano, H., et al.\ 2017, \pasj, 69, L5
\bibitem[Fukui et al.(2018a)]{fuk18a} Fukui, Y., Torii, K., Hattori, Y., et al.\ 2018a, \apj, 859, 166
\bibitem[Fukui et al.(2018b)]{fuk18b} Fukui, Y., Kohno, M., Yokoyama, K., et al.\ 2018b, \pasj, 70, S41
\bibitem[Fukui et al.(2018c)]{fuk18c} Fukui, Y., Kohno, M., Yokoyama, K., et al.\ 2018c, \pasj, 70, S44
\bibitem[Fukui et al.(2018d)]{fuk18d} Fukui, Y., Ohama, A., Kohno, M., et al.\ 2018d, \pasj, 70, S46
\bibitem[Fukui et al.(2019)]{fuk19} Fukui, Y., Tokuda, K., Saigo, K., et al.\ 2019, \apj, 886, 14
\bibitem[Fukui et al.(2020)]{fuk20} Fukui, Y., Inoue, T., Hayakawa, T., et al.\ 2020, \pasj, doi:10.1093/pasj/psaa079
\bibitem[Furukawa et al.(2009)]{fur09} Furukawa, N., Dawson, J. R., Ohama, A., Kawamura, A., Mizuno, N., Onishi, T., \& Fukui, Y. 2009, ApJL, 696, L115 
\bibitem[Galv{\'a}n-Madrid et al.(2013)]{gal13} Galv{\'a}n-Madrid, R., Liu, H.~B., Zhang, Z.-Y., et al.\ 2013, \apj, 779, 121
\bibitem[Gilbert \& Graham(2007)]{gil07} Gilbert, A.~M., \& Graham, J.~R.\ 2007, \apj, 668, 168
\bibitem[Gilden(1984)]{gil84} Gilden, D.~L.\ 1984, \apj, 279, 335
\bibitem[Ginsburg et al.(2016)]{gin16} Ginsburg, A., Henkel, C., Ao, Y., et al.\ 2016, \aap, 586, A50
\bibitem[Ginsburg et al.(2018)]{gin18} Ginsburg, A., Bally, J., Barnes, A., et al.\ 2018, \apj, 853, 171
\bibitem[Gong et al.(2017)]{gon17} Gong, Y., Fang, M., Mao, R., et al.\ 2017, \apjl, 835, L14
\bibitem[Gutermuth et al.(2008)]{gut08} Gutermuth, R.~A., Myers, P.~C., Megeath, S.~T., et al.\ 2008, \apj, 674, 336
\bibitem[Habe \& Ohta(1992)]{hab92} Habe, A., \& Ohta, K.\ 1992, \pasj, 44, 203
\bibitem[Hacar et al.(2017)]{hac17} Hacar, A., Tafalla, M., \& Alves, J.\ 2017, \aap, 606, A123
\bibitem[Harada et al.(2019)]{har19} Harada, R., Onishi, T., Tokuda, K., et al.\ 2019, \pasj, 71, 44
\bibitem[Hasegawa et al.(1994)]{has94} Hasegawa, T., Sato, F., Whiteoak, J.~B., et al.\ 1994, \apjl, 429, L77
\bibitem[Haworth et al.(2015)]{haw15} Haworth, T.~J., Shima, K., Tasker, E.~J., et al.\ 2015, \mnras, 454, 1634
\bibitem[Haworth et al.(2018)]{haw18} Haworth, T.~J., Glover, S.~C.~O., Koepferl, C.~M., et al.\ 2018, New Astronomy Reviews, 82, 1
\bibitem[Hayashi et al.(2018)]{hay18} Hayashi, K., Sano, H., Enokiya, R., et al.\ 2018, \pasj, 70, S48
\bibitem[Hayashi et al.(2020)]{hay20} Hayashi, K., Yoshiike, S., Enokiya, R., et al.\ 2020, \pasj, doi:10.1093/pasj/psaa054
\bibitem[Heitsch et al.(2008)]{hei08} Heitsch, F., Hartmann, L. W., \& Burkert, A. 2008, \apj, 683, 786
\bibitem[Hennebelle et al.(2008)]{hen08} Hennebelle, P., Banerjee, R., V\'azquez-Semadeni, E., Klessen, R. S., \& Audit, E.  2008, \aap, 486, L43
\bibitem[Henshaw et al.(2013)]{hen13} Henshaw, J.~D., Caselli, P., Fontani, F., et al.\ 2013, \mnras, 428, 3425
\bibitem[Higuchi et al.(2014)]{hig14} Higuchi, A.~E., Chibueze, J.~O., Habe, A., et al.\ 2014, \aj, 147, 141
\bibitem[Hillenbrand(1997)]{hil97} Hillenbrand, L.~A.\ 1997, \aj, 113, 1733
\bibitem[Harayama et al.(2008)]{hir08} Harayama, Y., Eisenhauer, F., \& Martins, F.\ 2008, \apj, 675, 1319
\bibitem[Homeier, \& Alves(2005)]{hom05} Homeier, N.~L., \& Alves, J.\ 2005, \aap, 430, 481
\bibitem[Hosokawa \& Inutsuka(2005)]{hos05} Hosokawa, T., \& Inutsuka, S.-. ichiro .\ 2005, \apj, 623, 917
\bibitem[Hunter et al.(1986)]{hun86} Hunter, J. H., Sandford, M. T., Whitaker, R. W., \& Klein, R. I. 1987, \apj, 305, 309
\bibitem[Hunter et al.(1998)]{hun98} Hunter, D.~A., Wilcots, E.~M., van Woerden, H., et al.\ 1998, \apjl, 495, L47
\bibitem[Inoue et al.(2018)]{ino18} Inoue, T., Hennebelle, P., Fukui, Y. et al. 2018, \pasj, 70, 53
\bibitem[Inoue \& Fukui (2013)]{ino13} Inoue, T., \& Fukui, Y. 2013, \apj, 774, 31
\bibitem[Inoue \& Inutsuka (2012)]{ino12} Inoue, T., \& Inutsuka, S. 2012, \apj, 759, 35
\bibitem[Inoue \& Inutsuka(2008)]{ino08} Inoue, T., \& Inutsuka, S. 2008, \apj, 687, 303
\bibitem[Inutsuka et al.(2015)]{inu15} Inutsuka, S.-. ichiro ., Inoue, T., Iwasaki, K., et al.\ 2015, \aap, 580, A49
\bibitem[Kakiuchi et al.(2018)]{kak18} Kakiuchi, K., Suzuki, T.~K., Fukui, Y., et al.\ 2018, \mnras, 476, 5629
\bibitem[Kandori et al.(2020)]{kan20} Kandori, R. et al. 2020, \apj, 892, 128
\bibitem[Kawamura et al.(2009)]{kaw09} Kawamura, A., Mizuno, Y., Minamidani, T., et al.\ 2009, \apjs, 184, 1
\bibitem[Kennicutt(1998)]{ken98} Kennicutt, R.~C.\ 1998, \apj, 498, 541
\bibitem[Kimura \& Tosa(1996)]{kim96} Kimura, T., \& Tosa, M.\ 1996, \aap, 308, 979
\bibitem[Kinoshita et al.(2020)]{kin20} Kinoshita, S.~W., Nakamura, F., Nguyen-Luong, Q., et al.\ 2020, \pasj, doi:10.1093/pasj/psaa053
\bibitem[Kobayashi et al.(2018)]{kob18} Kobayashi, M.~I.~N., Kobayashi, H., Inutsuka, S.-. ichiro ., et al.\ 2018, \pasj, 70, S59
\bibitem[Kohno et al.(2018a)]{koh18a} Kohno, M., Torii, K., Tachihara, K., et al.\ 2018a, \pasj, 70, S50
\bibitem[Kohno et al.(2018b)]{koh18b} Kohno, M., Tachihara, K., Fujita, S., et al.\ 2018b, \pasj, 126
\bibitem[Kohno et al.(2020)]{koh20} Kohno, M., Tachihara, K., Torii, K., et al.  2020, PASJ in press (doi:10.1093/pasj/psaa015)
\bibitem[Koo et al.(1994)]{koo94} Koo, B.-C., Lee, Y., Fuller, G.~A., et al.\ 1994, \apj, 429, 233
\bibitem[Koyama \& Inutsuka(2002)]{koy02} Koyama, H. \& Inutsuka, S. 2002, \apj, 564, L97
\bibitem[Kroupa(2001)]{kro01} Kroupa, P.\ 2001, \mnras, 322, 231
\bibitem[Krumholz et al.(2009)]{kru09} Krumholz, M.~R., Klein, R.~I., McKee, C.~F., et al.\ 2009, Science, 323, 754
\bibitem[Krumholz et al.(2012)]{kru12} Krumholz, M.~R., Klein, R.~I., \& McKee, C.~F.\ 2012, \apj, 754, 71
\bibitem[Kudryavtseva et al.(2012)]{kud12} Kudryavtseva, N., Brandner, W., Gennaro, M., et al.\ 2012, \apjl, 750, L44
\bibitem[Kuwahara et al.(2020)]{kuw20} Kuwahara, S., Torii, K., Mizuno, N., et al.\ 2020, \pasj, doi:10.1093/pasj/psaa017
\bibitem[Lattanzio et al.(1985)]{lat85} Lattanzio, J.~C., Monaghan, J.~J., Pongracic, H., et al.\ 1985, \mnras, 215, 125
\bibitem[Lattanzio \& Henriksen(1988)]{lat88} Lattanzio, J.~C., \& Henriksen, R.~N.\ 1988, \mnras, 232, 565
\bibitem[Lee et al.(1997)]{lee97} Lee, H.-G., Koo, B.-C., Park, Y.-S., et al.\ 1997, \pasj, 49, 639
\bibitem[Lelli et al.(2014)]{lel14} Lelli, F., Verheijen, M., \& Fraternali, F.\ 2014, \aap, 566, A71
\bibitem[Li et al.(2018)]{li18} Li, Q., Tan, J.~C., Christie, D., et al.\ 2018, \pasj, 70, S56
\bibitem[Lim et al.(2020)]{lim20} Lim, W., Nakamura, F., Wu, B., et al.\ 2020, \pasj, doi:10.1093/pasj/psaa035
\bibitem[Liow \& Dobbs(2020)]{lio20} Liow, K.~Y. \& Dobbs, C.\ 2020, \mnras, doi:10.1093/mnras/staa2857
\bibitem[Lockman et al.(2012)]{loc12} Lockman, F.~J., Free, N.~L., \& Shields, J.~C.\ 2012, \aj, 144, 52
\bibitem[Longmore et al.(2013a)]{lon13a} Longmore, S.~N., Bally, J., Testi, L., et al.\ 2013a, \mnras, 429, 987
\bibitem[Longmore et al.(2013b)]{lon13b} Longmore, S.~N., Kruijssen, J.~M.~D., Bally, J., et al.\ 2013b, \mnras, 433, L15
\bibitem[Looney et al.(2006)]{loo06} Looney, L.~W., Wang, S., Hamidouche, M., et al.\ 2006, \apj, 642, 330
\bibitem[Loren(1976)]{lor76} Loren, R.~B.\ 1976, \apj, 209, 466
\bibitem[Loren(1977)]{lor77} Loren, R.~B.\ 1977, \apj, 218, 716
\bibitem[Luisi et al.(2017)]{lui17} Luisi, M., Anderson, L.~D., Balser, D.~S., et al.\ 2017, \apj, 849, 117
\bibitem[McKee \& Ostriker(2007)]{mck07} McKee, C.~F., \& Ostriker, E.~C.\ 2007, \araa, 45, 565
\bibitem[Menten et al.(2005)]{men05} Menten, K.~M., Pillai, T., \& Wyrowski, F.\ 2005, Massive Star Birth: A Crossroads of Astrophysics, 227, 23
\bibitem[McKee \& Tan(2003)]{mck03} McKee, C.~F., \& Tan, J.~C.\ 2003, \apj, 585, 850
\bibitem[Moffat et al.(2004)]{mof04} Moffat, A.~F.~J., Poitras, V., Marchenko, S.~V., et al.\ 2004, \aj, 128, 2854
\bibitem[Molinari et al.(2011)]{mol11} Molinari, S., Bally, J., Noriega-Crespo, A., et al.\ 2011, \apjl, 735, L33
\bibitem[Motte et al.(2003)]{mot03} Motte, F., Schilke, P., \& Lis, D.~C.\ 2003, \apj, 582, 277 
\bibitem[Motte et al.(2014)]{mot14} Motte, F., Nguy{\^e}n Luong, Q., Schneider, N., et al.\ 2014, \aap, 571, A32
\bibitem[Motte et al.(2018a)]{mot18a} Motte, F., Nony, T., Louvet, F., et al.\ 2018a, Nature Astronomy, 2, 478
\bibitem[Motte et al.(2018b)]{mot18b} Motte, F., Bontemps, S., \& Louvet, F.\ 2018b, \araa, 56, 41
\bibitem[Muraoka et al.(2020)]{mur20} Muraoka, K., Kondo, H., Tokuda, K., et al.\ 2020, arXiv:2009.05804
\bibitem[Murray \& Rahman(2010)]{mur10} Murray, N. \& Rahman, M.\ 2010, \apj, 709, 424
\bibitem[Nagasawa \& Miyama(1987)]{nag87} Nagasawa, M., \& Miyama, S.~M.\ 1987, Progress of Theoretical Physics, 78, 1250
\bibitem[Nakamura et al.(2012)]{nak12} Nakamura, F., Miura, T., Kitamura, Y., et al.\ 2012, \apj, 746, 25
\bibitem[Nakamura et al.(2014)]{nak14} Nakamura, F., Sugitani, K., Tanaka, T., et al.\ 2014, \apjl, 791, L23
\bibitem[Nguyen Luong et al.(2011)]{ngu11} Nguyen Luong, Q., Motte, F., Schuller, F., et al.\ 2011, \aap, 529, A41
\bibitem[Nishimura et al.(2018)]{nis18} Nishimura, A., Minamidani, T., Umemoto, T., et al.\ 2018, \pasj, 70, S42
\bibitem[Nishimura et al.(2020)]{nis20} Nishimura, A., Fujita, S., Kohno, M., et al.\ 2020, arXiv:2008.05939
\bibitem[Ntormousi et al.(2016)]{nto16} Ntormousi, E., Hennebelle, P., Andr\'e, P., \& Masson, J. 2016, \aap, 589, 24
\bibitem[Ohama et al.(2010)]{oha10}Ohama, A., et al. 2010, ApJ, 709, 975
\bibitem[Ohama et al.(2018a)]{oha18a} Ohama, A., Kohno, M., Hasegawa, K., et al.\ 2018a, \pasj, 70, S45
\bibitem[Ohama et al.(2018b)]{oha18b} Ohama, A., Kohno, M., Fujita, S., et al.\ 2018b, \pasj, 70, S47
\bibitem[Oort(1954)]{oor54} Oort, J.~H.\ 1954, \bain, 12, 177
\bibitem[Ostriker 1964]{ost64} Ostriker, J. 1964, \apj, 140, 1056
\bibitem[Peretto et al.(2013)]{per13} Peretto, N., Fuller, G. A., Duarte-Cabral, A. et al. 2013, \aap, 555, 112
\bibitem[Peretto et al.(2014)]{per14} Peretto, N., Fuller, G. A., Andr\'e, P. 2014, \aap, 561, 83
\bibitem[Portegies Zwart et al.(2010)]{por10} Portegies Zwart, S.~F., McMillan, S.~L.~W., \& Gieles, M.\ 2010, \araa, 48, 431
\bibitem[Renaud et al.(2015)]{ren15} Renaud, F., Bournaud, F., \& Duc, P.-A.\ 2015, \mnras, 446, 2038
\bibitem[Rho et al.(2006)]{rho06} Rho, J., Reach, W.~T., Lefloch, B., et al.\ 2006, \apj, 643, 965
\bibitem[Roelfsema et al.(1989)]{roe89} Roelfsema, P.~R., Goss, W.~M., \& Geballe, T.~R.\ 1989, \aap, 222, 247
\bibitem[Saigo et al.(2017)]{sai17} Saigo, K., Onishi, T., Nayak, O., et al.\ 2017, \apj, 835, 108
\bibitem[Sakre et al.(2020)]{sak20} Sakre, N., Habe, A., Pettitt, A.~R., et al.\ 2020, \pasj, doi:10.1093/pasj/psaa059
\bibitem[Sancisi(1974)]{san74} Sancisi, R.\ 1974, Galactic Radio Astronomy, 60, 115
\bibitem[Sano et al.(2018)]{san18} Sano, H., Enokiya, R., Hayashi, K., et al.\ 2018, \pasj, 70, S43
\bibitem[Sano et al.(2020)]{san20} Sano, H., Tsuge, K., Tokuda, K., et al.\ 2020, \pasj, doi:10.1093/pasj/psaa045  
\bibitem[Schmidt(1959)]{sch59} Schmidt, M.\ 1959, \apj, 129, 24
\bibitem[Serabyn et al.(1993)]{ser93} Serabyn, E., Guesten, R., \& Schulz, A.\ 1993, \apj, 413, 571
\bibitem[Shima et al.(2018)]{shi18} Shima, K., Tasker, E.~J., Federrath, C., et al.\ 2018, \pasj, 70, S54
\bibitem[Smith(1980)]{smi80} Smith, J.\ 1980, \apj, 238, 842
\bibitem[Sofue et al.(2019)]{sof19} Sofue, Y., Kohno, M., Torii, K., et al.\ 2019, \pasj, 71, S1
\bibitem[Solomon \& Sanders(1980)]{sol80} Solomon, P.~M., \& Sanders, D.~B.\ 1980, Giant Molecular Clouds in the Galaxy, 41
\bibitem[Spitzer(1968)]{spi68} Spitzer, L.\ 1968, Nebulae and Interstellar Matter, 1
\bibitem[Stod\'olkiewicz(1963)]{sto63} Stod\'olkiewicz, J. S. 1963, AcA, 13, 30
\bibitem[Stolte et al.(2006)]{sto06} Stolte, A., Brandner, W., Brandl, B., et al.\ 2006, \aj, 132, 253
\bibitem[Stolte et al.(2008)]{sto08} Stolte, A., Ghez, A.~M., Morris, M., et al.\ 2008, \apj, 675, 1278
\bibitem[Stolte et al.(2014)]{sto14} Stolte, A., Hu{\ss}mann, B., Morris, M.~R., et al.\ 2014, \apj, 789, 115
\bibitem[Stone(1970a)]{sto70a} Stone, M.~E.\ 1970, \apj, 159, 277
\bibitem[Stone(1970b)]{sto70b} Stone, M.~E.\ 1970, \apj, 159, 293
\bibitem[Stutz \& Gould(2016)]{stu16} Stutz, A.~M., \& Gould, A.\ 2016, \aap, 590, A2
\bibitem[Suwannajak et al.(2014)]{suw14} Suwannajak, C., Tan, J.~C., \& Leroy, A.~K.\ 2014, \apj, 787, 68
\bibitem[Suzuki et al.(2015)]{suz15} Suzuki, T.~K., Fukui, Y., Torii, K., et al.\ 2015, \mnras, 454, 3049
\bibitem[Tachihara et al.(2018)]{tac18} Tachihara, K., Gratier, P., Sano, H., et al.\ 2018, \pasj, 70, S52
\bibitem[Takahira et al.(2014)]{tak14} Takahira, K., Tasker, E.~J., \& Habe, A.\ 2014, \apj, 792, 63
\bibitem[Takahira et al.(2018)]{tak18} Takahira, K., Shima, K., Habe, A., et al.\ 2018, \pasj, 70, S58
\bibitem[Tan(2000)]{tan00} Tan, J.~C.\ 2000, \apj, 536, 173
\bibitem[Tanaka(2018)]{tan18} Tanaka, K.\ 2018, \apj, 859, 86
\bibitem[Tasker \& Tan(2009)]{tas09} Tasker, E.~J., \& Tan, J.~C.\ 2009, \apj, 700, 358
\bibitem[Tasker(2011)]{tas11} Tasker, E.~J.\ 2011, \apj, 730, 11
\bibitem[Tokuda et al.(2019)]{tok19} Tokuda, K., Fukui, Y., Harada, R., et al.\ 2019, \apj, 886, 15
\bibitem[Tokuda et al.(2020)]{tok20} Tokuda, K., Muraoka, K., Kondo, H., et al.\ 2020, \apj, 896, 36
\bibitem[Tomisaka (2014)]{tom14} Tomisaka, K. 2014, \apj, 785, 24
\bibitem[Torii et al.(2010)]{tor10} Torii, K., Kudo, N., Fujishita, M., et al.\ 2010, \pasj, 62, 675
\bibitem[Torii et al.(2011)]{tor11} Torii, K., Enokiya, R., Sano, H., et al.\ 2011, \apj, 738, 46
\bibitem[Torii et al.(2015)]{tor15} Torii, K., Hasegawa, K., Hattori, Y., et al.\ 2015, \apj, 806, 7
\bibitem[Torii et al.(2017a)]{tor17a} Torii, K., Hattori, Y., Hasegawa, K., et al.\ 2017a, \apj, 835, 142
\bibitem[Torii et al.(2017b)]{tor17b} Torii, K., Hattori, Y., Hasegawa, K., et al.\ 2017b, \apj, 840, 111
\bibitem[Torii et al.(2018a)]{tor18a} Torii, K., Fujita, S., Matsuo, M., et al.\ 2018a, \pasj, 70, S51
\bibitem[Torii et al.(2018b)]{tor18b} Torii, K., Hattori, Y., Matsuo, M., et al.\ 2018b, \pasj, 121
\bibitem[Torii et al.(2020)]{tor20} Torii, K., Tokuda, K., Tachihara, K., et al.\ 2019, arXiv:1907.07358
\bibitem[Tsuboi et al.(2015a)]{tsu15a} Tsuboi, M., Miyazaki, A., \& Uehara, K.\ 2015a, \pasj, 67, 90
\bibitem[Tsuboi et al.(2015b)]{tsu15b} Tsuboi, M., Miyazaki, A., \& Uehara, K.\ 2015b, \pasj, 67, 109
\bibitem[Tsuge et al.(2019)]{tsu19} Tsuge, K., Sano, H., Tachihara, K., et al.\ 2019, \apj, 871, 44
\bibitem[Tsuge et al.(2020)]{tsu20a} Tsuge, K., Fukui, Y., Tachihara, K., et al.\ 2020a, \pasj, doi:10.1093/pasj/psaa033
\bibitem[Tsuge et al.(2020b)]{tsu20b} Tsuge, K., Tachihara, K., Fukui, Y., et al.\ 2020b, arXiv e-prints, arXiv:2005.04075
\bibitem[Uehara et al.(2019)]{ueh19} Uehara, K., Tsuboi, M., Kitamura, Y., et al.\ 2019, \apj, 872, 121
\bibitem[Vaidya et al.(2013)]{vai13} Vaidya, B., Hartquist, T. W., \& Falle, S. A. E. G. 2013, \mnras, 433, 1258
\bibitem[Vallee, \& Avery(1990)]{val90} Vallee, J.~P., \& Avery, L.~W.\ 1990, \aap, 233, 553
\bibitem[van den Bergh \& Tammann(1991)]{van91} van den Bergh, S. \& Tammann, G.~A.\ 1991, \araa, 29, 363
\bibitem[V\'azquez-Semadeni et al.(2007)]{vaz07} V\'azquez-Semadeni, E. et al. 2007, \apj, 657, 870
\bibitem[Walawender et al.(2008)]{wal08} Walawender, J., Bally, J., Francesco, J.~D., et al.\ 2008, Handbook of Star Forming Regions, Volume I, 346
\bibitem[Wang et al.(2004)]{wan04} Wang, J.-J., Chen, W.-P., Miller, M., et al.\ 2004, \apjl, 614, L105
\bibitem[Whitmore et al.(2010)]{whi10} Whitmore, B.~C., Chandar, R., Schweizer, F., et al.\ 2010, \aj, 140, 75
\bibitem[Whitmore et al.(2014)]{whi14} Whitmore, B.~C., Brogan, C., Chandar, R., et al.\ 2014, \apj, 795, 156
\bibitem[Whitmore \& Schweizer(1995)]{whi95} Whitmore, B.~C., \& Schweizer, F.\ 1995, \aj, 109, 960
\bibitem[Whitney et al.(2004)]{whi04} Whitney, B.~A., Indebetouw, R., Babler, B.~L., et al.\ 2004, \apjs, 154, 315
\bibitem[Whitworth et al.(1994)]{whi94a} Whitworth, A. P., Bhattal, A. S., Chapman, S. J., Disney, M. J., \& Turner, J. A. 1994a, \mnras, 268, 291
\bibitem[Whitworth et al.(1994)]{whi94b} Whitworth, A. P., Bhattal, A. S., Chapman, S. J., Disney, M. J., \& Turner, J. A. 1994b, \aap, 290, 421
\bibitem[Whitworth et al.(2018)]{whi18} Whitworth, A., Lomax, O., Balfour, S., et al.\ 2018, \pasj, 70, S55
\bibitem[Wilson et al.(2000)]{wil00} Wilson, C.~D., Scoville, N., Madden, S.~C., et al.\ 2000, \apj, 542, 120
\bibitem[Wolfire \& Cassinelli(1987)]{wol87} Wolfire, M.~G., \& Cassinelli, J.~P.\ 1987, \apj, 319, 850
\bibitem[Wu et al.(2018)]{wu18} Wu, B., Tan, J.~C., Nakamura, F., et al.\ 2018, \pasj, 70, S57
\bibitem[Xu et al.(2020)]{xu20} Xu, J.-L., Xu, Y., Jiang, P., et al.\ 2020, \apjl, 893, L5
\bibitem[Xue, \& Wu(2008)]{xue08} Xue, R., \& Wu, Y.\ 2008, \apj, 680, 446
\bibitem[Zavagno et al.(2010)]{zav10} Zavagno, A., Russeil, D., Motte, F., et al.\ 2010, \aap, 518, L81
\bibitem[Zinnecker \& Yorke(2007)]{zin07} Zinnecker, H., \& Yorke, H.~W.\ 2007, \araa, 45, 481
\end{thebibliography}
\end{document}